\documentclass[trackchanges,twocolumn]{aastex7}

\usepackage{hyperref}
\usepackage{makecell}
\usepackage{booktabs}
\usepackage{listings}
\usepackage{comment}
\usepackage{ulem}
\usepackage{multirow}

\newcommand{\vadd}[1]{\textcolor{black}{#1}}
\newcommand{\jtadd}[1]{\textcolor{black}{#1}}

\begin{document}

\title{Figuring Out Gas \& Galaxies In Enzo (FOGGIE). XIV. The Observability of Emission from Accretion and Feedback in the Circumgalactic Medium with Current and Future Instruments}
\shorttitle{FOGGIE XIV. CGM Emission Observability}

\author[orcid=0009-0000-7559-7962,sname='Saeedzadeh']{Vida Saeedzadeh}
\affiliation{Center for Astrophysical Sciences, William H.\ Miller III Department of Physics \& Astronomy, Johns Hopkins University, 3400 N.\ Charles Street, Baltimore, MD 21218}
\email[show]{vsaeedz1@jh.edu}  

\author[0000-0002-7982-412X]{Jason Tumlinson}
\affiliation{Space Telescope Science Institute, 3700 San Martin Dr., Baltimore, MD 21218}
\affiliation{Center for Astrophysical Sciences, William H.\ Miller III Department of Physics \& Astronomy, Johns Hopkins University, 3400 N.\ Charles Street, Baltimore, MD 21218}
\email{tumlinson@stsci.edu}

\author[0000-0003-1455-8788]{Molly S.\ Peeples}
\affiliation{Space Telescope Science Institute, 3700 San Martin Dr., Baltimore, MD 21218}
\affiliation{Center for Astrophysical Sciences, William H.\ Miller III Department of Physics \& Astronomy, Johns Hopkins University, 3400 N.\ Charles Street, Baltimore, MD 21218}
\email{molly@stsci.edu}

\author[0000-0002-2786-0348]{Brian W.\ O'Shea}
\affiliation{Department of Computational Mathematics, Science, and Engineering, Michigan State University, East Lansing, MI, US}
\affiliation{Department of Physics and Astronomy, Michigan State University, East Lansing, MI, US}
\affiliation{Facility for Rare Isotope Beams, Michigan State University, East Lansing, MI 48824, USA}
\affiliation{Institute for Cyber-Enabled Research, 567 Wilson Road, Michigan State University, East Lansing, MI 48824}
\email{bwoshea@msu.edu}

\author[0000-0003-1785-8022]{Cassandra Lochhaas}
\affiliation{Center for Astrophysics, Harvard \& Smithsonian, 60 Garden St., Cambridge, MA 02138}
\affiliation{NASA Hubble Fellow}
\email{clochhaas@cfa.harvard.edu}

\author[0000-0002-0646-1540]{Lauren Corlies}
\affiliation{University of California Observatories/Lick Observatory, Mount Hamilton, CA 95140, USA}
\email{lcorlies@ucsc.edu}

\author[orcid=0000-0001-7813-0268]{Cameron W.\ Trapp}
\affiliation{Center for Astrophysical Sciences, William H.\ Miller III Department of Physics \& Astronomy, Johns Hopkins University, 3400 N.\ Charles Street, Baltimore, MD 21218}
\email{ctrapp2@jhu.edu}  

\author[0000-0002-6804-630X]{Britton D.\ Smith}
\affiliation{Institute for Astronomy, University of Edinburgh, Royal Observatory, EH9 3HJ, UK}
\email{Britton.Smith@ed.ac.uk}

\author[0000-0002-0355-0134]{Jessica K.\ Werk}
\affil{University of Washington, Department of Astronomy, Seattle, WA 98195, USA}
\email{jwerk@uw.edu}

\author[0000-0003-4804-7142]{Ayan Acharyya}
\affiliation{INAF - Astronomical Observatory of Padova, vicolo dell’Osservatorio 5, IT-35122 Padova, Italy}
\email{ayan.acharyya@inaf.it}

\author[0000-0001-7472-3824]{Ramona Augustin}
\affiliation{Leibniz-Institut f{\"u}r Astrophysik Potsdam (AIP), An der Sternwarte 16, 14482 Potsdam, Germany}
\email{raugustin@aip.de}

\author[0000-0003-0724-4115]{Andrew J.\ Fox}
\affiliation{AURA for ESA, Space Telescope Science Institute, 3700 San Martin Dr., Baltimore, MD 21218}
\email{afox@stsci.edu}

\author[0000-0001-9158-0829]{Nicolas Lehner}
\affiliation{Department of Physics, University of Notre Dame, Notre Dame, IN 46556}
\email{nlehner@nd.edu}

\author[0000-0002-1685-5818]{Anna C.\ Wright}
\affiliation{Center for Computational Astrophysics, Flatiron Institute, 162 Fifth Avenue, New York, NY 10010}
\email{awright@flatironinstitute.org}

\begin{abstract}

Observing the circumgalactic medium (CGM) in emission lines from ionized gas enables direct mapping of its spatial and kinematic structure, offering new insight into the gas flows that regulate galaxy evolution. Using the high-resolution Figuring Out Gas \& Galaxies In Enzo (FOGGIE) simulations, we generate mock emission-line maps for six Milky Way–mass halos. Different \vadd{lines (e.g., H$\alpha$, \ion{O}{6})} trace distinct CGM phases and structures, highlighting the need for multi-species observations. We quantify the observable CGM mass fraction as a function of instrument spatial resolution and surface brightness sensitivity, finding that sensitivity is the dominant factor limiting detectability across all ions. At fixed sensitivity, higher spatial resolution reveals more structures; at fixed spatial resolution, higher sensitivity recovers a higher percentage of the total mass. We construct emissivity-weighted projected velocity maps and comparing line-of-sight velocities between \vadd{emission lines}. \ion{O}{6} shows the largest kinematic deviation from \vadd{H$\alpha$}, while \ion{Mg}{2} and \ion{Si}{2} most closely follow \vadd{H$\alpha$} velocities. Distinguishing these phases out to 50 kpc from the galaxy center requires spectral resolution better than 30 km s$^{-1}$ for most ion pairs. Separating inflowing from outflowing gas based on projected kinematics requires high spectral resolution: at 30 km s$^{-1}$, more than 80\% of gas above the emission detection threshold can be distinguished kinematically, but this fraction drops to $< 40$\% with a resolution of 200 km s$^{-1}$. Our results provide predictions for future instruments, showing that recovering the multiphase structure and kinematics of the CGM in emission will require both high sensitivity and fine kinematic resolution. 

\end{abstract}


\section{Introduction} 
\label{sec:intro}

The circumgalactic medium (CGM) is the region around galaxies through which they acquire gas for star formation, the destination for outflows from supernovae and AGN, and the site of recycling of recent outflows into future accretion \citep{Tumlinson2017, 2023ARA&A..61..131F}. Since it was first noticed in the \jtadd{1950s \citep{1956ApJ...124...20S}}, the CGM has been studied intensively to understand its role in fueling galactic accretion and star formation. 
Primarily from absorption-line observations, 
we have good evidence that the CGM bears a significant fraction of galactic baryons \citep{Werk2014} and metals \citep{2014ApJ...786...54P}: \vadd{(a)} the material observed is usually consistent with being bound to the host galaxy (at least at low redshift, {\jtadd{for $L^*$ galaxies and more massive systems}) \citep{2016ApJ...833..259B}; \vadd{(b)} there is a range of metallicities indicating diverse origins in accretion, feedback, and recycling \citep{Lehner2013,prochaska2017}; and \vadd{(c)} quenched galaxies show a distinct behavior with less overall gas and a different range of ionization states \citep{2019MNRAS.484.2257Z}. 

Most of the detailed information we have about the CGM comes from absorption-line measurements using distant QSOs (or sometimes stars) as background sources. These techniques date back to the 1960s, shortly after the discovery of QSOs themselves \citep{1969ApJ...156L..63B}. Most of these absorber observations were obtained without prior knowledge of the foreground galaxy's redshift, mass, or separation from the sightline. Only in the last 15 years has it become possible, with large-area galaxy surveys and efficient UV/optical spectrographs, to choose a particular set of galaxies and efficiently probe the CGM as a function of mass, redshift and other properties. These observations excel at measuring CGM gas \jtadd{column densities} even down to very low mass density by combining sensitive spectroscopy from ground and space-based telescopes. 

Yet even state-of-the-art surveys of the CGM using the largest available samples of QSO absorbers \citep{Tumlinson2013, Nielsen2013} are limited to tens or hundreds of objects, and in most cases each galaxy collected in a sample is probed by only one background QSO. There are exceptions in the nearby Universe, where galaxies subtending large areas on the sky can be probed by $> 10$ background objects  \citep[e.g.;][]{Bowen2016,krishnarao2022}, with up to $\sim 50$ in the case of the nearby M31 \citep{Lehner2020, Lehner2025}. But generally, the maps of the CGM that we can compare to theoretical models or hydrodynamical simulations are highly averaged - over azimuthal bands in radius, and usually across galaxy mass. Careful attention to galaxy orientation with high-quality imaging has allowed some groups to study the azimuthal dependence of CGM absorption, which shows a preference for the minor axis of disks in a way that suggests galactic winds \citep{2012ApJ...760L...7K, 2014ApJ...784..108B, 2018ApJ...866...36L, 2021MNRAS.502.3733W, 2025ApJ...986..190S}. The resulting averages are informative, but they lose the rich 3D structure and dynamic balance that we expect is really there. 

To make progress beyond the 1D picture, we must strive to obtain projected 2D distributions of the real 3D CGM. There are several approaches to this. Strong gravitational lensing can provide highly-extended background galaxies that are useful for probing the CGM of foreground galaxies in 2D \citep[e.g.,][]{2020MNRAS.491.4442L, Augustin2021, Bordoloi2022Nature}. However the most promising technique of all is to map the CGM of galaxies with emission from the gas itself, making background sources of all kinds irrelevant and permitting 2D maps of the CGM gas of individual galaxies with \vadd{much} less ambiguity, with spectra providing a third dimension. Recognizing the potential of emission observations to advance our understanding of the baryon cycle through galaxies, the Astro2020 Decadal Survey listed ``Mapping the Circumgalactic and Intergalactic Medium in Emission" as one of the key Discovery Areas for this decade. 

Many pioneering observations of CGM emission already exist from UV and optical measurements. The first feasible spectroscopic searches for circumgalactic \ion{O}{6} were made with NASA's Far Ultraviolet Spectroscopic Explorer (FUSE), which launched in 1999. In the first search for emission around external galaxies with FUSE, \cite{2003ApJ...591..821O} detected \ion{O}{6} emission above the disk of the edge-on spiral NGC 4631, coinciding with emerging feedback seen in H$\alpha$ imaging. \cite{2021ApJ...916....7C} revisited these FUSE data and reported four \ion{O}{6} detections in NGC~4631 and NGC~891. With archival FUSE data, \cite{2024AJ....168...11K} were able to detect \ion{O}{6} emission in the prototypical M82 galactic superwind. FUSE also detected \ion{O}{6} emission in a local ``Lyman break'' galaxy Haro 11 \citep{2007ApJ...668..891G}. The Solar Blind Channel on the Hubble Advanced Camera for Surveys (ACS/SBC) enables a differential imaging technique for \ion{O}{6} by subtracting two filter bandpasses that cover \ion{O}{6} and the nearby continuum. In their pioneering deployment of this method, \cite{Hayes2016} identified \ion{O}{6} emission extended over $\sim 20$ kpc surrounding the disk of a star-bursting galaxy. With similar techniques, \cite{2024AJ....168...11K} mapped \ion{O}{6} in the Makani superwind and related it to lower ion emission mapped in the optical with Keck Cosmic Web Imager. \cite{kcwi2018} and \cite{2018Natur.562..229W} used VLT/MUSE to find that nearly every high-redshift galaxy has a Ly$\alpha$ halo. 

In addition to these CGM/halo emission line detections with existing instrumentation, there are now operating and planned instruments that are purpose-built and optimized for this faint light. Operational instruments include the FIREBALL-2 balloon payload, last flown in 2023 \citep{2025ApJS..278...58M}, the Circumgalactic H$\alpha$ Spectrograph \cite[CHaS,][]{2024ApJ...974..161M}, and the Dragonfly Spectral Line Mapper \citep{2025PASP..137h4103C}. Planned instruments include the Aspera CubeSat that plans to map \ion{O}{6} in a sample of nearby galaxies \citep{Aspera2021}, the proposed JUNIPER CubeSat \citep{witt2025juniper} \jtadd{ and MAGPIE  Pioneer concept \citep{MAGPIE2025} that are optimized for multiple low- and high-ionization UV ions. Larger facility-class instruments include the ELT/HARMONI IFU \citep{Harmoni2014} and the NASA Habitable Worlds Observatory that is expected to fly a UV multi-object spectrograph and/or IFU. } 

In light of increasing prospects for multi-ion maps using UV ions (observed from space, below the UV atmospheric cutoff, or from the ground at significant redshift), we consider the requirements and benefits of observing CGM emission in a key set of UV diagnostic lines. In particular, we analyze a high-resolution set of simulations from the FOGGIE suite \citep{Peeples2019, Simons2020, Wright2024} to assess the detectability of multiphase gas. This paper builds on the foundation of previous work with FOGGIE to predict CGM emission at $z = 3$ \citep{Corlies2020} and lower redshifts \citep{Lochhaas2025}; see also similar work by \citet{2019MNRAS.489.2417A} based on RAMSES simulations and \citet{2019ApJ...877....4L} in the EAGLE simulations. Our primary goal is to assess the observational feasibility of separating galactic inflows and outflows from disk gas, and to assess the instrument performance limits that must be achieved to make this feasible.  

In Section~\ref{sec:methods} we describe our simulations and analysis methods. In Section~\ref{sec:theory} we describe the multi-ion emission-line results. Section~\ref{sec:mass} evaluates how much CGM mass can be detected as a function of observational limits. Section~\ref{sec:kinematics} investigates how inflows, outflows, and disk can be separated kinematically. Section~\ref{sec:instruments} places these results in the context of specific instruments, current and planned. Section~\ref{sec:summary} summarizes our conclusions.

\section{Simulations and Methods} \label{sec:methods}

In this Section, we explain the hydrodynamical simulation used for analysis (Section \ref{sec:simulation}), how we defined and removed the \ion{H}{1} disk (Section \ref{sec:disk_removal}) and how the gas emissivity is calculated (Section \ref{sec:emissivity_cal}). 

\subsection{Simulations}\label{sec:simulation}

For the analysis in this paper we use the FOGGIE (Figuring Out Gas \& Galaxies In Enzo) simulations. FOGGIE uses the Enzo Adaptive Mesh Refinement (AMR) code \citep{Bryan2014,BrummelSmith2019} which enables customized placement of high-resolution grids according to an adaptive, multi-criteria refinement scheme. These refinement schemes are designed to optimize the simulations for resolving the circumgalactic medium (CGM) at levels not usually achieved in cosmological simulations.
In this study, we analyze all six FOGGIE halos at $z = 0.5$. These halos were originally selected to be Milky Way–mass halos at $z = 0$, and they undergo no major mergers after $z = 2$ \citep[see][]{Wright2024}\footnote{One exception: Squall has a 2:1 merger at $z = 0.7$.}. Because the growth of the FOGGIE halo is dominated by minor mergers over most of cosmic time, their total mass evolves slowly, and they already reside in the Milky Way mass range at $z = 0.5$. 


\vadd{We base our analysis at $z = 0.5$ because, within the FOGGIE simulations, the CGM of these halos exhibits a broader range of spatial and kinematic substructure at intermediate redshifts than in their $z = 0$ snapshots.
In FOGGIE, the CGM of these halos at $z = 0$ shows 
less pronounced small-scale CGM structure compared to earlier epochs
\citep{Augustin2021,Lochhaas2025,Ticoras2026}. Selecting halos at $z = 0.5$ therefore ensures that a wider range of CGM morphologies is present, enhancing the diversity of physical conditions probed.
Our goal is not to model the CGM of Milky Way–mass galaxies at all redshifts directly, nor to predict the detailed redshift evolution of CGM emission properties. Rather, we use this intermediate-redshift snapshot as a testbed for assessing the instrumental requirements needed to detect and interpret multiphase CGM emission.}

As described in \citet{Peeples2019}, FOGGIE uses a ``forced refinement'' technique, in which a comoving cubic region 200 kpc/$h$ comoving on a side is centered on the galactic center of mass as it moves through the domain. Within this moving box, cells are refined until the maximum cell size in the box is 1100 pc \vadd{comoving}. \cite{Corlies2020} found that this refinement scheme improved the sampling of the density, temperature, and metallicity fields of CGM gas, resolving smaller structures and even the internal ionization structures of small clouds.  Similar forced-refinement techniques have been used to good effect in other studies that are optimized to resolve the CGM \citep{vandeVoort2019, Hummels2019}.

Another ``cooling refinement'' criterion was first introduced in \cite{Simons2020}: within the refinement box, cells are further refined if their cooling time is shorter than their local sound-crossing time. These cells reach sizes as small as 274 pc \vadd{comoving} (or 183 pc at $z = 0.5$). This combined refinement approach ensures that resolution is concentrated where it is most physically necessary, and results in 90–99\% of the CGM mass being resolved according to the cooling time criterion \citep[][]{Simons2020}. Because the AMR scheme does not impose a minimum gas mass per cell, the median CGM cell mass in the halos studied here is $\sim 44$ $M_{\odot}$ and some cells reach as low as $\leq 1$ $M_{\odot}$ (see also \citealt{Lochhaas2023}). Similar results have been seen in an independent implementation of cooling refinement by the MEGATRON project \citep{2025arXiv251005667C}. With forced and cooling refinement in place, \cite{Simons2020} could investigate the effects of highly-resolved ram-pressure stripping on dwarf satellites, and \cite{Lochhaas2021} could show that non-thermal, turbulent motions are a key source of dynamical support for halos that generally never reach the canonical virial temperature. \cite{Augustin2025} showed that small clumps are ubiquitous and often show internal ionization structure. FOGGIE methods have also been used to examine the Milky Way's high-velocity sky \citep{Zheng2020}, the assembly of the stellar halo from numerous satellites \citep{Wright2024}, the evolution of angular momentum in the CGM \citep{Simons2024}, and the stochastic variations of internal metallicity distributions \citep{Acharyya2024}. Basic properties of the six FOGGIE halos are given in Table~\ref{tab:halos}. 

The background cosmology for all simulations is a $\Lambda$CDM universe with cosmological parameters consistent with the \citealt{Planck2016} results: $\Omega_m = 0.285$, $\Omega_{\Lambda} = 0.715$, $\Omega_{b} = 0.0461$, $H_0 =  69.5$ km s$^{-1} \rm Mpc^{-1}$ and $\sigma _8 = 0.82$. 


\begin{deluxetable*}{lcccc}
\tabletypesize{\normalsize}  
\tablecaption{The halo mass $M_{vir}$, virial radius $R_\mathrm{vir}$, halo gas mass $M_{gas}$ and stellar mass $M_*$ of each of the six FOGGIE galaxies analyzed in this paper at $z=0.5$.}
\label{tab:halos}
\tablehead{
\colhead{Halo} &
\colhead{$M_{\mathrm{vir}}$} &
\colhead{$R_{\mathrm{vir}}$} &
\colhead{$M_{\mathrm{gas}}$} &
\colhead{$M_{\ast}$} \\
\colhead{} &
\colhead{[10$^{12}$ M$_{\odot}$]} &
\colhead{[kpc]} &
\colhead{[10$^{10}$ M$_{\odot}$]} &
\colhead{[10$^{11}$ M$_{\odot}$]}
}
\startdata
Tempest   & 0.43 & 132 & 1.99 & 0.46 \\
Maelstrom & 0.81 & 163 & 5.09 & 0.90 \\
Squall    & 0.68 & 154 & 2.99 & 0.80 \\
Blizzard  & 0.94 & 172 & 3.52 & 1.30 \\
Hurricane & 1.38 & 195 & 6.29 & 1.90 \\
Cyclone   & 1.41 & 197 & 5.24 & 2.02 \\
\enddata
\end{deluxetable*}

\begin{deluxetable*}{lcc}
\tabletypesize{\normalsize}  
\tablecaption{Rest-frame wavelengths and characteristic temperature ranges of emission lines analyzed in this study.\footnote{Temperature ranges are shown assuming collisional ionization.}}
\label{tab:emission_lines}
\tablehead{
\colhead{Line} &
\colhead{Wavelength} &
\colhead{Temperature} \\
\colhead{} &
\colhead{[\AA]} &
\colhead{[K]}
}
\startdata
H$\alpha$        & 6563            & $\sim 1 \times 10^{4}$ \\
\ion{Mg}{2}      & 2796            & $< 5 \times 10^{4}$ \\
\ion{Si}{2}      & 1260            & $< 5 \times 10^{4}$ \\
\ion{C}{3}       & 1910            & $5 \times 10^{4} < T < 10^{5}$ \\
\ion{Si}{3}      & 1207            & $5 \times 10^{4} < T < 10^{5}$ \\
\ion{C}{4}       & 1548            & $10^{5} < T < 5 \times 10^{5}$ \\
\ion{Si}{4}      & 1394            & $10^{5} < T < 5 \times 10^{5}$ \\
\ion{O}{6}       & 1032, 1038      & $10^{5} < T < 10^{6}$ \\
\enddata
\end{deluxetable*}

The FOGGIE simulations implement simple gas physics, including radiative cooling with a self-shielding approximation for high density gas, star formation, and thermal feedback from supernovae. The simulation includes metallicity-dependent cooling and a metagalactic UV background \citep{Haardt2012}, using the Grackle chemistry and cooling library \citep{Smith2017} with self-shielding \citep{emerick19}. Star formation occurs in gas with a comoving number density exceeding $\sim ~ 0.1 ~ \rm cm^{-3}$. The minimum star particle mass starts at $1000$ M$_\odot$ at high redshift and then ramps up to 10,000 M$_\odot$ between $z = 1$ and $z = 2$ to reduce the total number of particles needed to simulate a Milky-Way-like host. Supernova feedback follows the implementation of \citealt{Cen&Ostriker2006} in which purely thermal energy is deposited into the 27 cells surrounding a star particle after 12 dynamical times have passed since its formation. Each star particle injects $1 \times 10^{-5} m_*c^2$ of energy and pushes $0.25m_*$ of mass to the surrounding gas over 12 dynamical times following particle creation ($m_*$ is the original star particle mass and $c$ is the speed of light). The total metal mass ejected is calculated as $0.025m_* (1-Z_*)+0.25Z_*$ to account for metal recycling. In this method all metals are tracked as a single combined field. In this study particular elemental abundances are calculated assuming solar abundances and ion fractions and ion masses are computed using the Trident package \citep{TridentRef}. 

The main analysis of the paper has been done on CGM gas defined as all gas inside a \vadd{100 physical kpc} box \vadd{at $z = 0.5$}, centered on the galaxy, corresponding to a 50  \vadd{physical} kpc projected distance from the galactic center. This region is smaller than the forced refinement track box and \vadd{therefore is} always resolved with cells $\leq 732$ physical pc \vadd{wide}. This spatial extent is motivated by observational limitations: CGM emission from gas at larger distances is typically too faint to detect with current instrumentation. Additionally, our interest lies in the interface region between the galaxy and its surrounding halo, where emission is brightest and gas dynamics are most active.

\subsection{Disk Removal}\label{sec:disk_removal}

For some analyses we want to remove the galactic disk and focus on isolated CGM emission. This is possible in simulations because we have the full 3D density information and so can label disk and/or CGM cells on that basis. This removal is guided by observational limits on disk detection, under the assumption that disk removal from real CGM emission images can be done in a similar fashion. Figure \ref{fig:HImaps} shows face-on \ion{H}{1} column density maps for the Squall halo within a 100 kpc box. The top panel includes all gas contributing to the \ion{H}{1} column density, while the bottom panel shows the result after removing the disk component.

As described in \citet{Trapp25a}, we define the \ion{H}{1} disk as a contiguous region with \ion{H}{1} number density above $\sim 2 \times 10^{-2}$ cm$^{-3}$. To avoid missing low-density regions in the disk associated with stellar feedback or dynamical effects, we fill both fully enclosed holes and ``donut'' holes that completely pierce the disk.
This disk removal method is based on the full 3D gas distribution in the simulation and does not depend on the chosen line of sight; it is also robust to removing misaligned disks and maintaining the disk-halo interface versus previous disk removal techniques \citep{Trapp25a}.
After disk removal, some \ion{H}{1} from in front of or behind the disk remains visible in projection within the central 20 kpc region. By subtracting the disk in this way, we focus our analysis on halo gas, isolating CGM features that are otherwise obscured by bright emission from the central disk. Figure \ref{fig:HImaps} shows projected \ion{H}{1} column density maps for Squall with and without the disk removed for comparison.

\jtadd{ Disk removal is simple in theory, as it is based on 3D density and kinematic fields that are drawn directly from simulations. When working with real data, the task will not be {\it disk removal} but rather {\it disk identification} - marking the spatial and kinematic location of disk gas for comparison to CGM. All datasets will naturally suffer projection effects that do not occur in the simulation, but there are proxies by which that might be accomplished. First, if emission maps in key ions cover enough of the disk with good kinematic resolution, its shape and rotation can be discerned directly from multi-channel emission data. In cases where HI 21 cm data is available with sufficient resolution, this position-velocity map may provide the best information on where the disk is located. Even in cases where the kinematic resolution is poor, HI emission with sufficient spatial resolution ($\lesssim 1$ kpc) will provide a   map of disk location to which CGM emission can be compared. Our simulations show that CGM emission extends many kpc beyond the disk, so that analyses based on UV emission can withstand some uncertainty about the disk boundary or ambiguity in its definition without serious loss of information. }

\subsection{Emissivity Calculation}\label{sec:emissivity_cal}
\begin{figure}[ht!]
\begin{center}
\includegraphics[width=0.9\columnwidth]{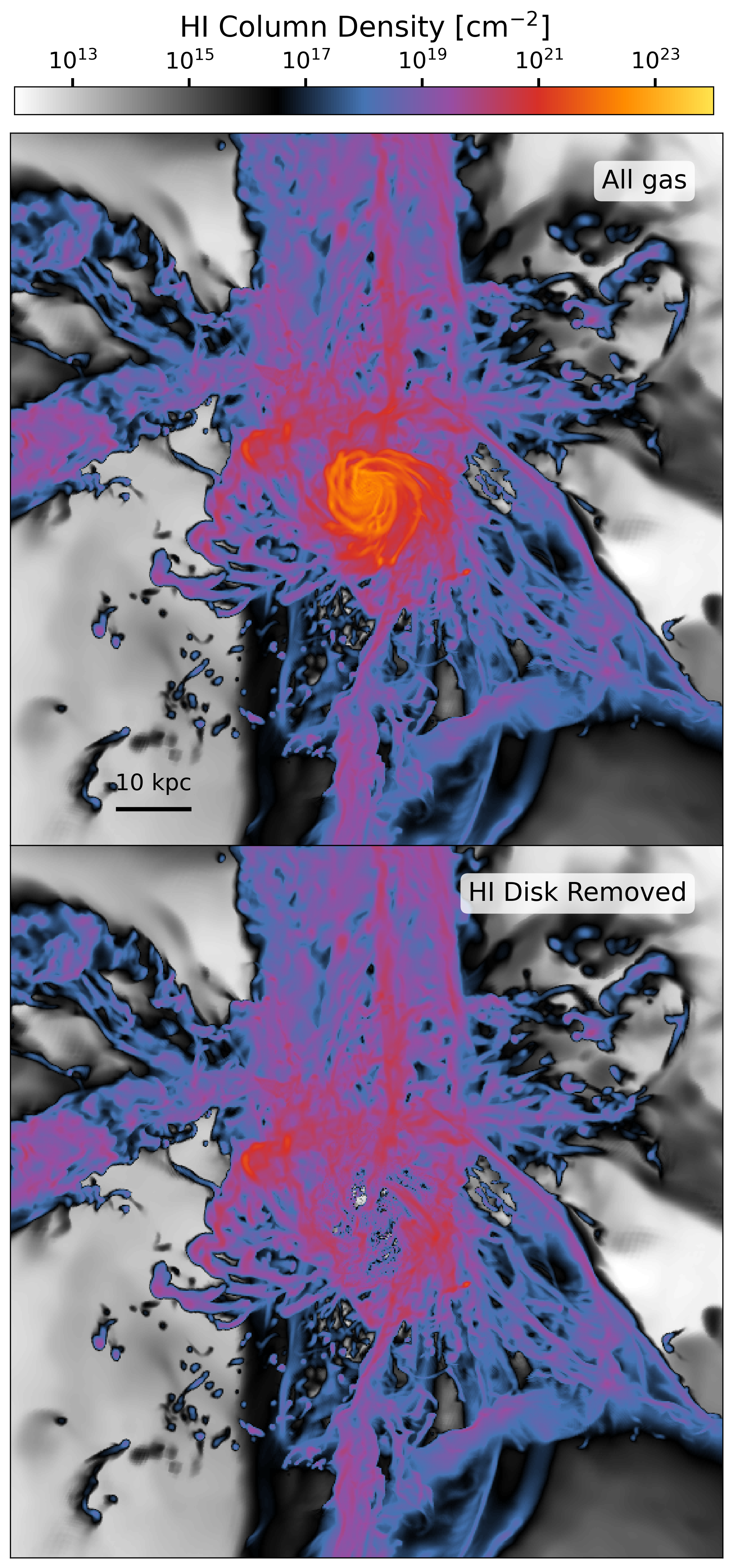}
\caption{Face-on \ion{H}{1} column density maps for the FOGGIE Squall halo within a 100 kpc box. \textit{Top:} All gas contributing to the \ion{H}{1} column density.
\textit{Bottom:} Map after removing the \ion{H}{1} disk component. Some \ion{H}{1} remains visible in the inner region after removing the disk; this gas is not part of the disk but lies in front of or behind it along the line of sight. This figure illustrates how removing the dense \ion{H}{1} disk reveals surrounding gas structures and helps isolate CGM features from the disk.}.
\label{fig:HImaps}
\end{center}
\end{figure}

Most of the emissivity calculations in this work follow the pipeline developed and described in detail by \citet{Corlies2020} and used in \citet{Lochhaas2025}. Here, we briefly summarize the key steps involved.

In the density and temperature regimes typical of the CGM, gas cools primarily via collisional excitation followed by radiative decay. As a result, the emissivity of a given line scales approximately with the square of the gas density ($n^2$). The brightest emission for a given line therefore originates from gas near the temperature at which the cooling curve peaks for that species \citep{Bertone2010}. To compute the line emissivity in each simulation cell, we post-process the outputs using the photoionization code \textsc{CLOUDY} \citep[version 23.01;][]{Chatzikos2023}. The emissivity for each line is drawn from CLOUDY tables that are parametrized by hydrogen number density ($n_{\rm H}$), gas temperature ($T$), and redshift. These emissivities assume solar metallicity. Each metal line emissivity is then scaled linearly by metallicity of each cell, which captures the dependence on metallicity produced by the Cloudy modeling.

To begin the post processing, we first constructed \textsc{CLOUDY} look-up tables of emissivity as a function of temperature ($\rm 10^3 < T < 10^8$ K with steps of $\Delta \log~ \rm T = 0.1$) and hydrogen number density ($\rm 10^{-5} < n_H < 10^2$ cm$^{-3}$ with steps of $\Delta \log ~ n_H = 0.5$) for each metal line using \textsc{CIAOLoop}\footnote{\url{https://github.com/brittonsmith/cloudy_cooling_tools.git}} \citep{Smith2009}. Solar metallicity and abundances are assumed in these calculations.

For each cell in the simulation, we interpolate the CLOUDY grid at the cell’s temperature and hydrogen number density to obtain the emissivity value. This interpolated value is then scaled by the cell's metallicity. We assume the gas is in ionization equilibrium and include both photo-ionization and collisional ionization processes. For the extragalactic ultraviolet background (EUVB), we adopt the \citet{Haardt2012} model at z = 0, which is the same evolving UVB used during the simulation runtime. 
We note that the choice of UVB model can introduce inherent uncertainty in the derived ionization states; for a quantitative discussion of the uncertainties related to UVB choice refer to \citet{Taira2025}.

Throughout this paper, we analyze emission from eight spectral lines: H$\alpha$,  \ion{Mg}{2}, \ion{Si}{2}, \ion{Si}{3}, \ion{C}{3}, \ion{C}{4}, and \ion{O}{6}. We choose H$\alpha$ over Ly$\alpha$ to avoid radiative transfer effects in optically thick gas, to get results for at least one line optimized for ground-based telescopes, and in consideration of the fact that UV instruments in low-Earth orbit will likely need to avoid or suppress Ly$\alpha$ to reduce scattered light. The specific rest-frame wavelengths used for each line are listed in Table \ref{tab:emission_lines}. All six FOGGIE halos are included in the full analysis. Figures presented in the paper are selected to be representative of the general trends observed across the full halo sample; while a single halo may be shown for illustrative purposes, the findings apply broadly and are not limited to the examples.

\begin{figure*}[ht!]
\begin{center}
\includegraphics[width=0.8\textwidth]{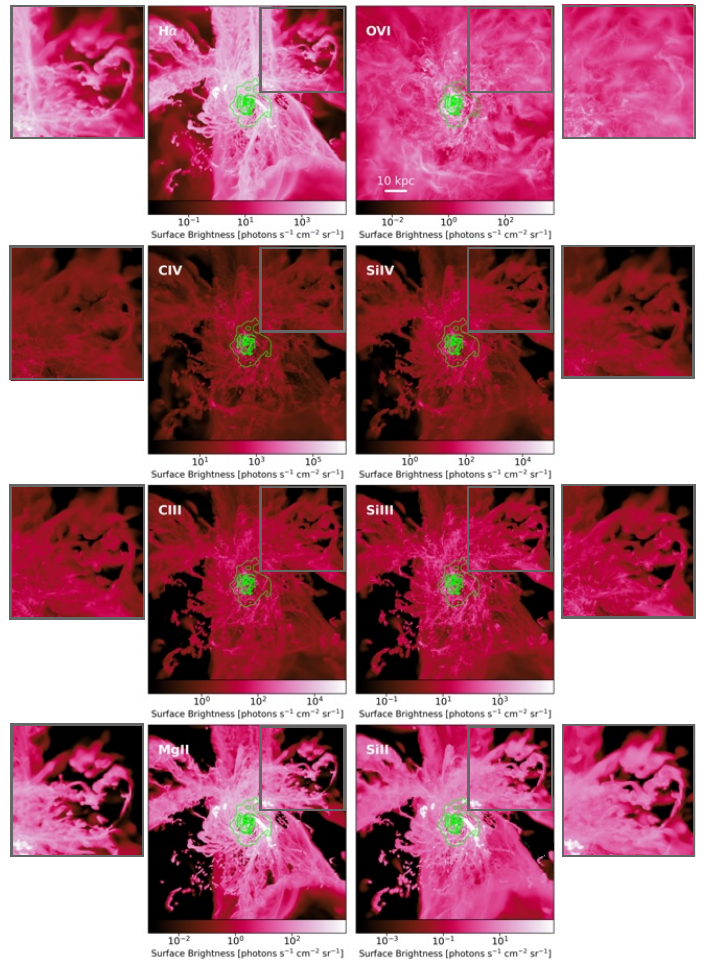}
\caption{Face-on surface brightness maps of the Squall halo for eight ions with emission lines in the UV/optical: H$\alpha$,  \ion{Mg}{2}, \ion{Si}{2}, \ion{Si}{3}, \ion{C}{3}, \ion{C}{4}, and \ion{O}{6}. The box sizes cover 100 kpc fields of view. \vadd{Each panel uses an independent color scale spanning seven orders of magnitude in surface brightness (from $10^{-8}$ to $10^{-1}$ of the maximum emission value in that panel), chosen to highlight the morphology of emission structures in each ion rather than to compare absolute surface brightness between ions.} The \ion{H}{1} disk has been removed to isolate CGM emission; green contours indicate where the disk was. These maps highlight the multiphase nature of the CGM, as different ions trace different physical structures and temperatures. The zoom-in panels emphasize key differences in morphology across ions. These comparisons demonstrate that no single emission line fully captures the CGM’s complexity. Each ion offers a distinct view of its structure and thermodynamic state. Observing multiple UV lines is essential for building a complete, multiphase picture of the circumgalactic medium.}
\label{fig:emissionmaps}
\end{center}
\end{figure*}

\section{Mapping the CGM in Multiphase Emission Lines} \label{sec:theory}

In this section we make predictions for H$\alpha$ and metal line emission and demonstrate the role instrumental spatial resolution and sensitivity play in detecting and understanding the CGM. 

Figure~\ref{fig:emissionmaps} presents face-on surface brightness maps for eight emission lines---H$\alpha$, \ion{O}{6}, \ion{C}{4}, \ion{Si}{4}, \ion{C}{3}, \ion{Si}{3}, \ion{Mg}{2}, and \ion{Si}{2}---for the Squall halo. Emission from the galactic \ion{H}{1} disk has been removed, and the green contours indicate the region where the disk was. These maps reveal a diverse range of emission morphologies across different ions, reflecting the multiphase structure of the CGM. Zoom-in panels are included to emphasize spatial differences and trace features that change significantly from one ion to another. \vadd{Each panel has its own colorbar, spanning seven orders of magnitude in surface brightness (from $10^{-8}$ to $10^{-1}$ of the maximum emission value in that panel), allowing dynamic range of each ion to be visualized. We note that the perceived morphology depends on the chosen color stretch; adopting individual color scales ensures that faint but extended structures are not artificially suppressed in intrinsically fainter lines.}

\ion{O}{6} emission, which traces warm and/or diffuse gas ($T \sim 10^{5-6}$ K), exhibits a smooth, volume-filling morphology that sets it apart from the filamentary or clumpy structures seen in lower ionization lines. Notably, the prominent extended loop seen in the H$\alpha$ zoom-in region is almost invisible in the \ion{O}{6} panel, emphasizing the distinct temperature and density regimes probed by these lines. Its diffuse distribution reflects its origin in lower-density, higher-temperature gas that permeates the CGM more uniformly.
Interestingly, a few bright clumps of \ion{O}{6} emission appear in the top-left region of these \ion{O}{6} emission maps that are not seen in any of the lower ionization lines. These localized enhancements are unique to \ion{O}{6} but common in all our simulations, and may trace transient or shock-heated structures in the hotter CGM phase. The differences between the \ion{O}{6} distribution and that of lower ions can play a key role in determining the nature of the gas in the CGM \citep{Ticoras2026}.

Moving to slightly lower ionization states, \ion{C}{4} and \ion{Si}{4} begin to trace some of the filamentary features seen in H$\alpha$. \ion{C}{4} shows both diffuse and structured emission, partially recovering the loop feature. \ion{Si}{4}, reveals the loop shape even more clearly, suggesting that it arises in gas at intermediate temperatures between the \ion{C}{4} and H$\alpha$ phases.

The intermediate ions \ion{C}{3} and \ion{Si}{3} also show interesting differences \vadd{compared to the higher ions. }
For instance, the loop structure is easier to recognize in \ion{C}{3} than in \ion{C}{4}. The clumpy emission \vadd{in the lower left part of maps} appears more diffuse in \ion{C}{4} and \ion{Si}{4} compared to the sharper, more compact structures in \ion{C}{3} and \ion{Si}{3}. These differences trace subtle variations in gas density and temperature that affect emission from these ions differently.

At the lowest ionization levels, \ion{Mg}{2} and \ion{Si}{2} show emission patterns more closely aligned with the H$\alpha$ distribution. However, important differences persist. For example, in the zoomed region, both \ion{Mg}{2} and \ion{Si}{2} trace parts of the loop seen in higher ions, but \ion{Mg}{2} displays distinct clumps in the region where the H$\alpha$ map shows a loop. \vadd{The emission structures in these two maps appear more compact and concentrated than in the higher-ionization lines.} On the right side of the disk, a filament visible in \ion{Si}{2} extends to the edge of the map, whereas in \ion{Mg}{2} it is truncated. These differences illustrate how even among low-ionization species, emission structure can vary depending on local gas properties.

Together, these maps demonstrate the value of observing the CGM in multiple UV emission lines. Each ion captures a different phase of the gas - ranging from hot and diffuse to cool and dense - and only by combining them can we build a comprehensive, multiphase picture of CGM structure and physical conditions.

\begin{figure*}[ht!]
\begin{center}
\includegraphics[width=\textwidth]{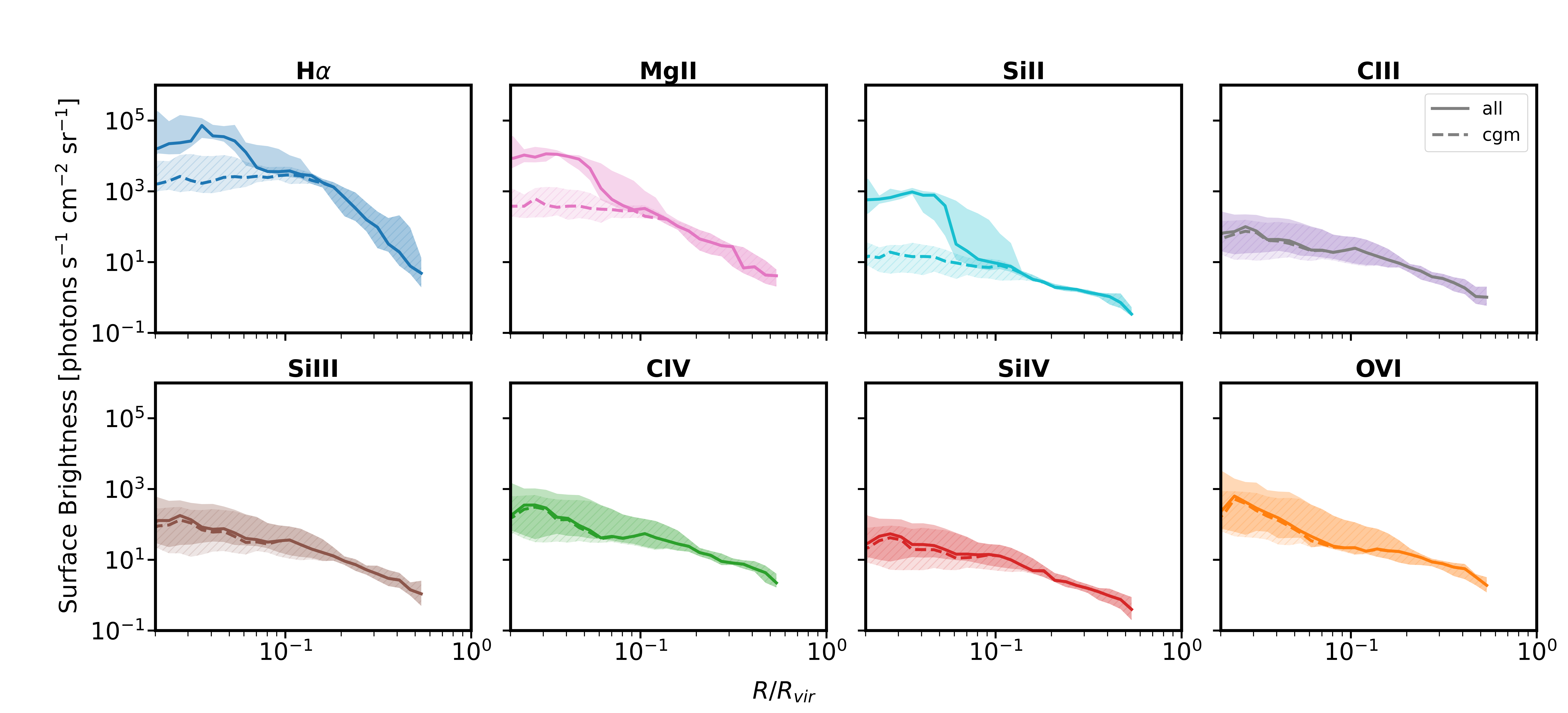}
\caption{Radial surface brightness profiles for eight ions across all FOGGIE halos (face-on, \vadd{0.18} kpc resolution).
Solid lines show median profiles including all gas; dashed lines show CGM-only emission after \ion{H}{1} disk removal. Shaded regions indicate the range of halo-to-halo variation. Most ions follow power-law profiles, while H$\alpha$, \ion{Mg}{2}, and \ion{Si}{2} show a break in slope around $0.1~R_{\rm vir}$—reflecting a transition from disk to CGM-dominated emission. CGM-only profiles flatten at small radii for low-ionization species.}
\label{fig:emissionprofiles}
\end{center}
\end{figure*}

Figure \ref{fig:emissionprofiles} presents radial surface brightness profiles for all eight emission lines analyzed in this study. Each panel shows the median surface brightness as a function of $R/R_{\rm vir}$, computed across all six FOGGIE halos \vadd{face-on}. Solid lines represent profiles including all gas (disk and CGM), while shaded regions indicate the scatter among halos. Dashed lines show the profiles after removing the \ion{H}{1} disk, isolating the CGM contribution. Note that in the CGM surface brightness profiles, only the galactic disk is removed; satellite galaxies are still included as part of the CGM. Removing satellites produces only minor changes to the profiles and does not affect our results.

Most ions (e.g., \ion{O}{6}, \ion{C}{4}, \ion{Si}{4}, \ion{C}{3}, \ion{Si}{3}) exhibit smooth power-law profiles extending \vadd{inward} from the halo outskirts. In contrast, H$\alpha$, \ion{Mg}{2}, and \ion{Si}{2} show a clear transition in slope around $R \sim 0.1~R_{\rm vir}$: from a steep power-law decline at large radii to an exponential rise toward a central plateau. This broken profile is consistent with recent observational findings by \citealt{Nielsen2024} for [OII] and [OIII] and reflects the transition between CGM and disk-dominated emission. After \ion{H}{1} disk removal, the central region becomes flattened for low-ionization lines, confirming that the steep inner profiles are driven by disk emission. These results support the interpretation that breaks in radial surface brightness profiles of lower ions mark a structural transition between CGM and galactic disk. The profiles also recall the findings of the AMIGA absorption line survey of M31, which found enhancements in low ions as sightlines approach the disk that are not reflected in the high ions \citep{2025arXiv250616573L}.


\begin{figure*}[ht!]
\begin{center}
\includegraphics[width=0.9\textwidth]{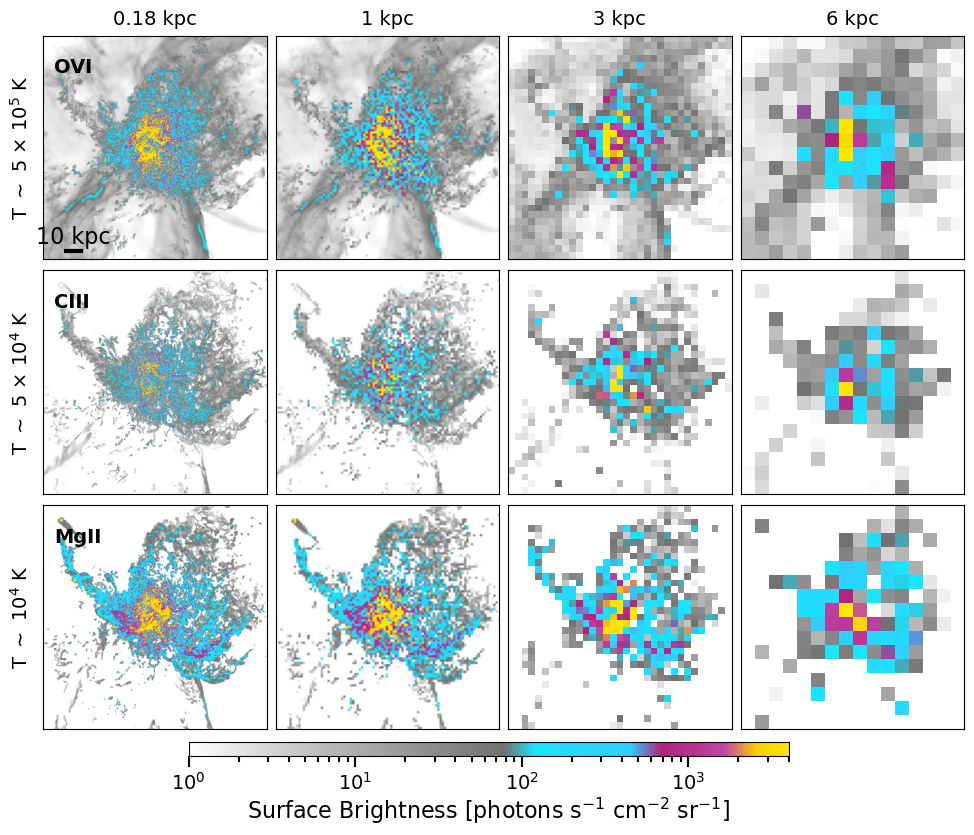}
\caption{Surface brightness maps for three UV emission lines for the Blizzard halo, shown edge-on. Each column corresponds to a different spatial resolution: \vadd{0.18} kpc, 1, 3 and 6 kpc. These correspond at $z = 0.5$ ($z = 0$, 10 Mpc) to $\sim$ 0.04 (5.6), 0.16 (21), 0.5 (62), and 1 (123) arcsec, respectively.
The galactic disk has been removed to isolate CGM emission. The colormap represents surface brightness in units of photons s$^{-1}$ cm$^{-2}$ sr$^{-1}$, with grayscale indicating regions below a representative instrument sensitivity threshold. Colored regions are potentially detectable by instruments with corresponding sensitivity. For reference, at $z = 0.5$, 6 kpc subtends $\sim 1$ arcsec, similar to the limits of seeing-limited ground-based observations, while a space-based telescope or AO-assisted ground-based telescope could reach $\sim 0.1$ arcsec, or $\lesssim$ 1 kpc spatial resolution. } 
\label{fig:mapswithspatialres}
\end{center}
\end{figure*}

\section{Estimating the CGM Mass in Emission} \label{sec:mass}

\begin{figure*}[ht!]
\begin{center}
\includegraphics[width=1\textwidth]{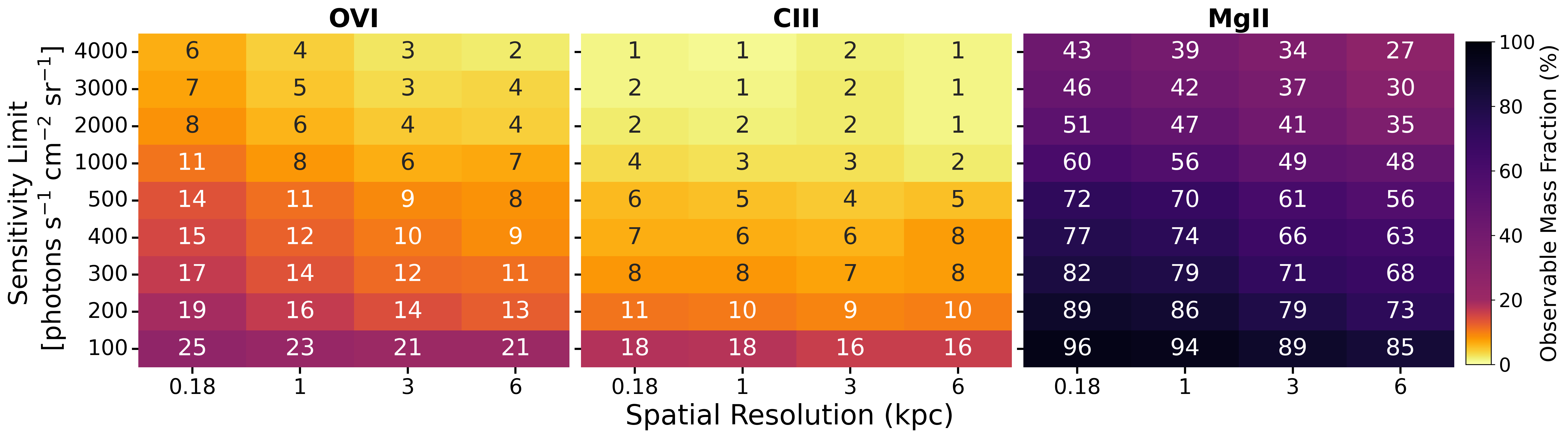}
\caption{Observable CGM mass fraction as a function of spatial resolution (x-axis) and surface brightness sensitivity (y-axis). 
The spatial resolutions of \vadd{0.18}, 1, 3, and 6 kpc, correspond at $z = 0.5$ ($z = 0$, 10 Mpc) to $\sim$ 0.04 (5.6), 0.16 (20), 0.5 (61), and 1 (120) arcsec, respectively.
These maps are for three representative emission lines, averaged over six FOGGIE halos and for edge-on projections (see Appendix \ref{sec:appendixA} for maps of all eight ions' face-on and edge-on results). Each cell shows the percentage of total CGM mass detectable above the given sensitivity limit at the specified resolution, with values color-coded and overlaid. As expected, observability decreases with lower sensitivity and coarser resolution, though sensitivity plays a stronger role: at fixed resolution, increasing the detection threshold from 100 to 500 reduces the detectable mass by more than 50\% in several ions. Achieving high sensitivity is critical for maximizing CGM detection, particularly for tracing diffuse, highly ionized gas.}
\label{fig:heatmap-massfrac}
\end{center}
\end{figure*}

\begin{figure}[ht!]
\begin{center}
\includegraphics[width=1\columnwidth]{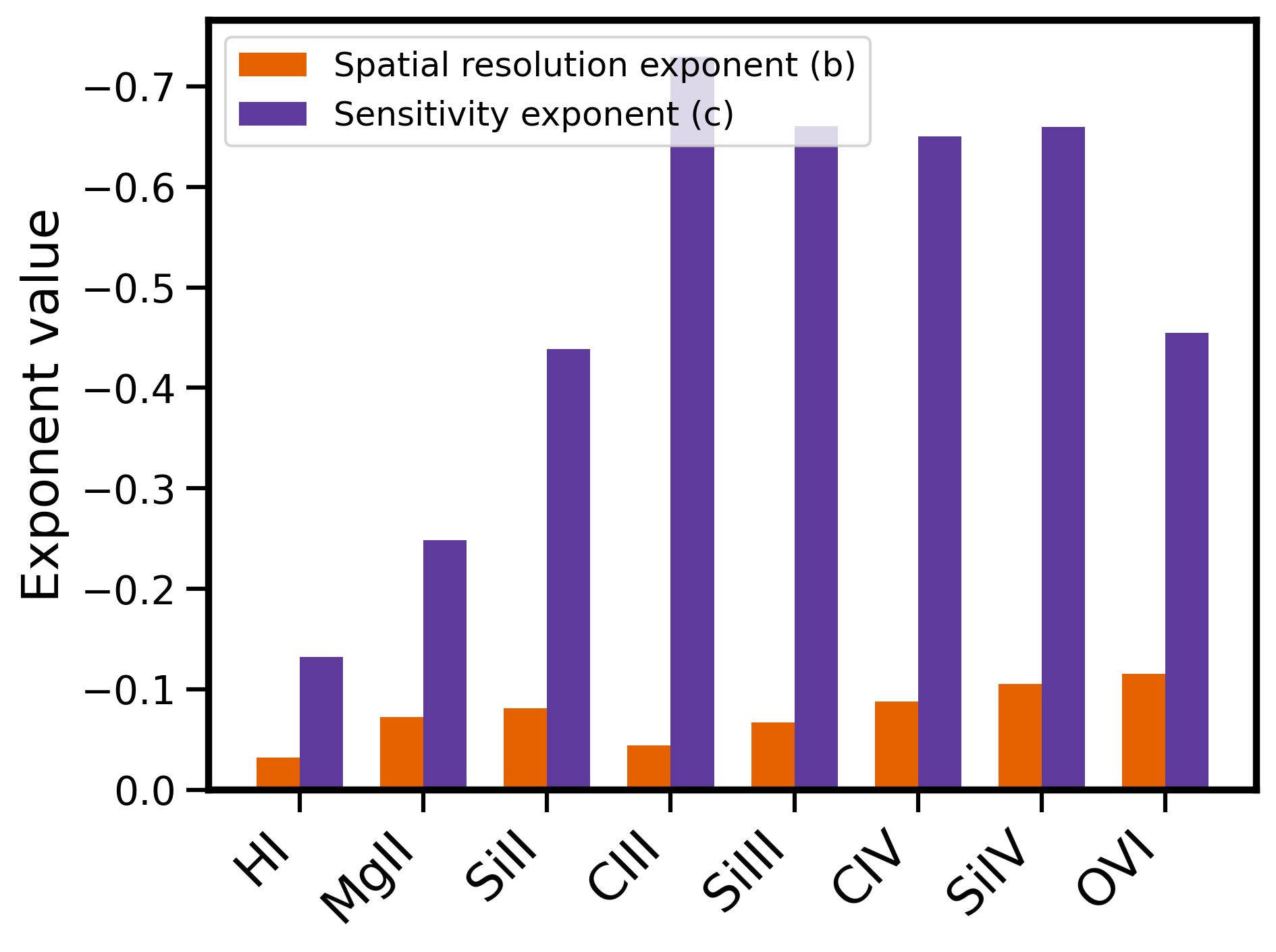}
\caption{Spatial resolution exponent (b) and sensitivity exponent (c) from the fitting formula (Eqn.~\ref{eqn:massfrac}). A larger value of the sensitivity exponent indicates a stronger influence of sensitivity on recovering the CGM mass for each ion.}
\label{fig:fit-massfrac}
\end{center}
\end{figure}

A complete understanding of the CGM requires observations across a wide range of low to high ionization states, each tracing different temperatures, densities, and spatial morphologies. In this section, we focus on the observability of these emission lines and examine how instrumental limitations, specifically spatial resolution and sensitivity, impact our ability to detect and map CGM gas in practice. All analysis has been done for the CGM after removing the \ion{H}{1} disk\footnote{Gas outside the 3D disk can still appear in projection at the disk's former location.}.

We begin by presenting edge-on surface brightness maps for the Blizzard halo in Figure~\ref{fig:mapswithspatialres}, for three representative UV emission lines, \ion{O}{6}, \ion{C}{3}, and \ion{Mg}{2}, capturing a wide range of CGM gas phases. \ion{O}{6} traces warm-hot diffuse gas ($T \sim 5 \times 10^5$ K), \ion{C}{3} traces intermediate temperature gas ($T \sim 5 \times 10^4$ K), and \ion{Mg}{2} traces the cold gas ($T \sim 10^4$ K) . While this figure focuses on one halo for illustration, the full analysis has been performed across all FOGGIE halos and eight emission lines, with consistent results. Throughout the paper, we select specific halos for display based on their clarity and the degree to which they are representative of patterns in the full set; these choices are made solely for illustrative purposes, such as when the viewing angle significantly affects the interpretation. The edge-on orientation is chosen here to better reveal the CGM structures surrounding the galaxy. 

In Figure~\ref{fig:mapswithspatialres} each row corresponds to a different ion, while each column represents increasing spatial resolution: 0.18 kpc, 1 kpc, 3 kpc, and 6  kpc. These correspond at $z = 0.5$ ($z = 0$, 10 Mpc) to approximately 0.04 (5.6), 0.16 (20), 0.5 (61), and 1 (120) arcseconds, respectively. The galactic disk has been excluded from all panels to focus on CGM emission. The colormap represents surface brightness in units of photons s$^{-1}$ cm$^{-2}$ sr$^{-1}$. Pixels below 100 photons s$^{-1}$ cm$^{-2}$ sr$^{-1}$  threshold appear in grayscale, while colored pixels indicate detectable emission. These thresholds mimic the sensitivity of current and possible future UV instruments.

At a sensitivity of $>100$ photons s$^{-1}$ cm$^{-2}$ sr$^{-1}$ (corresponding to regions shown in blue, purple, and yellow), extended CGM structures such as filaments and streams are visible out to distances of $\sim$50 kpc in all three ions in the highest resolution column. These features are preserved at 1 kpc resolution but begin to disappear at coarser resolutions as the brighter regions are averaged down with fainter areas. This fading occurs because surface brightness \vadd{of the peripheral pixels} is only slightly above the sensitivity threshold - around 100 photons s$^{-1}$ cm$^{-2}$ sr$^{-1}$, and they are surrounded by regions with much lower surface brightness. As the pixel size increases, each pixel averages over a larger area. This means bright regions are blended with adjacent, fainter regions. Although the total number of photons within a larger pixel may increase, the surface brightness (i.e., photons per unit area) often decreases, because the increase in area outweighs the gain in photon counts. If a large pixel includes more faint gas than bright gas, the surface brightness of that pixel may drop below the instrument’s detection limit—causing previously visible structures to vanish in the map.

At a higher threshold of $\sim$ 500 photons s$^{-1}$ cm$^{-2}$ sr$^{-1}$ (purple and yellow), the extended CGM structures are no longer visible. An instrument limited to this sensitivity would primarily detect the denser regions near the galaxy center. At the highest resolution, some features can still be seen around the central region. For Mg II, extended bright structures out to $\sim$30 kpc remain detectable at this threshold, whereas intermediate ions such as C III are fainter and show less extended structure beyond the center. Comparable structures are visible at 1 kpc resolution, though much of the fine detail disappears by 3 kpc, even though the radial extent of detection remains similar to the \vadd{0.18} kpc maps. At the poorest spatial resolution of 6 kpc, only a handful of bins are detectable. At this scale, detailed structures vanish entirely, but the overall radial extent of the detectable emission remains comparable to that seen at very high spatial resolution.

At a sensitivity threshold of $2000~\rm photons~s^{-1}~cm^{-2}~sr^{-1}$ (yellow), the detectable O VI emission is similar to that at 500 photons s$^{-1}$ cm$^{-2}$ sr$^{-1}$  for the \vadd{0.18} and 1 kpc maps. However, in the lower-resolution maps (3 and 6 kpc), detection is restricted to very small, bright clumps near the galactic center. The same trend is seen for C III. In contrast, the extent of Mg II emission changes significantly between sensitivities of 500 and 2000 photons s$^{-1}$ cm$^{-2}$ sr$^{-1}$: at 2000, detectable emission is limited to the central regions. At coarse resolutions of 3 and 6 kpc, only a few pixels remain above the detection threshold.

In addition to how structure and extent of detectable ions change by changing sensitivity and spatial resolution of an instrument, we can also explore the amount of mass that is observable. Figure \ref{fig:heatmap-massfrac} presents heatmaps of the observable CGM mass fraction for the same representative ions: \ion{O}{6}, \ion{C}{3}, \ion{Mg}{2};
the heat map for all eight emission lines: H$\alpha$, \ion{O}{6}, \ion{C}{4}, \ion{Si}{4}, \ion{C}{3}, \ion{Si}{3}, \ion{Mg}{2}, and \ion{Si}{2} can be found in Appendix \ref{sec:appendixA}.
The mass of each ion is computed using \textsc{Trident} \citep{TridentRef}.
These maps indicate the fraction of CGM by mass that is emitting in each line that would be detectable within a 100 kpc field of view. The analysis uses an edge-on projection and the values are averaged over all six halos. The general trends described below hold for individual halos and for face-on projections (see Figure~\ref{fig:heatmap-massfrac-face} in Appendix \ref{sec:appendixA}). Each panel shows the percentage of each ion's mass in the CGM that would be detectable by an instrument with a given combination of surface brightness sensitivity (y-axis, in photons s$^{-1}$ cm$^{-2}$ sr$^{-1}$) and spatial resolution (x-axis, in kpc). The values are color-coded and also explicitly labeled in each cell.

The observable mass fraction decreases toward the upper-right corner of each panel, i.e., under conditions of low sensitivity and coarse resolution. At fixed sensitivity (e.g., 100 $\rm photons ~s^{-1} cm^{-2} sr^{-1}$), changing the spatial resolution from \vadd{0.18} to 6 kpc leads to a gradual reduction in observable mass, but this decline is modest for most ions. In contrast, at fixed resolution (e.g., 1 kpc), changing the sensitivity limit from 100 to 500 $\rm photons ~s^{-1} cm^{-2} sr^{-1}$ results in a much steeper drop in detectable mass, especially for ions tracing hotter, more diffuse phases such as \ion{O}{6}.

To further quantify the effects of sensitivity and spatial resolution on CGM detectability, we fit a simple power-law model to the observable mass fraction:

\begin{equation}
\label{eqn:massfrac}
    f(x,y) = ax^by^c
\end{equation}

where $f(x, y)$ is the fraction of CGM mass observable in a given ion, $x$ is the spatial resolution, and $y$ is the surface brightness sensitivity. The exponents $b$ and $c$ represent how strongly each parameter influences detectability: higher values indicate that small changes in that parameter produce larger changes in observable mass.

Figure~\ref{fig:fit-massfrac} presents the best-fit values of $b$ and $c$ for each ion. In all cases, the sensitivity exponent $c$ is larger than the spatial resolution exponent $b$, indicating that sensitivity has a stronger impact on CGM observability. The fitting function reproduces the heatmap values with a mean absolute error (MAE) $\leq$ 7\% for each ion, demonstrating the robustness of the model.

These results highlight the need for emission mapping instruments to prioritize sensitivity---particularly below the 500 $\rm photons ~s^{-1} cm^{-2} sr^{-1}$ threshold---to maximize the detection of CGM emission across a wide range of ionization states. Sensitivity gains are especially critical for uncovering the more diffuse and extended components of the warm-hot CGM traced by higher ions.

It is important to note that the observable emission-line strengths presented in this section are likely lower limits. The FOGGIE simulations at some stages of their evolution under-predict column densities of these ions by $\sim 1$--2 dex compared to quasar absorption line observations \citep[e.g.,][]{Tumlinson2011, Lehner2020}. Since emission scales with density \vadd{squared}, the resulting surface brightness---and thus the inferred detectable mass---may also be underestimated by a similar amount. This discrepancy could arise from several factors, including lower simulated CGM densities or metallicities, differences in temperature structure, or missing physical processes such as AGN-driven ionization (see also \citealp{Wright2024} for a thorough discussion of FOGGIE's current star formation and feedback implementations). Therefore, while these maps and fits provide useful insight into how instrument parameters shape detectability, the absolute values of observable CGM mass fractions should be interpreted as lower bounds rather than precise predictions. Relative comparisons (FOGGIE-to-FOGGIE) can be considered reliable, but may be discrepant with findings from other simulations. In any case our finding that sensitivity plays a more dominant role than spatial resolution remains robust and is independent of the specific column density normalization in FOGGIE.

\begin{figure*}[ht!]
\begin{center}
\includegraphics[width=0.60\textwidth]{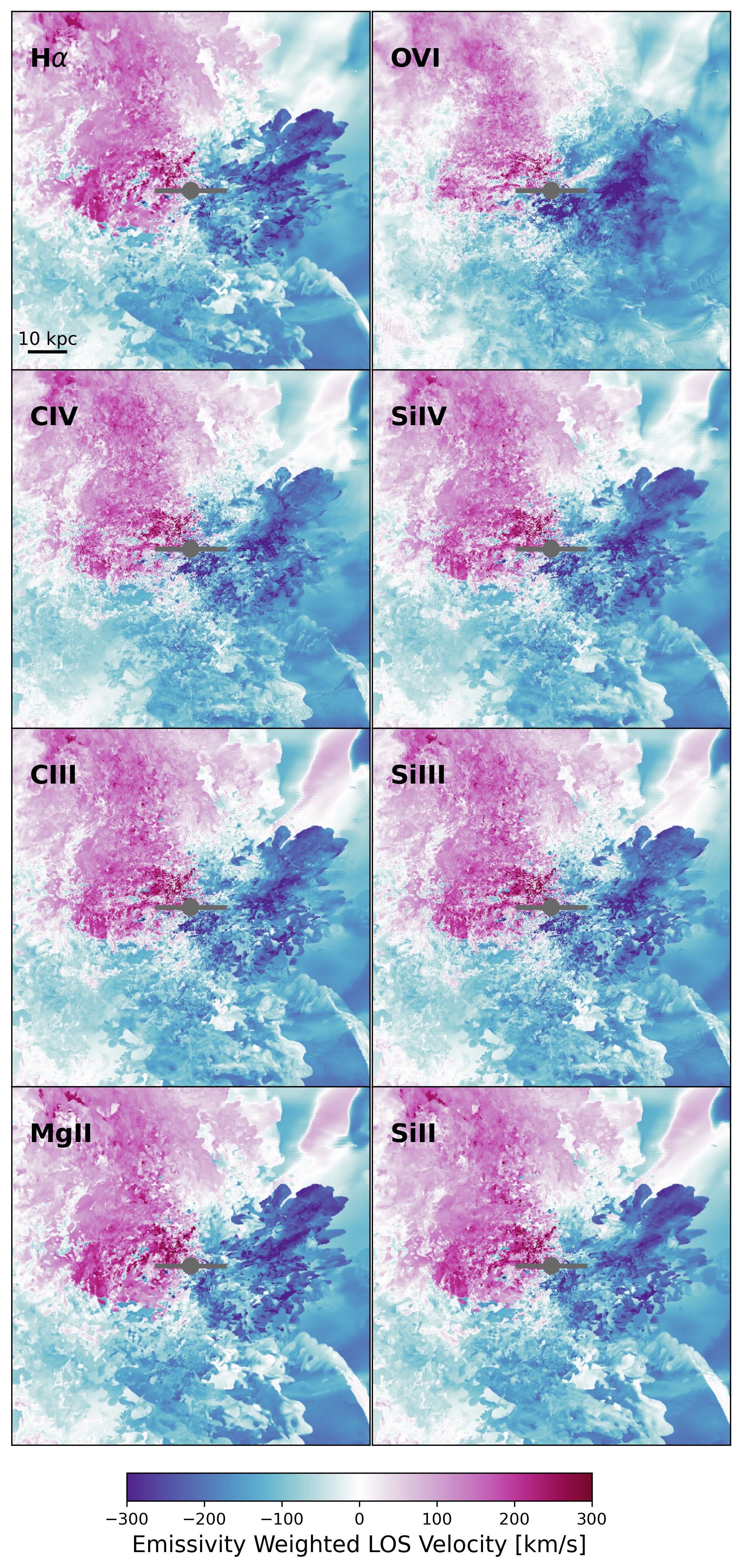}
\caption{Emissivity-weighted projected line of sight velocity maps for eight ions for the Cyclone halo in an edge-on view and a field of view of 100 kpc. The galactic disk has been removed in all panels to isolate the kinematics of CGM gas. The gray line shows where the disk was.}
\label{fig:vlosmaps}
\end{center}
\end{figure*}

\section{Resolving CGM Kinematics in Emission} \label{sec:kinematics}
\begin{figure*}[ht!]
\centering
\includegraphics[width=0.7\textwidth]{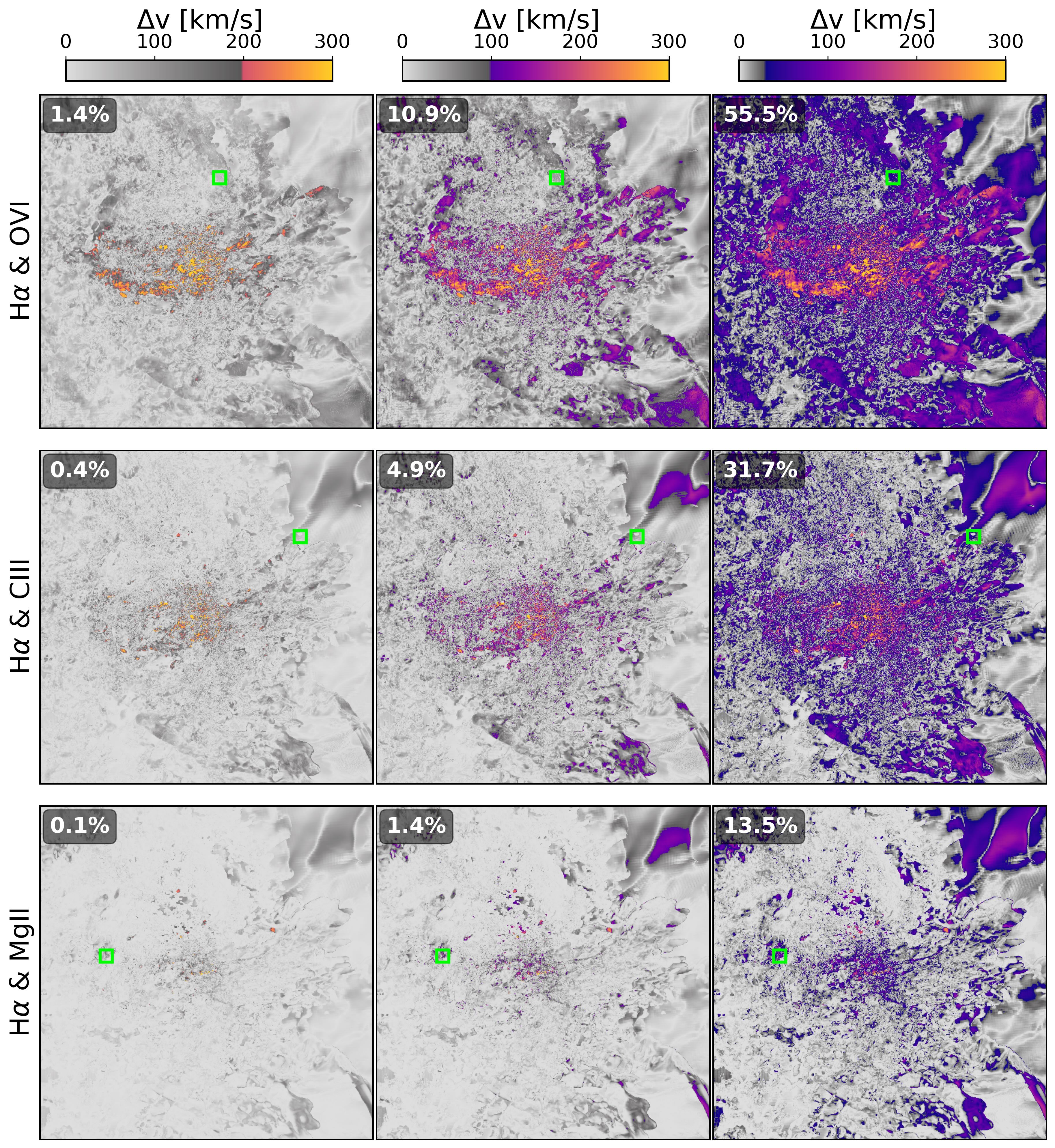}
\vspace{0.5cm} 
\includegraphics[width=0.7\textwidth]{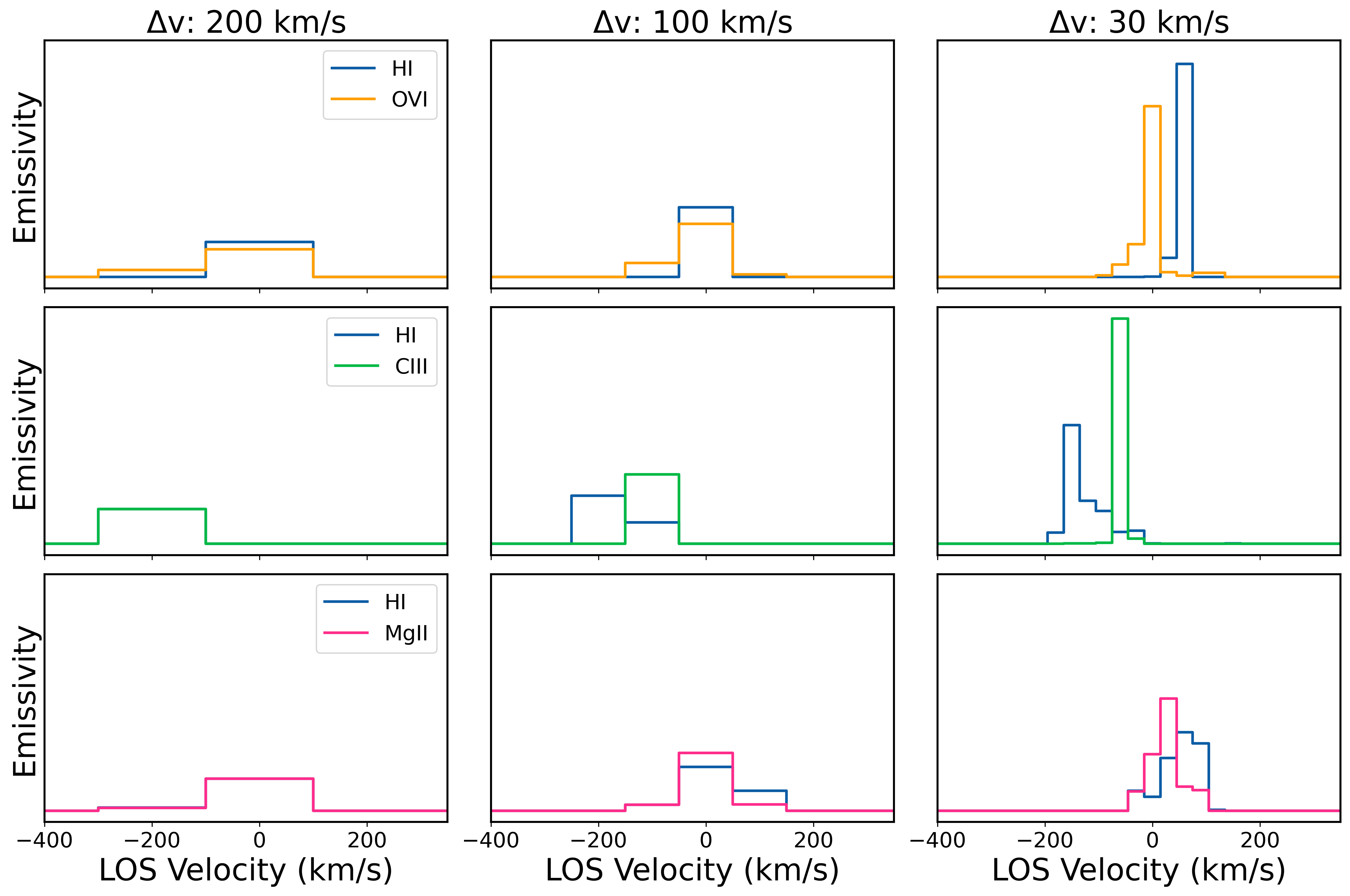}
\caption{Velocity differences between H$\alpha$ and \ion{O}{6}, \ion{C}{3}, and \ion{Mg}{2} in the Cyclone halo. \textit{Top:} Maps show where emissivity-weighted velocity differences exceed kinematic resolution thresholds (200, 100, 30 km s$^{-1}$). Gray pixels are unresolved; colored pixels are kinematically distinguishable. Percentages in the top-left of each panel indicate the fraction of the area with resolved kinematic differences (i.e., colored pixels divided by total area).  \textit{Bottom:} Normalized velocity profiles along the selected sightlines (marked by green boxes in the top panels). Each column corresponds to the same sightline across resolution levels. These profiles highlight how improved spectral resolution enables clearer separation between the velocity structures of different ions.}
\label{fig:dv_vlos}
\end{figure*}

\subsection{Separating Different Gas Phases}\label{sec:phases}

Advancing our understanding of the CGM requires more than just detecting its structures; we must also disentangle its dynamics. In this section, we focus on the instrument capabilities necessary to resolve the kinematics of gas inflow and outflow around galaxies.

In this analysis and discussion, we will consider both ``kinematic'' resolution and ``spectral'' resolution as distinct concepts. The latter generally refers to a fixed property or specification of an instrument, which is generally expressed as $R$, and defined as $R = \lambda_0 / \Delta \lambda$ for a line centered at wavelength $\lambda_0$ observed with a dispersion of $\Delta \lambda$. For a real instrument, this quantity can vary with wavelength or across a field of view, but generally does not vary from observation to observation. By contrast, we use ``kinematic resolution'' to describe the final, delivered velocity separation between two components that can be reliably distinguished in a final, reduced dataset. Unlike ``spectral resolution'', which is a specification of the instrument, kinematic resolution depends on properties of the observation such as noise, instrumental broadening, intrinsic non-Gaussian line profiles, or other unknown effects. In many observations the relevant spectral and kinematic resolution will be similar, but we maintain the distinction in any case to retain the idea that our analysis relates more closely to the final, delivered kinematic resolution and not the original instrument specification. 

In making these simulated maps at different spatial resolutions, we have not accounted for realistic image performance or point-spread functions (PSFs) which should vary widely from instrument to instrument or even across a field in a single instrument. Instead we study the idealized maps resampled with varying levels of binning or pixelation. To simulate a particular instrument's real performance we would need to convolve these maps with a PSF that depends on wavelength, field position, and other properties, and resample, but this complication is beyond the scope of our current investigation. For present purposes the ``spatial resolution'' here is only an approximation to real instrument performance.

Figure \ref{fig:vlosmaps} shows projected maps of the emissivity-weighted average velocity along line of sight for eight ions: H$\alpha$, \ion{O}{6}, \ion{C}{4}, \ion{Si}{4}, \ion{C}{3}, \ion{Si}{3}, \ion{Mg}{2}, and \ion{Si}{2} for the Cyclone halo in an edge-on view. The galactic disk has been removed, as described in Section \ref{sec:methods}, in all panels to isolate the kinematics of the CGM. The gray horizontal line shows where the disk was.

A general trend across all ions is the transition from blue to pink in the projected maps, indicating large-scale radial rotation of gas around the galactic center. This pattern is highlighted by the choice of an edge-on viewing angle.
Although these maps reveal broad velocity structures and trends, distinguishing differences between ions by eye is challenging.
Nevertheless, some patterns emerge.
High-ionization species such as \ion{O}{6} exhibit relatively smooth velocity gradients, consistent with diffuse, large-scale flows, such as the inflows that \citet{Lochhaas2025} finds that \ion{O}{6} emission tends to trace. In contrast, lower-ionization lines like \ion{Mg}{2} and \ion{Si}{2} trace more irregular, small-scale velocity structures, often associated with clumps and filaments \cite[see also][]{Augustin2021}. Other ions such as \ion{C}{4}, \ion{Si}{4}, and \ion{C}{3} show a mix of these patterns, highlighting their role in bridging different thermal phases.

\begin{figure*}[ht!]
\begin{center}
\includegraphics[width=1\textwidth]{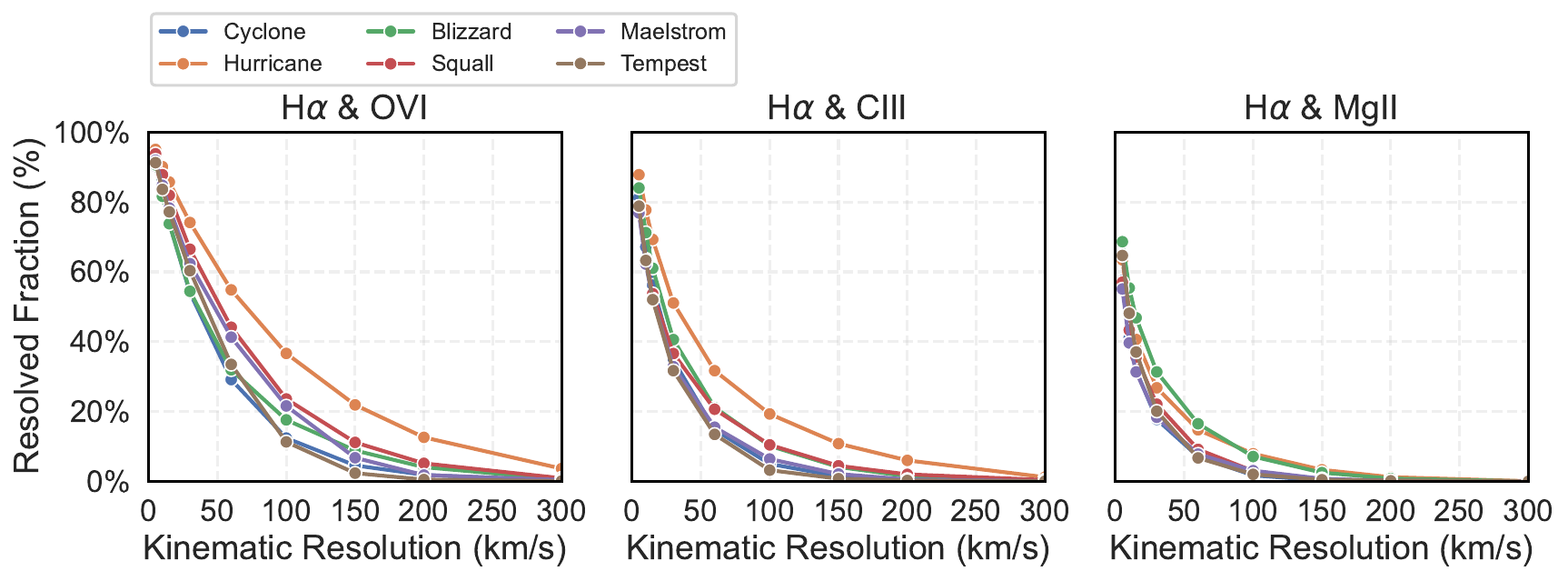}
\caption{Resolved fraction of CGM surface area as a function of kinematic resolution for velocity differences between \ion{H}{1} and three representative ions: \ion{O}{6} (left), \ion{C}{3} (middle), and \ion{Mg}{2} (right). Each line corresponds to one of the six FOGGIE halos. The resolved fraction indicates the percentage of pixels where the velocity difference between the two ions exceeds the given resolution threshold and would therefore be distinguishable by an instrument. O VI shows the largest kinematic differences from \ion{H}{1}. \ion{C}{3} requires finer resolution ($<$ 30 km s$^{-1}$) for broad coverage, while \ion{Mg}{2} remains largely unresolved even at 30 km s$^{-1}$, reflecting its close kinematic alignment with \vadd{H$\alpha$}.}
\label{fig:ions-vdiff}
\end{center}
\end{figure*}

To quantify the kinematic differences between gas traced by different lines, Figure \ref{fig:dv_vlos} presents a detailed comparison between the projected velocities of \ion{H}{1} and three representative species (\ion{O}{6}, \ion{C}{3}, and \ion{Mg}{2}) and examines how the ability to distinguish these differences depends on spectral resolution of an instrument.  We start by explaining the top three: the plots in these rows show spatial maps of absolute line-of-sight velocity differences between \ion{H}{1} and the other ions (row 1: \ion{H}{1}–\ion{O}{6}, row 2: \ion{H}{1}–\ion{C}{3}, row 3: \ion{H}{1}–\ion{Mg}{2}). These maps are computed by subtracting the \ion{H}{1} projected velocity map from the projected velocity map of each ion on a pixel-by-pixel basis, and taking the absolute value. Each column corresponds to a different kinematic resolution: 200 km s$^{-1}$, 100 km s$^{-1}$, and 30 km s$^{-1}$. Colored pixels indicate regions where the velocity difference exceeds the resolution threshold and would therefore be distinguishable by an instrument that achieves that kinematic resolution, while gray pixels mark unresolved regions. The percentages on the top left show how much of the surface area in each map is resolved. These resolved percentages are shown for all six FOGGIE halos in Figure \ref{fig:ions-vdiff}.

The comparison reveals that velocity differences between \ion{H}{1} and \ion{O}{6} are the largest among the three ion pairs. Even at 200 km s$^{-1}$ resolution, as shown in Figure \ref{fig:ions-vdiff}, a larger portion of the CGM shows resolvable differences between \ion{H}{1} and \ion{O}{6} compared to other ions. This reflects the fact that \ion{O}{6} primarily traces hot, diffuse gas \citep{Lochhaas2025} with distinct kinematic behavior from the colder structures traced by \ion{H}{1}. In contrast, \ion{C}{3} shows smaller differences from \ion{H}{1}, with only limited areas resolvable at 200 km s$^{-1}$, and broader coverage at 100 km s$^{-1}$ and 30 km s$^{-1}$. The \ion{H}{1}–\ion{Mg}{2} pair shows the smallest differences; even at 30 km s$^{-1}$ resolution, less than 30\% of the mass is resolved (see right panel of Figure \ref{fig:ions-vdiff}), indicating that these ions are correlated and trace similar cold gas structures with nearly identical kinematics. To resolve more than 60\% of the map area for \ion{H}{1}–\ion{Mg}{2}, kinematic resolution finer than 5 km s$^{-1}$ would be required.

The bottom panel of Figure \ref{fig:dv_vlos} shows normalized emissivity vs line-of-sight velocity for illustrative lines of sight (highlighted in green in the top maps). Each column corresponds to one kinematic resolution bin (200, 100, and 30 km s$^{-1}$ from left to right), while each row compares \ion{H}{1} (blue) to \ion{O}{6} (orange), \ion{C}{3} (green), and \ion{Mg}{2} (pink), respectively. At 200 km s$^{-1}$ resolution, the velocity profiles for \ion{H}{1} and the comparison ion are nearly indistinguishable in each case, therefore they are shown in grey color. However, as the resolution improves to 100 and then 30 km s$^{-1}$, distinct features emerge in the profiles, especially for \ion{O}{6} and \ion{C}{3}. The \ion{Mg}{2} and \ion{H}{1} profiles remain closely aligned even at the highest resolution, reinforcing the result that these ions share tightly coupled kinematics; \citet{Augustin2025} also show that the gas traced by these ions is closely aligned spatially.

Figs. \ref{fig:dv_vlos} and \ref{fig:ions-vdiff} emphasize the critical role of kinematic resolution in disentangling the kinematics of the multiphase CGM. High-ionization species like \ion{O}{6} exhibit distinct velocity structures compared to neutral hydrogen, while lower-ionization lines such as \ion{Mg}{2} often trace cold gas dynamics very similar to \ion{H}{1}. The ability to resolve these differences depends strongly on instrumental resolution: coarse resolutions obscure important velocity offsets, especially between phases, whereas finer resolutions are necessary to capture the diversity of motions across ionization states. These results underscore that recovering the full kinematic picture of the CGM requires not only spatial mapping but also sufficient kinematic precision to separate the dynamics of different phases.

\begin{figure*}[ht!]
\begin{center}
\includegraphics[width=0.95\textwidth]{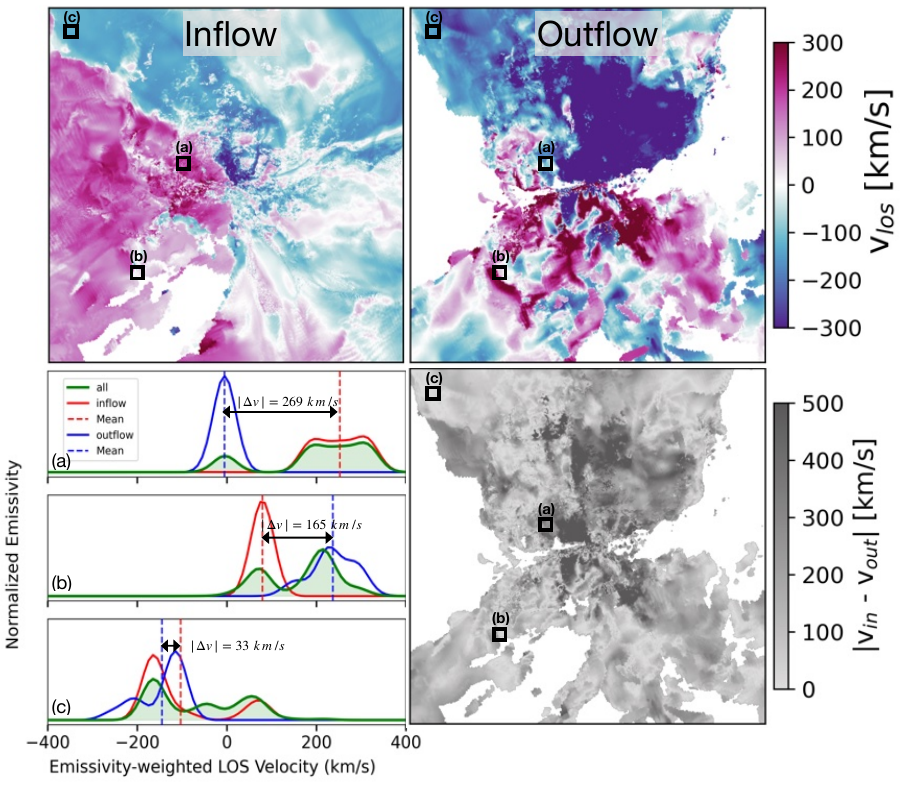}
\caption{ Top panels: Emissivity-weighted line-of-sight velocity maps for O VI in the Maelstrom halo (edge-on view), showing only gas classified as inflow (with a center-directed $v_{\mathrm{rad}} \leq -100$ km s$^{-1}$ in the frame of the galaxy's center of mass; left panel) or outflow ($v_{\mathrm{rad}} \geq 200$ km s$^{-1}$; right panel), based on 3D radial velocities. Bottom left panels (a, b, c): Emissivity-weighted line-of-sight velocity profiles for three selected sightlines (black squares labeled a–c in the top and bottom maps), showing distributions for all gas (green), inflowing gas (red), and outflowing gas (blue). Dashed lines mark the emissivity-weighted mean velocity of each component. The labeled $|\Delta v|$ indicates the absolute difference between the inflow and outflow mean velocities along each sightline. Bottom-right: Grayscale map showing the absolute difference in emissivity-weighted projected velocity between inflow and outflow at each pixel.
This figure illustrates the kinematic signatures of inflowing and outflowing gas, with velocity separations ranging from tens to hundreds of km s$^{-1}$.
\label{fig:vlosdiff-p1}} 
\end{center}
\end{figure*}
\begin{figure*}[ht!]
\begin{center}
\includegraphics[width=0.95\textwidth]{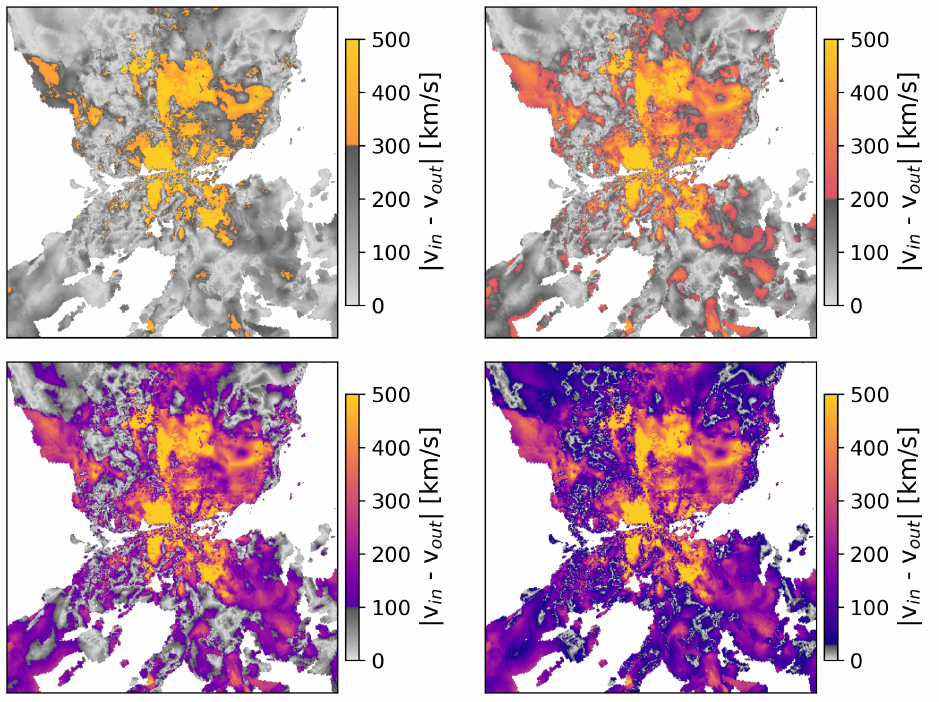}
\caption{ Projected velocity maps for inflowing and outflowing \ion{O}{6}-emitting gas in the Maelstrom halo (edge-on). Each panel shows the same projected map of pixel-by-pixel velocity differences between inflowing and outflowing O VI-emitting gas in the Maelstrom halo (as in the bottom-right panel of Figure \ref{fig:vlosdiff-p1}), but filtered by different kinematic resolution thresholds: 300, 200, 100, and 30 km s$^{-1}$ (top-left to bottom-right). The grayscale background represents the full $|\langle v_{\mathrm{in}} \rangle - \langle v_{\mathrm{out}} \rangle|$ values. Colored pixels highlight regions where the velocity difference exceeds the corresponding kinematic resolution, indicating that an instrument with that velocity resolution could distinguish inflow from outflow in those areas. Gray regions fall below the resolution threshold and would be kinematically unresolved.
The maps demonstrate that increasingly finer kinematic resolution enables inflow–outflow separation across larger fractions of the halo, emphasizing the importance of $\leq$30 km s$^{-1}$ velocity resolution for resolving CGM gas flows in emission.}
\label{fig:vlosdiff-p2}
\end{center}
\end{figure*}

\subsection{Separating Inflow From Outflow From Disk}\label{sec:flows}

\begin{figure}[ht!]
\begin{center}
\includegraphics[width=1\columnwidth]{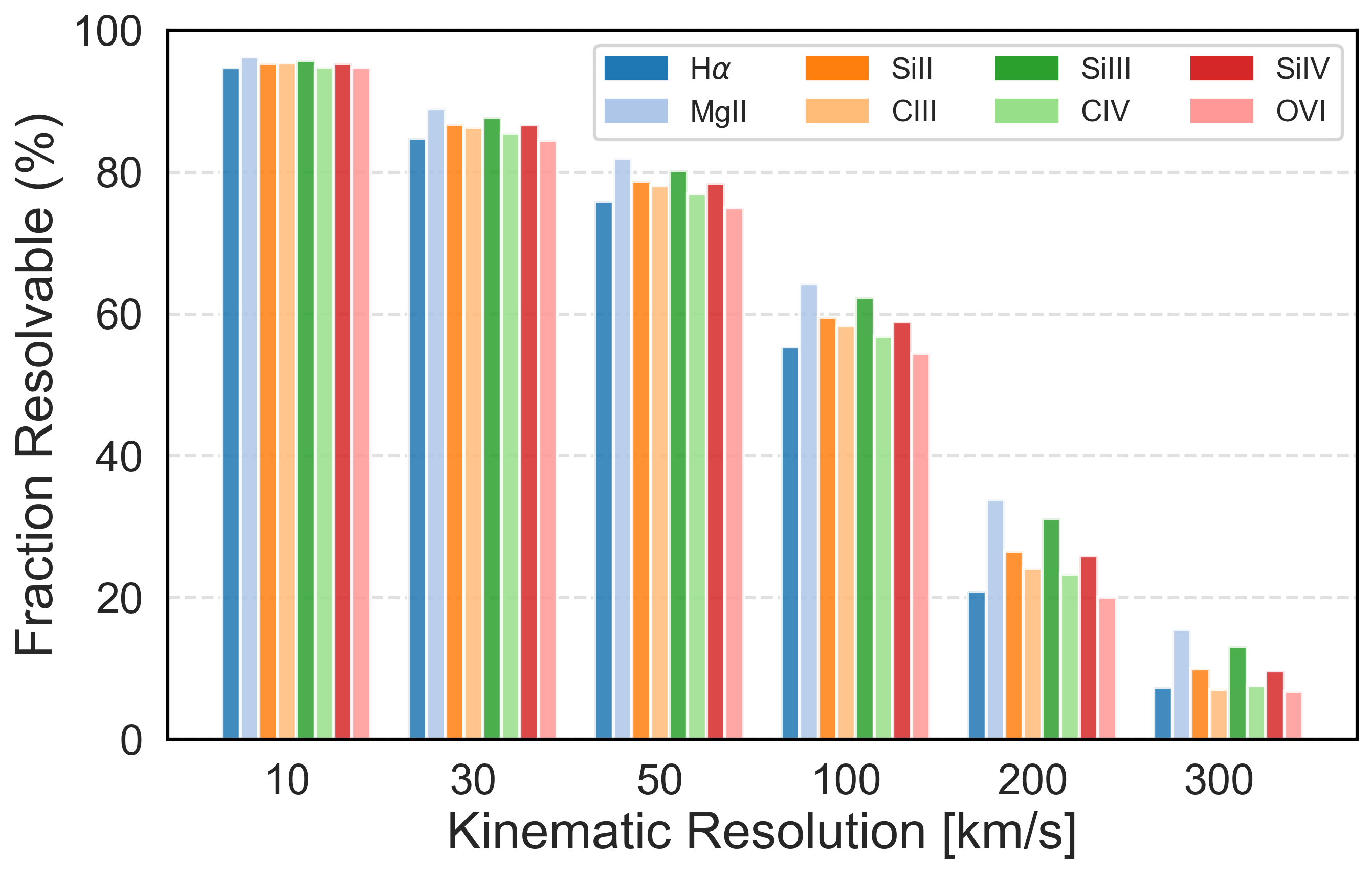}
\caption{Fraction of projected area where inflows and outflows are kinematically distinguishable as a function of kinematic resolution for eight UV emission lines. Values represent the percentage of pixels from Figure~\ref{fig:vlosdiff-p2} in which the absolute difference between inflow and outflow projected velocities exceeds the given kinematic resolution threshold. At low resolution (300 km s$^{-1}$), fewer than 20\% of regions are distinguishable, while $\sim$60\% can be resolved at 100 km s$^{-1}$. A resolution of $< 30$ km s$^{-1}$ is required to distinguish inflow from outflow across more than 80\% of the CGM area in all ions.}
\label{fig:vlosdiff-barchart}
\end{center}
\end{figure}

\begin{figure*}[ht!]
\begin{center}
\includegraphics[width=0.9\textwidth]{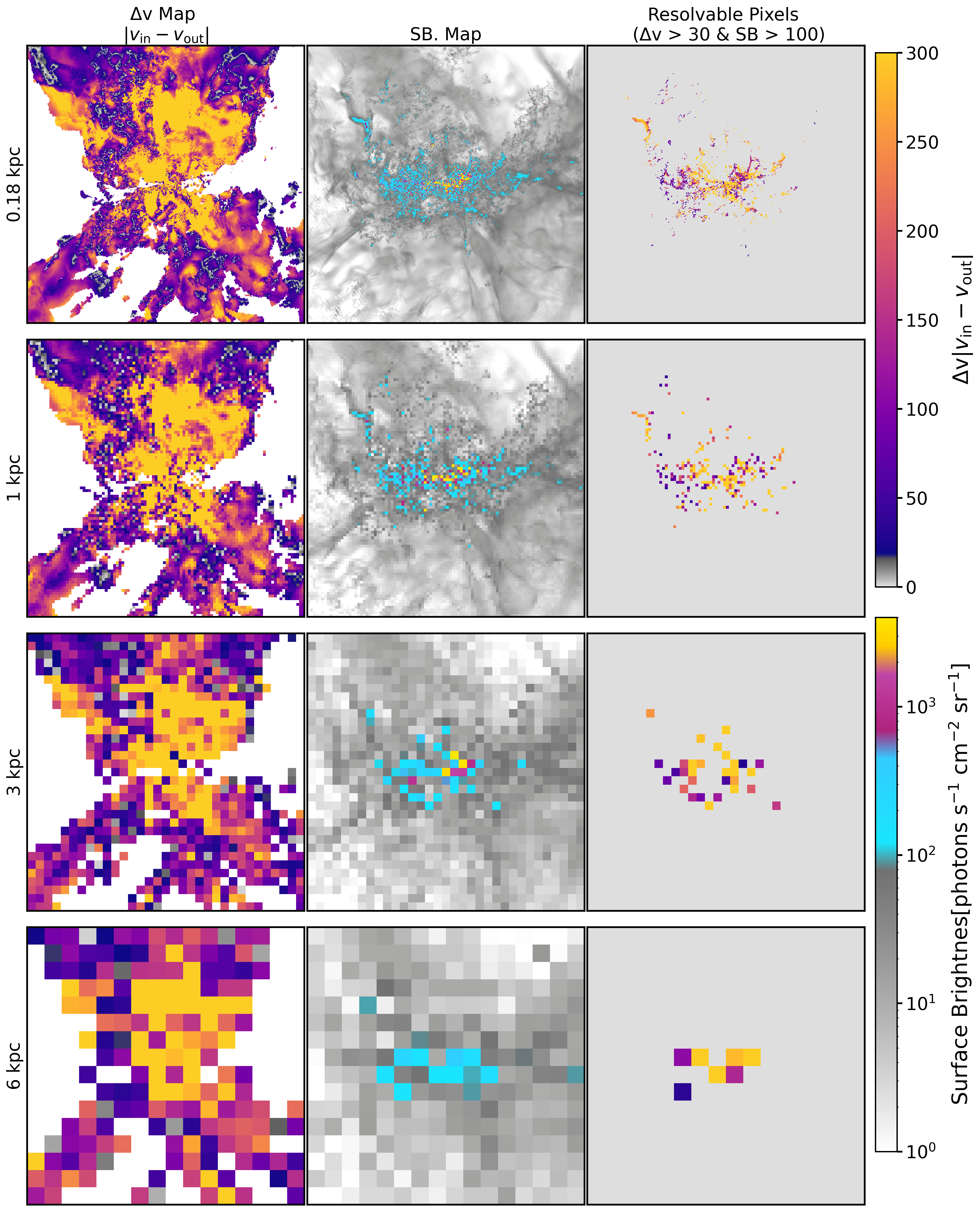}
\caption{Combined effects of spectral and spatial resolution on CGM kinematic observability for the Maelstrom halo using \ion{O}{6} emission.Each row corresponds to a different spatial resolution (\vadd{0.18}, 1, 3, and 6 kpc, which correspond at $z = 0.5$ ($z = 0$) to $\sim$ 0.04 (5.6), 0.16 (20), 0.5 (61), and 1 (120) arcsec, respectively.). Left column: Maps of absolute projected velocity differences ($\Delta v$) between inflowing and outflowing gas (defined as $v_{\rm rad} < -100$ km s$^{-1}$ and $v_{\rm rad} > 200$ km s$^{-1}$, respectively). Gray pixels are unresolved ($\Delta v < 30$ km s$^{-1}$); colored pixels indicate increasing $\Delta v$ detectable at different kinematic resolutions: blue-purple (30–100 km s$^{-1}$), purple-orange (100–200 km s$^{-1}$), and orange-yellow ($>$200 km s$^{-1}$). Middle column: Surface brightness maps showing detectable regions above a sensitivity threshold of 100 photons s$^{-1}$ cm$^{-2}$ sr$^{-1}$ (cyan, purple and yellow). Right column: Pixels that are resolved in both surface brightness and velocity (using $\Delta v = 30$ km s$^{-1}$ and SB$ = 100 $photons s$^{-1}$ cm$^{-2}$ sr$^{-1}$ thresholds), color-coded by velocity difference. This panel shows how improved kinematic resolution enables separation of inflows and outflows in regions already observable in emission. At 30 km s$^{-1}$ resolution, over 90\% of the emission-resolved gas can be kinematically distinguished; at 200 km s$^{-1}$, this fraction drops to $\sim$40–50\%.}
\label{fig:dv_em_res}
\end{center}
\end{figure*}
Beyond distinguishing between different ionization phases, resolving CGM kinematics is also essential for uncovering the underlying motions of the gas like disentangling inflowing gas from outflowing gas. One of the persistent challenges in interpreting CGM observations is separating these opposing gas motions along the line of sight. High kinematic resolution offers a pathway to tackle this challenge by capturing the subtle velocity differences.

To explore this further, we leverage the full 3D velocity information available in the simulations to define inflow and outflow in the CGM. We define inflow as gas with a 3D radial velocity $v_{\mathrm{rad}} \leq -100$ km s$^{-1}$ and outflow as gas with $v_{\mathrm{rad}} \geq 200$ km s$^{-1}$. The inflowing velocity cut of $-100$ km s$^{-1}$ captures gas moving at at least one-third of the free-fall velocity within 50 kpc of the galaxy centers in the halos studied here, while the outflowing velocity cut of $200$ km s$^{-1}$ approximately corresponds to the escape velocity for these halos. These cuts are chosen to isolate coherent radial flows rather than turbulent or random gas motions \citep{Lochhaas2021}.

The top two panels of Figure \ref{fig:vlosdiff-p1} present emissivity-weighted projected velocity maps for O VI in the Maelstrom halo viewed edge-on. The left panel includes only gas classified as inflow, while the right panel includes only outflowing gas. In both projections, only gas meeting the respective inflow or outflow criteria contributes to the emissivity-weighted velocity at each pixel.

The bottom-right panel of Figure \ref{fig:vlosdiff-p1} shows the pixel-by-pixel difference between these two velocity maps. Specifically, each pixel in this grayscale map represents the difference between the emissivity-weighted line-of-sight velocity of outflowing gas and that of inflowing gas at the same location. Darker shades correspond to larger differences in projected velocity between the two components. To help interpret these differences, the bottom-left panel shows three example lines of sight (labeled a, b, c). For each one, the sub-panels illustrate several components: the green histogram shows the velocity distribution of all CGM gas along that sightline (before any classification); the red curve shows the emissivity-weighted velocity distribution of inflowing gas\footnote{Note that the inflow and outflow are defined using 3D velocity information.}; and the blue curve shows the same for outflowing gas. The dashed vertical lines mark the emissivity-weighted \textit{mean} velocities of the red and blue components, respectively. The absolute difference between these two mean values is reported directly on the panel (e.g., $|\Delta v| = 269$ km s$^{-1}$ for line of sight ``a''), and corresponds to the value shown in the grayscale map at the same location. Together, these panels highlight how kinematically distinct inflowing and outflowing gas can appear in projection, and demonstrate the potential for using velocity differences to observationally separate the two flows, provided sufficient kinematic resolution. They also show how differently the velocity components from inflowing and outflowing gas can appear in a velocity profile, highlighting that additional information (such as geometric context) is required to determine which velocity corresponds to which flow of gas.

To assess how well CGM kinematics can be studied observationally, Figure \ref{fig:vlosdiff-p2} builds on the bottom-right panel of Figure \ref{fig:vlosdiff-p1} by incorporating realistic kinematic resolution thresholds. Each panel shows the same map of pixel-by-pixel velocity differences between inflowing and outflowing gas, but now applies a resolution cut to simulate instrumental sensitivity. The color scale remains the same as in Figure \ref{fig:vlosdiff-p1}, representing $|v_{\mathrm{in}}  - v_{\mathrm{out}} |$. In these maps only pixels where the velocity difference exceeds a given threshold are colored; the rest are shown in gray. Colored pixels therefore indicate regions where an instrument with that kinematic resolution could resolve the kinematic separation between inflows and outflows.

The four panels represent different kinematic resolution limits: 300, 200, 100, and 30 km s$^{-1}$, shown from top-left to bottom-right. At the poorest resolution (300 km s$^{-1}$), very few regions meet the criterion for detectability. As the resolution improves to 200 km s$^{-1}$ more structure becomes apparent, especially near the central regions. At 100 km s$^{-1}$, significant portions of the halo show separable inflow and outflow signatures. By 30 km s$^{-1}$, much of the O VI-bright CGM becomes distinguishable, revealing widespread, spatially resolved kinematic differences between the two flow directions.

This analysis demonstrates that resolving the dynamics of CGM gas flows is only possible with high kinematic resolution. For much of the halo, a resolution finer than $\sim$ 30 km s$^{-1}$ is required to capture the full range of O VI-emitting gas motions. These findings emphasize the need for next-generation spectrographs to combine both sensitivity to faint emission and fine velocity resolution to accurately trace CGM gas flows in emission.

While Figs. \ref{fig:vlosdiff-p1} and \ref{fig:vlosdiff-p2} provide illustrative examples for O VI, Figure \ref{fig:vlosdiff-barchart} extends this analysis to all eight ions to quantify the fraction of the projected CGM area that is kinematically resolvable as a function of instrumental resolution. We find that the resolvable fraction increases steeply with improving kinematic resolution. At 300 km s$^{-1}$, less than 20\% of the projected area of inflow/outflow overlap can be distinguished reliably. This fraction rises to $\sim$ 40\% at 200 km s$^{-1}$ and $\sim$ 60\% at 100 km s$^{-1}$. Achieving $\geq$80\% kinematic separation requires a resolution of 30 km s$^{-1}$ or better. While there are small variations between ions, all ions follow a similar trend. 
Note that the absolute values may vary depending on the halo or viewing angle. The trend itself is robust.
Instruments aiming to constrain CGM kinematics should reach kinematic resolutions of $\sim$30 km s$^{-1}$ or better to resolve the complexity of multiphase gas motions traced by UV emission lines.

Figure~\ref{fig:dv_em_res} illustrates the combined effects of spatial and kinematic resolution on the observability of CGM flows. Each row corresponds to a different spatial resolution (\vadd{0.18}, 1, 3, and 6 kpc). The left column shows maps of the absolute differences in emissivity-weighted line-of-sight velocities between inflowing and outflowing gas, as defined above. The color scale indicates the magnitude of the velocity difference ($\Delta v$), with gray regions denoting areas where the velocity difference falls below the threshold of 30 km s$^{-1}$. Colored regions represent areas that would be resolved at increasing levels of kinematic resolution.

The middle column shows surface brightness maps of \ion{O}{6} emission for the same spatial resolutions, identical in style to those in Figure \ref{fig:mapswithspatialres}. Colored pixels indicate regions with surface brightness above a sensitivity threshold of 100 $\rm photons\,s^{-1}\,cm^{-2}\, sr^{-1}$. Gray regions fall below the sensitivity limit and are unresolved in terms of surface brightness.

The right column combines these two criteria — kinematic and spatial resolution — and shows the subset of pixels that are resolved in both velocity ($\Delta v > 30$ km s$^{-1}$) and surface brightness ($\rm SB > 100~photons \, s^{-1} \, cm^{-2}\,sr^{-1}$). These maps represent the CGM gas that would be both detectable and kinematically distinguishable by an instrument with those capabilities. Pixels are color-coded by their $\Delta v$ value, illustrating how changes in kinematic resolution would impact the observable fraction: for instance, only the yellow/orange pixels would remain resolved if the kinematic resolution were limited to 200 km s$^{-1}$.

This figure emphasizes that even if an instrument is sensitive enough to detect faint CGM emission, its kinematic resolution determines whether the dynamics of that gas can actually be measured. For example, at \vadd{0.18} kpc spatial resolution and a surface brightness threshold of 100 photons s$^{-1}$ cm$^{-2}$ sr$^{-1}$, an instrument with 30 km s$^{-1}$ spectral resolution would resolve over 90\% of the detectable gas kinematically, while an instrument with only 200 km s$^{-1}$ resolution would capture only about 40–50\% of it.

\subsection{Analysis of Spatially and Kinematically Resolved CGM}

\jtadd{The previous sections have explored how CGM inflows and outflows populate the spatial and kinematic phase space that is accessible to instruments with varying resolution. Yet this simulation-based analysis does not itself specify or prescribe how analysis of observed data will separate the accretion and feedback components as applied to observed data. There are already numerous data analysis techniques applied to spatially resolved spectroscopy. Three-dimensional ($x$,$y$,$v$) datacubes and 2D channel maps derived from them have long been used in radio astronomy to explore environments from interstellar turbulence \citep[e.g.,][]{2000ApJ...537..720L}, to mainstream star-forming disks \citep[e.g.,][]{2008AJ....136.2563W}, to extraplanar halo gas like that considered here \citep[e.g.,][]{2011A&A...526A.118H}. These techniques have more recently been applied to newer ground-based integral-field spectrographs in the optical and near-infrared, exemplified by surveys like DUVET using KCWI \citep{2022MNRAS.511.5782R}. These velocity-resolved datacubes help trace kinematically distinct structures that would not otherwise be separable. } 

\jtadd{The amount of information contained in these datacubes can be large, and it generally grows as the spatial and kinematic resolution get finer. Various kinds of tools have been developed to extract patterns from these datasets. At their simplest these methods use channel maps directly to seek coherent spatial-kinematic structures \cite[e.g.,][]{2019Natur.574..643R, chas2022}. Beyond simply identifying kinematically distinct structures, extracting physical information about accretion and outflows from velocity-resolved datacubes requires more sophisticated modeling frameworks. Multi-component kinematic decomposition techniques can statistically separate overlapping components and constrain their relative contributions, distinguishing smooth accretion flows from turbulent outflows based on their distinct kinematic signatures in position-velocity space\citep[e.g.,][]{2022MNRAS.511.5782R,2022ApJ...941..163R}. More physically motivated approaches include forward modeling that compares observed channel maps to synthetic observations from simulations, allowing constraints on mass loading factors, opening angles, and velocity structures of galactic winds \citep[e.g.,][]{2021ApJ...919..112D}. For absorption-line studies, kinematic modeling of multi-phase gas has been used to infer cloud geometries and distinguish inflowing from outflowing material based on covering fractions and velocity centroids \citep[e.g.,][]{2014ApJ...794..156R, 2014ApJ...784..108B, 2012MNRAS.426..801B}. Our goal is not to review or evaluate these interpretive frameworks - instead we are quantifying the spatial, and especially the kinematic resolution, needed for there to be enough information for these models to operate on. With simulations, we can tag and follow gas flows with arbitrary precision. Our finding is that the ability to distinguish outflow from inflow is poor with resolution above $\sim 100$ km s$^{-1}$ but rises rapidly below that level. \vadd{To illustrate how improved spectral resolution affects the appearance of spatial–kinematic structures, we present example velocity channel maps in Appendix~\ref{sec:appendixc04}. These maps demonstrate how coherent structures that are blended at coarse velocity resolution become distinguishable when the data are resolved into finer velocity bins.} These findings can guide the collection and interpretation of new data and the design of future instruments, which can then use or extend these modeling techniques to extract physical insights.}

\section{Current and Future Emission Mapping Instruments \label{sec:instruments}}

Faint, extended emission poses strong challenges to our telescopes and instrumentation: at 1000 photons cm$^{-2}$ s$^{-1}$ sr$^{-1}$, for a source subtending 1 square arcsecond, a telescope of 8 meters in diameter will collect {\it one} source photon in 100 seconds of observation. Achieving a signal-to-noise ratio SNR = 5, with only Poisson shot noise accounted for, requires 2500 photons cm$^{-2}$ s$^{-1}$ sr$^{-1}$, larger spatial bins on the sky, a larger telescope, or a brighter source. 

Simulations can assist with guiding requirements for instruments optimized for different kinds of observations. Simulations can also stimulate and inform the development of new instrument concepts by examining how physical outcomes depend on hypothetical instrument parameters. With detailed analysis of the FOGGIE simulations, we are able to quantify how sensitivity, wavelength coverage, and access to multiple ions recover the multiphase structure of the CGM in emission line maps. High spatial resolution (on the order of $\sim$1 kpc) enables detection of small-scale features, constraining the morphology and clumpiness of the gas. Sensitivity to low surface brightness ($\lesssim 500$ photons cm$^{-2}$ s$^{-1}$ sr$^{-1}$) can reveal faint, extended structures and therefore recover a larger fraction of the ionized gas mass. Finally, kinematic resolution plays a critical role in completing our understanding of the CGM. While sensitivity determines whether CGM emission can be detected at all, kinematic resolution governs whether we can disentangle its dynamics. Distinguishing different velocity components, such as inflowing vs. outflowing gas, and recovering kinematics across ionization phases becomes increasingly feasible at kinematic resolutions of $\lesssim 30$ km s$^{-1}$. 

Building on these benchmarks, this section reviews a set of current and future instruments that have been or are intended to be applied to this problem. Figure~\ref{fig:instruments} puts these instruments within this parameter space, using instrument properties listed in Table~\ref{tab:instrument-table}. 
\vadd{We include two categories of instruments: Space-based instruments/ satellite concepts and ground-based instruments }

The instruments already being used for CGM emission studies in Figure~\ref{fig:instruments} and Table~\ref{tab:instrument-table} are the Multi Unit Spectroscopic Explorer \citep[MUSE,][]{MUSE2008} the Keck Cosmic Web Imager Integral Field Spectrograph \citep[KCWI,][]{kcwi2018}, the Circumgalactic H$\alpha$ Spectrograph at the MDM Observatory \citep{2024ApJ...974..161M}, and the Dragonfly Telephoto Array \citep{Abraham2014,Lokhorst2019,chen2024}. Instruments operating in optical wavelength on large telescopes have the advantage of high spatial resolution and large collecting areas; conversely the suite of ions they can access in the rest frame (e.g., \ion{H}{1}, \ion{O}{2}) limits the temperature range of the gas they probe. The UV ions we focus on here are available at higher redshift and have been used to good effect with these powerful ground-based IFUs \citep[e.g.,][]{2019NatAs...3..822M, Kusakabe2024, 2025ApJ...986...87H}. 

\vadd{Here we first describe the quantities listed in Table~\ref{tab:instrument-table} before comparing instrument capabilities. The second column lists the observed-frame wavelength coverage of each instrument as defined by its design specifications. Third column lists the lines available to each instrument, out of all emission lines considered in this study (see Table~\ref{tab:emission_lines}). This does not imply that the instrument is limited to observing only the listed lines; rather, we include only the subset of lines studied in this paper that are accessible to each facility.
Then for each accessible line, we compute the corresponding redshift range using
$z = \frac{\lambda_{\rm obs}}{\lambda_{\rm rest}} - 1 $,
and report in column 4 the range of redshifts over which that line can be observed given the instrument’s wavelength coverage.
}

\vadd{The angular resolution listed in Table~\ref{tab:instrument-table} corresponds to the delivered point-spread function (PSF) full width at half maximum (FWHM) on the sky in a representative observing configuration appropriate for extended, low-surface brightness CGM emission (e.g., seeing-limited wide-field mode for ground-based facilities and diffraction-limited performance for space-based concepts; see Table comments). These values reflect the effective spatial resolution for diffuse emission and are not determined by detector pixel scale or IFU slice width.
For each emission line, the corresponding physical spatial resolution (in kpc) is computed at the lowest redshift of the accessible redshift range listed for that line and instrument ($z_{\rm min}$), using the adopted \citet{Planck2016} cosmology. For entries with $z_{\rm min} = 0 $, we evaluate the conversion at  $z = 0.001$. By evaluating the physical resolution at $z_{\rm min}$, we report the finest physical scale achievable for that line with a given instrument; at higher redshift within the accessible range, the physical resolution becomes correspondingly coarser.}

\vadd{The spectral resolution is reported as the instrumental resolving power, $R = \lambda_0 / \Delta \lambda$, evaluated at the observed wavelength of each emission line at the lowest redshift of its accessible range $z_{\rm min}$. The corresponding instrumental velocity FWHM of one spectral resolution element is given by $\Delta v = c/R $.
For instruments offering multiple gratings, slicers, or operational modes, we adopt a representative configuration commonly used for extended, low-surface brightness emission, as summarized in Table~\ref{tab:instrument-table} comments.}


\vadd{The surface brightness (SB) limits listed in Table~\ref{tab:instrument-table} are derived from published emission-line flux or surface brightness sensitivities for each facility, as reported in the references listed in the table comments. Where surface brightness sensitivities are directly reported per arcsec$^{2}$, those values are adopted and converted only in units. For instruments that report point-source line-flux limits, we convert these to surface brightness limits by assuming the emission is distributed uniformly over an area of one square arcsecond on the sky.
All surface brightness limits are quoted per spectral resolution element corresponding to the instrumental line-spread function FWHM (i.e., using the same $\Delta v$ listed in the table), and are expressed in observed-frame units of photons cm$^{-2}$ s$^{-1}$ sr$^{-1}$. When sensitivity is provided in erg s$^{-1}$ cm$^{-2}$ arcsec$^{-2}$, we convert to photon surface brightness units at the observed wavelength of each emission line using the photon energy $E_{\rm ph}=hc/\lambda_{\rm obs}$. Exposure-time scaling is assumed to follow background-limited behavior, with flux limits proportional to $t^{-1/2}$, allowing rescaling to the 10 hr and 100 hr values reported in the table.
These limits therefore represent observed surface brightness sensitivities per arcsec$^{2}$ per spectral resolution element in the representative configuration adopted for each instrument. They are intended for comparative assessment of instrumental capabilities and do not account for additional systematic effects such as sky-subtraction residuals or correlated noise in reduced data products.
For instruments where sensitivities are reported for extended emission over larger spatial scales or in survey-level form, we convert or reinterpret these values to an effective surface brightness per arcsec$^{2}$ using the published assumptions, in order to enable consistent comparison across instruments.}


\vadd{Looking at these surface brightness limits, it} is perhaps surprising that for faint CGM emission, telescopes at the opposite end of the size scale can compete with 8-meter-class facilities on the ground. In space, the full range of diagnostic UV ions is available at $z<1$, offering the largest practical set of physical diagnostics. Sky backgrounds are substantially darker in the UV than in the optical. And, for diffuse emission, spatial resolution adequate to resolve kiloparsec-scale structures can be achieved even by meter-scale telescopes. This region of the trade space has been pioneered by the Aspera SmallSat \citep{Aspera2021}, now in development, followed by proposed missions such as the Juniper CubeSat \citep{witt2025juniper} and the MAGPIE Pioneer concept \citep{MAGPIE2025}. Further down the road there is the prospect for future flagship concepts like the Habitable Worlds Observatory (HWO), which apply extreme sensitivity at fine spatial resolution. 

Figure~\ref{fig:instruments} presents a set of 2D comparisons for individual characteristics and a 3D plot that brings together all of the critical parameters discussed above. \vadd{We show these comparisons for two representative emission lines: H$\alpha$ in the optical and the \ion{O}{6} 1032,1038 doublet in the UV. We select \ion{O}{6} because it is the only UV line that can be observed by all instruments in Table~\ref{tab:instrument-table} with ultraviolet capability.} 
\vadd{The} error bars indicate the range of surface brightness limits corresponding to exposure times of $T_{\rm exp} = 10$ and 100 hours, while the marker denotes the mean value.

The first panel plots spatial resolution on the y-axis, converted from angular resolution to physical kpc by assuming the lowest redshift targets \vadd{
accessible to each instrument. For the H$\alpha$ panels in the right column, the minimum redshift is $z_{\rm min}=0$ for all instruments. As noted in the caption of Table~\ref{tab:instrument-table}, we adopt $z=0.001$ when converting to physical units in these cases, which results in very small spatial resolutions ($<0.1$ kpc).}

The second panel shows the spectral resolution of each instrument in km s$^{-1}$.
\vadd{The third panel shows the number of lines accessible to each instrument from the list of lines studied in this work. Basically we count the number of lines listed in third column of Table~\ref{tab:instrument-table}, with OVI doublet counted as two lines. This serves as a proxy for how well each instrument can probe the multiphase structure and mass budget of the CGM.The higher number of lines covered, a better estimation of multiphase CGM mass budget. }

The final panel combines these parameters into a single 3D view: spatial resolution, spectral resolution, and sensitivity are placed along the three axes, while \vadd{number of lines} is encoded in the marker size (\vadd{larger symbols correspond to a larger number of accessible emission lines}). In all panels, current instruments are represented with star symbols and upcoming/future missions with circles. Gray shaded regions in the 2D panels, and a gray box in the 3D panel, indicate the parameter space identified in this study as critical for detecting faint, multiphase CGM emission and for resolving its underlying kinematics.

\begin{deluxetable*}{llcccccccc}
\tablecaption{Instrumental parameters for CGM emission lines\label{tab:instrument-table}. }
\tabletypesize{\footnotesize}
\tablehead{
\colhead{Instrument} &
\colhead{Wavelength Coverage} &
\colhead{Line} &
\colhead{$z_{\rm min}$--$z_{\rm max}$\tablenotemark{a}} &
\multicolumn{2}{c}{Spatial Resolution} &
\multicolumn{2}{c}{Spectral Resolution} &
\multicolumn{2}{c}{SB Limits\tablenotemark{e}}
\\
\colhead{} &
\colhead{(\AA)} &
\colhead{} &
\colhead{} &
\colhead{$\theta_{\rm PSF}$\tablenotemark{b}} &
\colhead{$r_{\rm phys}$\tablenotemark{c}} &
\colhead{$R$\tablenotemark{d}} &
\colhead{$\Delta v$\tablenotemark{d}} &
\colhead{10 hr} &
\colhead{100 hr}
\\
\colhead{} &
\colhead{} &
\colhead{} &
\colhead{} &
\colhead{(arcsec)} &
\colhead{(kpc)} &
\colhead{--} &
\colhead{(km s$^{-1}$)} &
\multicolumn{2}{c}{(ph cm$^{-2}$ s$^{-1}$ sr$^{-1}$)}
}
\startdata
\multicolumn{10}{c}{\textbf{Space-based instruments / satellite concepts}} \\[2pt]
\hline
\multirow{7}{*}{HWO} &
\multirow{7}{*}{900--3000} &
\ion{Mg}{2}~2798   & 0--0.07 & 0.05 & 0.0011 & 15000 & 20 & 7550 & 2397 \\
& & \ion{Si}{2}~1260   & 0--1.38 & 0.05 & 0.0011 & 15000 & 20   & 3400 & 1079\\
& & \ion{Si}{3}~1207   & 0--1.49 & 0.05 & 0.0011 & 15000 & 20  & 3260 & 1034 \\
& & \ion{C}{3}~1910   & 0--0.57 & 0.05 & 0.0011 & 15000 & 20  & 5160 & 1636  \\
& & \ion{Si}{4}~1394    & 0--1.15 & 0.05 & 0.0011 & 15000 & 20  & 3770 & 1194 \\
& & \ion{C}{4}~1548    & 0--0.94 & 0.05 & 0.0011 & 15000 & 20  & 4190 & 1326 \\
& & \ion{O}{6}~1032,1038     & 0--1.91 & 0.05 & 0.0011 & 15000 & 20  & 2800 & 887 \\
\hline
\multirow{1}{*}{Aspera} &
\multirow{1}{*}{1030--1040} &
\ion{O}{6}~1032,1038  & 0--0.01 & 45 & 0.964 & 2000 & 150 & 3123  & 987 \\
\hline
\multirow{3}{*}{Juniper} &
\multirow{3}{*}{1025--1929} &
\ion{C}{3}~1910  & 0--0.01 & 20 & 0.429 & 7500 & 40 & 4167 & 1319 \\
& & \ion{C}{4}~1548    & 0--0.25 & 20 & 0.429 & 7500 & 40 & 3392 & 1074 \\
& & \ion{O}{6}~1032,1038   & 0--0.87 & 20 & 0.429 & 7500 & 40 & 2256 & 713\\
\hline
\multirow{6}{*}{MAGPIE} &
\multirow{6}{*}{980--2095} &
\ion{Si}{2}~1260   & 0--1.38 & 4.5 & 0.09 & 2100 & 143 & 1686 & 495 \\
& & \ion{Si}{3}~1207   & 0--1.38 & 4.5 & 0.09 & 2012 & 149 & 1615 & 474 \\
& & \ion{C}{3}~1909   & 0--0.57 & 4.5 & 0.09 & 1909 & 157 & 2554 & 750 \\
& & \ion{Si}{4}~1394   & 0--1.15 & 4.5 & 0.09 & 2323 & 129 & 1865 & 547 \\
& & \ion{C}{4}~1548     & 0--0.94 & 4.5 & 0.09 & 2580 & 116 & 2071 & 608 \\
& & \ion{O}{6}~1032,1038  & 0--1.91 & 4.5 & 0.09 & 1720 & 174 & 1381 & 405 \\
\hline
\multicolumn{10}{c}{\textbf{Ground-based instruments}} \\[2pt]
\hline
\multirow{8}{*}{MUSE} &
\multirow{8}{*}{4650--9300} &
H$\alpha$ 6563 & 0--0.42 & 0.4 & 0.0086 & 2500 & 120 & 1096 & 346 \\
& & \ion{Mg}{2}~2798    & 0.66--2.33 & 0.4 & 2.872 & 1770 & 170 & 778& 246 \\
& & \ion{Si}{2}~1260    & 2.69--6.38 & 0.4 & 3.253 &  1770 & 170 & 778& 246  \\
& & \ion{Si}{3}~1207  & 2.85--6.71 & 0.4 & 3.204 & 1770 & 170 & 778& 246 \\
& & \ion{C}{3}~1910  & 1.43--3.87 & 0.4 & 3.468 & 1770 & 170 & 778& 246  \\
& & \ion{Si}{4}~1394    & 2.34--5.67 & 0.4 & 3.354 & 1770 & 170& 778& 246  \\
& & \ion{C}{4}~1548     & 2.00--5.01 & 0.4 & 3.434 & 1770 & 170& 778& 246  \\
& & \ion{O}{6}~1032,1038     & 3.51--8.01 & 0.4 & 2.996 &  1770 & 170 & 778& 246  \\
\hline
\multirow{8}{*}{KCWI} &
\multirow{8}{*}{3500--10500} &
H$\alpha$ 6563 & 0--0.60 & 0.6 & 0.0129 & 3000 & 100 & 1030 & 325 \\
& & \ion{Mg}{2}~2798    & 0.25--2.76 & 0.6 & 2.420 & 3000 & 100 & 547 & 173 \\
& & \ion{Si}{2}~1260    & 1.78--7.33 & 0.6 & 5.201 & 3000 & 100 & 547 & 173 \\
& & \ion{Si}{3}~1207   & 1.90--7.70 & 0.6 & 5.177 & 3000 & 100 & 547 & 173 \\
& & \ion{C}{3}~1910   & 0.83--4.50 & 0.6 & 4.691 & 3000 & 100 & 547 & 173 \\
& & \ion{Si}{4}~1394    & 1.51--6.53 & 0.6 & 5.213 & 3000 & 100 & 547 & 173 \\
& & \ion{C}{4}~1548    & 1.26--5.78 & 0.6 & 5.145 & 3000 & 100 & 547 & 173 \\
& & \ion{O}{6}~1032,1038     & 2.39--9.17 & 0.6 & 5.011 & 3000 & 100 & 547 & 173 \\
\hline
\multirow{1}{*}{CH$\alpha$S} &
\multirow{1}{*}{4000--9000} &
H$\alpha$ & 0--0.37 & 1.3 & 0.0279 & 15000 & 20 & 1580 & 475 \\
\hline
\multirow{1}{*}{Dragonfly} &
\multirow{1}{*}{5007--6563} &
H$\alpha$ & 0--0.2 & 2.8 & 0.0600 &750 & 400 & 2450 & 775 \\
\enddata

\tablenotetext{a}{Accessible redshift range for each line is computed from the wavelength coverage via $z=\lambda_{\rm obs}/\lambda_{\rm rest}-1$. }
\tablenotetext{b}{Angular resolution is the delivered PSF FWHM in a representative configuration appropriate for extended, low-surface brightness CGM emission (see note below).}
\tablenotetext{c}{Physical spatial resolution is computed at $z_{\rm min}$ using \citet{Planck2016} cosmology; for entries with $z_{\rm min}=0$, we adopt $z=0.001$ for calculation.}
\tablenotetext{d}{Spectral resolution is reported as resolving power $R$ evaluated at the observed wavelength of each line at $z_{\rm min}$. The corresponding instrumental velocity FWHM of one spectral resolution element is given by $\Delta v = c/R$.}
\tablenotetext{e}{Surface brightness limits are quoted per arcsec$^{2}$ per spectral resolution element (consistent with $\Delta v$) in observed-frame units and scaled to 10 hr and 100 hr assuming background-limited behavior ($\propto t^{-1/2}$). For concept missions such as HWO, we adopt representative SB sensitivities based on current design estimates, assuming emission averaged over a 1\arcsec $\times$ 1\arcsec. For wide-field instruments (Dragonfly, CH$\alpha$S), values represent effective surface brightness sensitivities reinterpreted from larger-scale measurements assuming uniform emission.}

\tablecomments{Representative observing configurations adopted: MUSE (WFM, seeing-limited); KCWI (representative moderate-resolution configuration appropriate for extended emission); CH$\alpha$S (standard configuration); Dragonfly (narrowband H$\alpha$ configuration); HWO, Juniper, Aspera, and MAGPIE use the design resolving powers and diffraction-limited (space) or design-reported performance described in their respective concept studies.}

\tablecomments{Sources: MUSE (\url{https://www.eso.org/sci/facilities/develop/instruments/muse.html}), KCWI (\url{https://www2.keck.hawaii.edu/inst/kcwi/primer.html}), CH$\alpha$S \citep{2022ApJ...941..185M}, HWO \citep{2019arXiv191206219T}, Aspera \citep{Chung2021}, Juniper \citep{witt2025juniper}, MAGPIE \citep{MAGPIE2025}, Dragonfly \citep{Lokhorst2019}.}
\end{deluxetable*}
\begin{figure*}[htbp]
\begin{center}
\includegraphics[width=0.95\linewidth]
{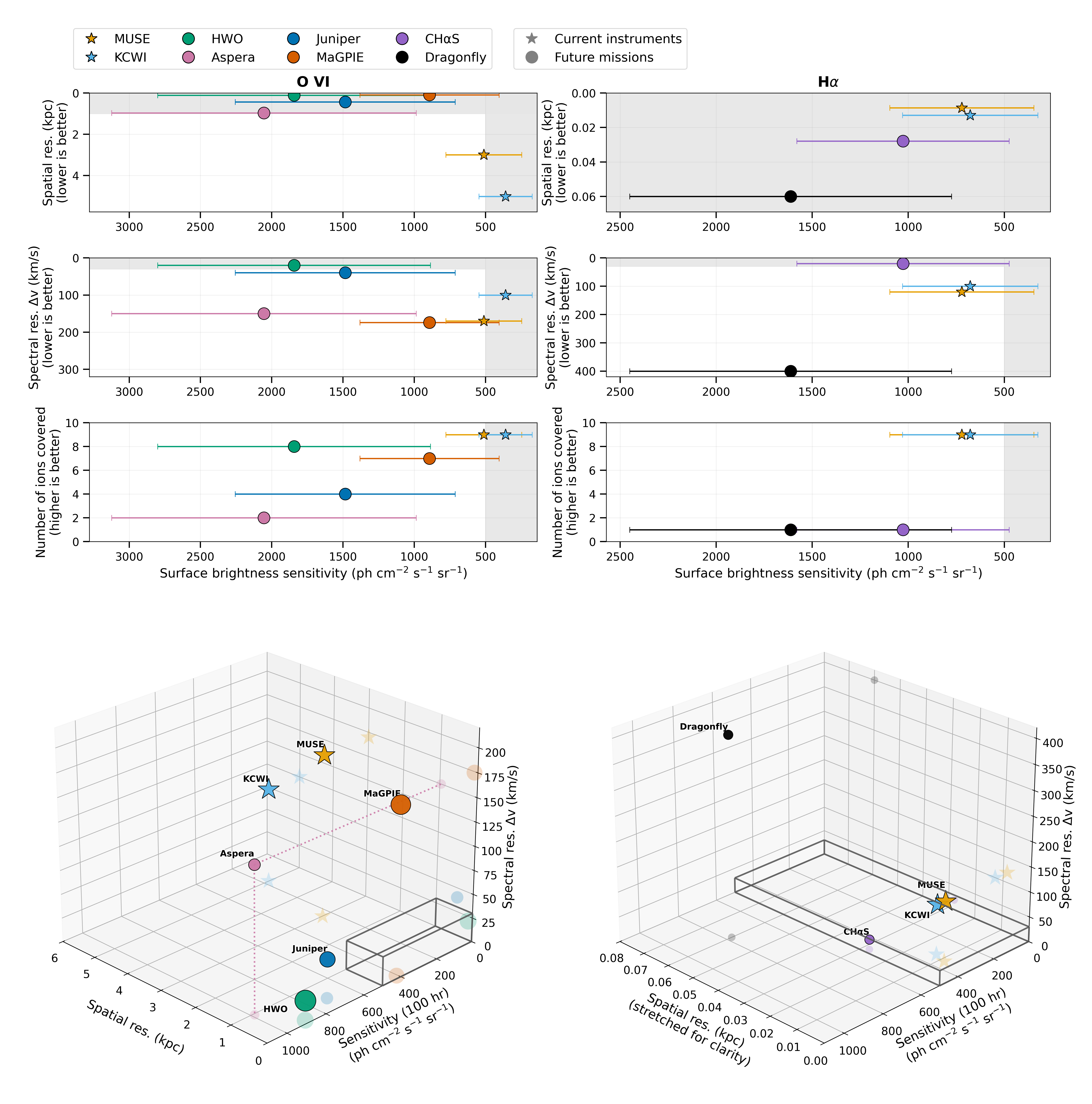}
\caption{Comparison of current (stars) and future (circles) instruments designed/suitable for CGM studies in emission across multiple performance planes.\vadd{The left column shows performance for O VI, while the right column shows H$\alpha$, including only the instruments capable of observing each line.} Sensitivity (photons s$^{-1}$ cm$^{-2}$ sr$^{-1}$) is shown on the x-axis of the three 2D panels. First panel: Spatial resolution versus sensitivity highlights mapping capability. Second panel: kinematic resolution versus sensitivity illustrates kinematic resolving power. Third panel: \vadd{Number of emission lines covered (from the list considered in this study)} versus sensitivity shows multiphase detection capability. Grey shaded regions mark regimes of particularly strong performance (sensitivity $\lesssim 500$ photons s$^{-1}$ cm$^{-2}$ sr$^{-1}$ combined with spatial resolution $<1$ kpc, kinematic resolution $<30$ km s$^{-1}$). Last panel: A 3D view combines all three axes, spatial resolution, sensitivity \vadd{(100 hr)}, and kinematic resolution, with marker size proportional \vadd{to the number of emission lines covered}, providing a holistic comparison of current and future instruments. The gray box corresponds to the shaded regions in the 2D panels. These performance thresholds serve as a guide for the design of next-generation instruments.}
\label{fig:instruments}
\end{center}
\end{figure*}
Starting with the first 2D panel, the gray shaded region on the right \vadd{side of each panel} marks the sensitivity required to detect faint, extended CGM emission. 
\vadd{
Considering first the left column, which shows the case for \ion{O}{6}, KCWI lies within the shaded region among current facilities, demonstrating its capability to probe the faint end of CGM emission. MUSE also approaches this regime for sufficiently long integration times. Among the upcoming missions, MAGPIE reaches the sensitivity requirement with high exposure time, while HWO, Juniper and Aspera remain outside the shaded region. Both HWO and Juniper approach the sensitivity threshold for exposure times of 100 hours but do not fully reach it.
In the H$\alpha$ panels (right column), both MUSE and KCWI reach the shaded sensitivity regime for long exposure times. CH$\alpha$S also falls within this region, while Dragonfly remains well outside it. Note that the sensitivities are different for KCWI and MUSE for UV vs H$\alpha$ lines because we used the lowest redshift they are available to these instruments.}

The shaded region at the top of the first two panels indicates the ideal spatial resolution, $\lesssim 1$ kpc, needed to resolve small-scale structures and to detect the morphology and clumpiness of the CGM. 
\vadd{In the \ion{O}{6} panels, current ground-based instruments such as MUSE and KCWI fall short of this threshold. This is largely because these facilities detect rest-frame UV emission only at higher redshifts, where the corresponding physical spatial scales are larger, resulting in coarser resolution in kpc. In contrast, all of the future facilities shown in the figure fall within the shaded region. The small-satellite concepts (Aspera, Juniper, MAGPIE) are designed to observe low-redshift galaxies, where nearby targets and smaller slicer scales allow sub-kpc physical resolution to be achieved. HWO stands out in particular because its angular resolution of $0.05''$ allows it to maintain excellent spatial resolution.
In the H$\alpha$ panels, all instruments lie within the shaded region. Because these observations target very low-redshift galaxies, the corresponding physical spatial scales are extremely small. As a result, the y-axis has been expanded to values below 0.08 kpc in order to display the data points separately.}


\vadd{The second panel shows the kinematic resolution on the y-axis, expressed as the velocity width of one spectral resolution element, $\Delta v$. The gray shaded region ($\lesssim 30$ km s$^{-1}$) marks the regime most favorable for resolving CGM kinematics. The values shown here correspond to the instrumental resolving powers listed in Table~\ref{tab:instrument-table}.
Considering the \ion{O}{6} panels, none of the current ground-based facilities reach this threshold. KCWI performs better than MUSE, achieving a spectral resolution of $\sim 100$ km s$^{-1}$ compared to $\sim170$ km s$^{-1}$ for MUSE. Although this is coarser than the ideal regime, it does not preclude useful kinematic measurements. For example, Figure~\ref{fig:vlosdiff-barchart} shows that with $\sim100$ km s$^{-1}$ resolution, roughly 60\% of the field of view can still be kinematically resolved.
Among the upcoming missions, only HWO and Juniper reach the shaded region. The Juniper CubeSat stands out for its high spectral resolution, enabling recovery of CGM kinematics at low redshift. HWO will likewise achieve the precision required to resolve velocity components across multiple phases. In contrast, Aspera and MAGPIE are less optimized for detailed kinematic studies.
In the H$\alpha$ panels, CH$\alpha$S reaches the shaded region with a spectral resolution of $\sim20$ km s$^{-1}$, making it well suited for detailed kinematic studies of ionized gas. KCWI also approaches this regime with $\sim100$ km s$^{-1}$ resolution, while MUSE operates at slightly coarser resolution ($\sim120$ km s$^{-1}$). Dragonfly, designed primarily for deep imaging rather than spectroscopy, has much lower spectral resolution ($\sim400$ km s$^{-1}$) and therefore is not optimized for resolving CGM kinematics.}

\vadd{The third 2D panel shows the number of emission lines from our study that are accessible to each instrument. A larger number indicates that an instrument can probe a broader range of CGM phases and therefore place stronger constraints on the total CGM mass budget. Here both panels in two columns should be considered together when comparing instruments.
Among current facilities, both MUSE and KCWI can observe all nine lines considered in this study, providing strong leverage for detecting multiphase CGM structure. Looking ahead, HWO stands out with the highest coverage among the proposed missions, enabling access to the largest suite of diagnostic ions. MAGPIE follows, targeting a subset of key ions that span both low- and high-ionization states. Juniper provides access to three of the nine lines in our list, while Aspera can observe only the \ion{O}{6} doublet (counted here as two lines). Finally, the H$\alpha$-focused instruments, CH$\alpha$S and Dragonfly, probe only the H$\alpha$ line. Despite this narrower wavelength coverage, such targeted observations can still provide valuable insights into the structure of the CGM. Aspera, for example, is intentionally designed to focus exclusively on \ion{O}{6}.}

Finally, the 3D panel places each instrument in a parameter space defined by spatial resolution, spectral resolution, and sensitivity, using the best available values from Table~\ref{tab:instrument-table}. Marker size corresponds to the wavelength range covered by each instrument. For clarity, each marker is mirrored as a shadow onto the spatial–sensitivity and sensitivity–spectral resolution planes, with a dotted guide line for Aspera to aid the reader’s eye. This visualization highlights where current instruments sit relative to the critical parameter space identified in this study, and how future facilities are expected to advance. Markers lying closer to the gray box indicate better overall performance. 
\vadd{Among current facilities, MUSE and KCWI occupy complementary regions of this space. Both instruments achieve sensitivities that allow the detection of faint CGM emission and can access a large number of emission lines at high redshift. However, their spatial resolution in physical units is limited when observing rest-frame UV emission, and their spectral resolution remains above the ideal regime for resolving detailed CGM kinematics. In the H$\alpha$ case, these instruments perform better in spatial resolution because of the small physical scales implied by the low-redshift conversion used in Table~\ref{tab:instrument-table}.
Looking ahead, several proposed missions move closer to the optimal region of parameter space. Juniper provides strong spectral resolution and is well suited for kinematic studies of nearby galaxies, while MAGPIE offers improved sensitivity and access to a broader set of UV emission lines. Aspera, designed specifically to target \ion{O}{6}, sacrifices wavelength coverage in favor of focused observations of this key diagnostic ion. 
Among these concepts, HWO achieves the highest spatial and spectral resolution and provides access to the largest suite of emission lines, placing it close to the optimal regime in multiple dimensions. However, its sensitivity is more moderate compared to some other concepts, reflecting current design expectations for detector performance at L2, where increased dark current from charged particles limits achievable surface-brightness sensitivity.
In contrast, missions such as MAGPIE and Juniper adopt coarser spatial sampling (larger effective spatial elements), which improves surface-brightness sensitivity for diffuse emission, albeit at the expense of spatial resolution. This highlights an inherent trade-off between spatial resolution and sensitivity in the design of CGM-focused instruments.}


This analysis also suggests that optical and UV instruments can combine to observe the same targets in the nearby universe, complementing one another and collecting a set of ionization diagnostics that spans the full available range. At low redshift, optical telescopes will be observing rest-frame optical lines (e.g., H$\alpha$, [\ion{O}{2}])\vadd{, while space-based UV facilities target diagnostic UV lines. Although these instruments differ significantly in aperture size and angular resolution, their achievable surface-brightness limits and the exposure times required to reach them can be comparable in some regimes, depending on instrument design and observing configuration.}
One particular use case would be to focus the smaller spatial resolution elements of large telescopes on interesting regions discovered in the CGM of a nearby galaxy, zooming in on small structures and examining their ionization and kinematics in detail. With current and future instrumentation in both space (Aspera, Juniper, HWO) and on the ground (e.g., E-ELT/HARMONI), there are excellent prospects for addressing CGM physics from multiple complementary instruments. 

\section{Summary and Conclusions\label{sec:summary}}

Observing the CGM in emission provides a powerful means of accessing the spatial and kinematic structure of gas that regulates galactic accretion and feedback. We have examined the detectability and kinematic separability of eight UV emission lines using the high-resolution FOGGIE simulations. Our key findings are:

\begin{itemize}
    \item A wide range of emission lines is essential to capture the multiphase structure of the CGM. High-ionization lines like \ion{O}{6} trace more diffuse emission from gas at $T \gtrsim 10^5$ K, while low-ion lines like H$\alpha$, \ion{Si}{2}, and \ion{Mg}{2} trace clumpier, denser structures. Intermediate ions (e.g., \ion{C}{4}, \ion{Si}{4}, \ion{C}{3}, \ion{Si}{3}) recover elements of both regimes. Observations restricted to one ion or a small set will miss structures traced by other ions, and multi-line observations are required to build a comprehensive picture of the multiphase CGM. 

    \item Surface brightness sensitivity at $500$ photons cm$^{-2}$ s$^{-1}$ sr$^{-1}$ or below is the primary requirement for revealing the full mass and spatial extent of CGM emission. As surface brightness limits rise above this threshold, the fraction of detectable gas drops sharply, especially for higher ionization lines like \ion{O}{6} and \ion{C}{4} (Figure~\ref{fig:fit-massfrac}). In contrast, low ions such as H$\alpha$ and \ion{Mg}{2} retain higher observable mass fractions at moderate sensitivity owing to a greater degree of clumping. These trends underscore the need for future instrumentation, especially in the UV, to prioritize sensitivity below $500$ photons cm$^{-2}$ s$^{-1}$ sr$^{-1}$ to uncover the faint, extended components of the CGM.

    \item Resolving CGM structures at the $\sim$ kpc scale is essential for mapping morphology and structures in the CGM. With coarser resolution ($\geq 3 - 6$ kpc), faint CGM features such as accreting filaments and cold clumps are smoothed out, even if their total flux (and mass) remains detectable. Figure~\ref{fig:mapswithspatialres} shows small-scale features are preserved at 1 kpc resolution but begin to blur or vanish at lower resolution. 
    
    \item Good kinematic resolution ($\Delta v \lesssim $ 30 km s$^{-1}$) is required to relate, or to distinguish, different phases of the gas when seen in projection. Typical velocity offsets between widely separated ion pairs, e.g., \ion{H}{1} and \ion{O}{6}, are only resolvable at kinematic resolutions finer than $\sim$ 30 km s$^{-1}$, and largely disappear at coarser resolution. In contrast, low-ionization species like Mg II show velocity structures closely aligned with \ion{H}{1}, and their kinematics remain indistinguishable even at 30 km s$^{-1}$ resolution. This highlights how kinematic resolution enables separation of multiphase gas motions in the CGM. 

    \item Distinguishing inflowing from outflowing gas also requires good kinematic resolution. Across all ions, the fraction of CGM area where inflow and outflow can be kinematically separated increases from $\leq$ 20\% at 300 km s$^{-1}$ to $\geq$ 90\% at 30 km s$^{-1}$. These results demonstrate that recovering the full dynamical picture of the CGM—including its multiphase structure and circulation patterns—requires spectrographs capable of achieving $\Delta v \leq $ 30 km s$^{-1}$.

\end{itemize}

The landscape of instrumentation intended to detect CGM emission is evolving rapidly. Our comparison of current and proposed \vadd{ground- and space-based} facilities (Figure~\ref{fig:instruments}) shows that no single current instrument is optimized for depth, spatial and kinematic resolution, and coverage of multiphase ions. Given the practical limitations of instrument design, it is likely that no instrument will optimize all three axes at once; our simulations therefore offer quantitative guidance about the trade-space along these three axes. For example, if kinematic resolution cannot be achieved at the desired level, designers should seek to cover a broad set of ions at good spatial resolution. Where covering multiple ions is not possible, deep surface brightness limits could be pursued with good spatial and/or kinematic resolution. The parameter space in Figure~\ref{fig:instruments} will likely continue to be populated by new and innovative instrument concepts developed to address this important problem. In this time of growth, our simulations also offer a quantitative basis for making tradeoffs and comparisons between instruments, and for computing how instruments with different optimizations can operate together in a complementary fashion to reveal the complex flows of the CGM in detail. 

\begin{acknowledgments}
We thank J.\ C.\ Howk for insightful discussion and suggestions.
VS was supported for this work in part by NASA via an Astrophysics Theory Program grant \#80NSSC24K0772, HST AR \#17549, and HST GO \#17093. CWT was supported for this work in part by JWST AR \#5486. VS and CWT were also supported in part by HST AR \#16151 and by NASA via a Theoretical and Computational Astrophysics Networks grant \#80NSSC21K1053.
BWO acknowledges support from NSF grants \#1908109 and \#2106575,
NASA ATP grants 80NSSC18K1105 and 80NSSC24K0772, and NASA TCAN grant 80NSSC21K1053. CL was supported by NASA through the NASA Hubble Fellowship grant \#HST-HF2-51538.001-A awarded by the Space Telescope Science Institute, which is operated by the Association of Universities for Research in Astronomy, Inc., for NASA, under contract NAS5-26555.
AA acknowledges support from the INAF Large Grant 2022 ``Extragalactic Surveys with JWST'' (PI Pentericci) and from the European Union – NextGenerationEU RFF M4C2 1.1 PRIN 2022 project 2022ZSL4BL INSIGHT.
RA acknowledges funding from the European Research Council (ERC) under the European Union's Horizon 2020 research and innovation programme (grant agreement 101020943, SPECMAP-CGM).

Computations described in this work were performed using the publicly-available \textsc{Enzo} code (\href{http://enzo-project.org}{http://enzo-project.org}), which is the product of a collaborative effort of many independent scientists from numerous institutions around the world. Their commitment to open science has helped make this work possible. The python packages {\sc matplotlib} \citep{hunter2007}, {\sc numpy} \citep{walt2011numpy}, {\sc rockstar} \citep{Behroozi2013a}, {\sc tangos} \citep{pontzen2018}, \textsc{scipy} \citep{scipy2020}, {\sc yt} \citep{ytpaper}, and {\sc Astropy} \citep{astropy2022} were all used in parts of this analysis or in products used by this paper. 
Resources supporting this work were provided by the NASA High-End Computing (HEC) Program through the NASA Advanced Supercomputing (NAS) Division at Ames Research Center and were sponsored by NASA's Science Mission Directorate.

\end{acknowledgments}

\begin{contribution}

VS led the analysis and writing of the paper. JT provided scientific guidance, refined the interpretation by suggesting additional diagnostics, and contributed to the writing. MSP, BWO, and JT are the principal investigators of the FOGGIE collaboration; they obtained funding for, developed, and ran the simulations used in this work. MSP and BWO also provided valuable feedback and insights throughout the discussion. CL contributed to the interpretation of the results and shared her experience with similar analyses. LC developed the pipeline for obtaining emissivities from Cloudy tables and provided additional insights based on related analyses. CWT developed the code used to remove the \ion{H}{1} disk and provided feedback on the draft. BDS wrote the CIAOLoop code that was essential for generating the Cloudy tables and developed the initial conditions for the FOGGIE simulations. JKW contributed through discussions and constructive feedback.
AA, RA, AJF, NL, and ACW provided insights through discussions during the analysis phase, as well as feedback on the final draft.

\end{contribution}

\appendix

\section{Surface Brightness Units and Equivalencies}\label{sec:appendixB}

Surface brightness of extended source emission is expressed in a variety of units without consistent usage in the existing literature. For line emission, the most common units are erg s$^{-1}$ cm$^{-2}$ arcsec$^{-2}$ and photons s$^{-1}$ cm$^{-2}$ sr$^{-1}$. The former are convenient when dealing with data or instruments with spatial resolution elements on the order of arcseconds.
The later, sometimes known as ``line units'' or ``LU'', result in quantities of order 100-10000 and can be compared without knowing the identity of the underlying line. The conversion between the two units requires knowing the {\it observed} wavelength of a given emission line. 

These two conversions are as follows. For a line observed at wavelength $\lambda _0$, the photon energy $E _{ph} = hc/\lambda _0$, where $h$ is Planck's constant \citep{Planck2016} and $c$ is the speed of light. Thus 
\begin{equation}
f_{\rm LU} = f_{\lambda} \times (1 \, {\rm photon} / E_{ph}) \times (4.25 \times 10^{10}\, {\rm arcsec ^2} \, {\rm sr}^{-1}) \label{eqn:fl_to_LU},
\end{equation}
where $f_{\rm LU}$ then has units of photons s$^{-1}$ cm$^{-2}$ sr$^{-1}$. 

Here are some worked examples for concreteness. For \ion{O}{6} observed at the rest frame, $\lambda _0$ = 1032 \AA. We can express the speed of light as $c = 2.9979 \times 10^{18}$ \AA\ s$^{-1}$. Each detected photon has energy $hc/\lambda _0$ or $E _{ph} = 1.93 \times 10^{-11}$ erg. By Equation~\ref{eqn:fl_to_LU},
\begin{equation}
f_{\rm LU}^{OVI} = 2.21 \times 10^{21} f_{\lambda}^{OVI} \label{eqn:b2}
\end{equation}
\cite{Hayes2016} reported \ion{O}{6} detections at $f_{\lambda} =$ 10$^{-17}$  erg s$^{-1}$ cm$^{-2}$ arcsec$^{-2}$. According to Equation~\ref{eqn:b2}, this value converts to 22064 photons s$^{-1}$ cm$^{-2}$ sr$^{-1}$. Going in the other direction, \cite{2003ApJ...591..821O} reported an \ion{O}{6} detection in NGC 4631 at $f_{\rm LU} = 8000$ photons s$^{-1}$ cm$^{-2}$ sr$^{-1}$. This converts to $f_{\lambda} = 3.63 \times 10^{-17}$  erg s$^{-1}$ cm$^{-2}$ arcsec$^{-2}$. For lines that are observed with some redshift, care must be taken to apply the correct ($1 + z$) factor to the photon energy. For a general rule of thumb, a reasonable approximation is that 200 LU converts to $10^{-19}$ erg s$^{-1}$ cm$^{-2}$ arcsec$^{-2}$ for \ion{O}{6} and $6\times 10^{-20}$ erg s$^{-1}$ cm$^{-2}$ arcsec$^{-2}$ for \ion{C}{4}. 

For our assumed $\Lambda$CDM cosmology, 1 arcsecond subtends an physical size of 8.54 kpc at $z = 2$, and 0.21 kpc at $z = 0.01$ (about $d = 40$ Mpc). These two regimes correspond to redshifted UV lines observed at sub-arcsecond spatial resolution, and rest-frame UV lines observed at $\sim 20$ arcsec resolution. These two very disparate extremes of telescope size and spatial resolution both yield physical resolutions near $\sim 1$ kpc (Figure~\ref{tab:instrument-table}). 

\section{CGM Mass Fraction Maps for All Ions}\label{sec:appendixA}

For completeness, we include here the full set of heatmaps analogous to Figure~\ref{fig:heatmap-massfrac}, showing the observable CGM mass fractions for all eight emission lines analyzed in this study. Figure~\ref{fig:heatmap-massfrac-edge} presents the results for edge-on galaxy projections, while Figure~\ref{fig:heatmap-massfrac-face} shows the corresponding face-on projections. These results reinforce the conclusion discussed in the main text: sensitivity has a more significant impact than spatial resolution on the ability to recover CGM mass across different ions.

\vadd{These heatmaps are intended to (i) quantify the relative leverage of sensitivity versus spatial resolution on the observable emitting mass, and (ii) provide a completeness reference for interpreting emission detections. For example, O VI emission traces only a small subset of the total O VI mass ($\leq$ 25\%), even at the deepest sensitivity thresholds considered here. The fraction of a given CGM phase that must be observed to enable specific scientific inferences is inherently goal-dependent and will be explored in future studies}

\begin{figure*}[ht!]
\begin{center}
\includegraphics[width=0.95\textwidth]{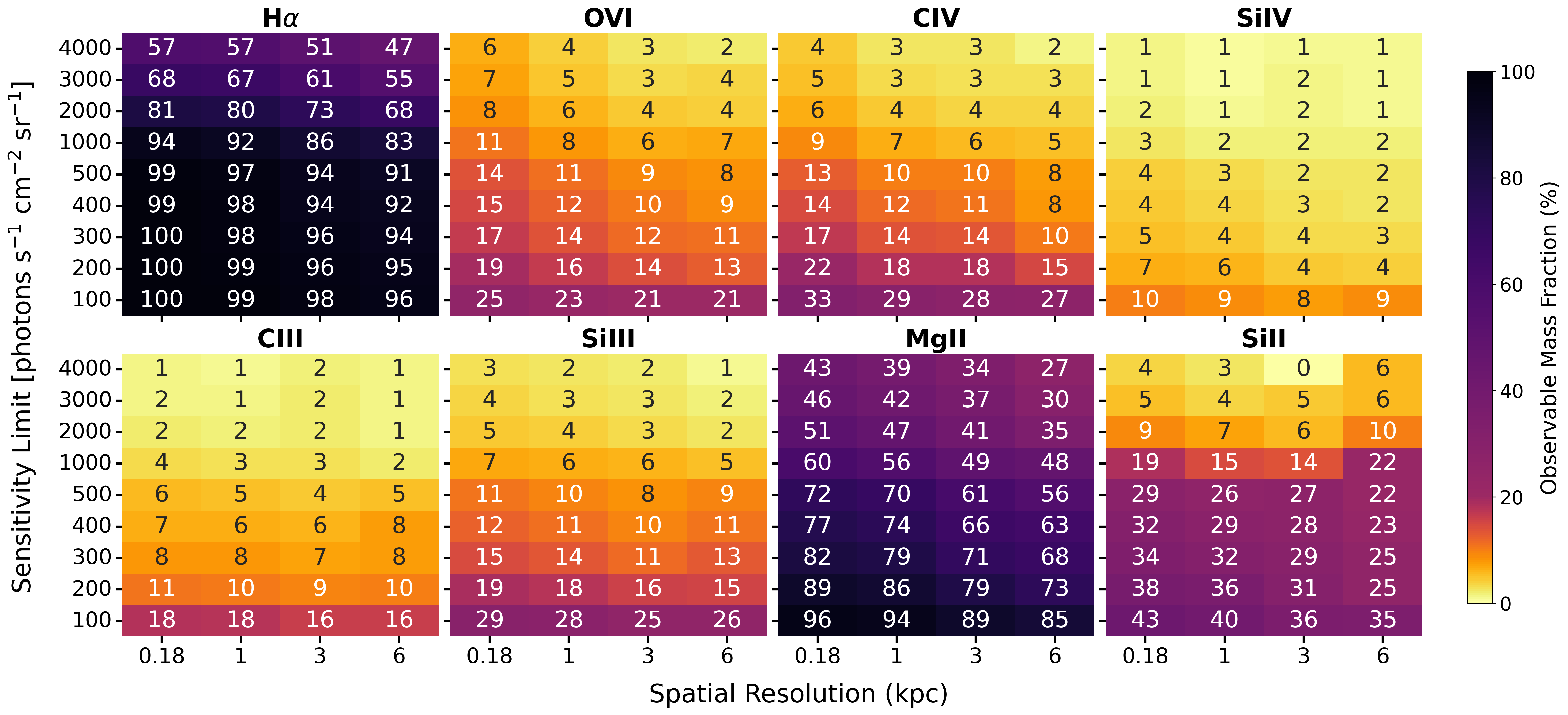}
\caption{Observable CGM mass fraction as a function of spatial resolution (x-axis) and surface brightness sensitivity (y-axis), for eight emission lines, averaged over six FOGGIE halos, edge-on. Each cell shows the percentage of total CGM mass detectable above the given sensitivity limit at the specified resolution, with values color-coded and overlaid. As expected, observability decreases with lower sensitivity and coarser resolution, though sensitivity plays a stronger role: at fixed resolution, increasing the detection threshold from 100 to 500 reduces the detectable mass by more than 50\% in several ions. These results emphasize that achieving high sensitivity is critical for maximizing CGM detection, particularly for tracing diffuse, highly ionized gas.}
\label{fig:heatmap-massfrac-edge}
\end{center}
\end{figure*}

\begin{figure*}[ht!]
\begin{center}
\includegraphics[width=0.95\textwidth]{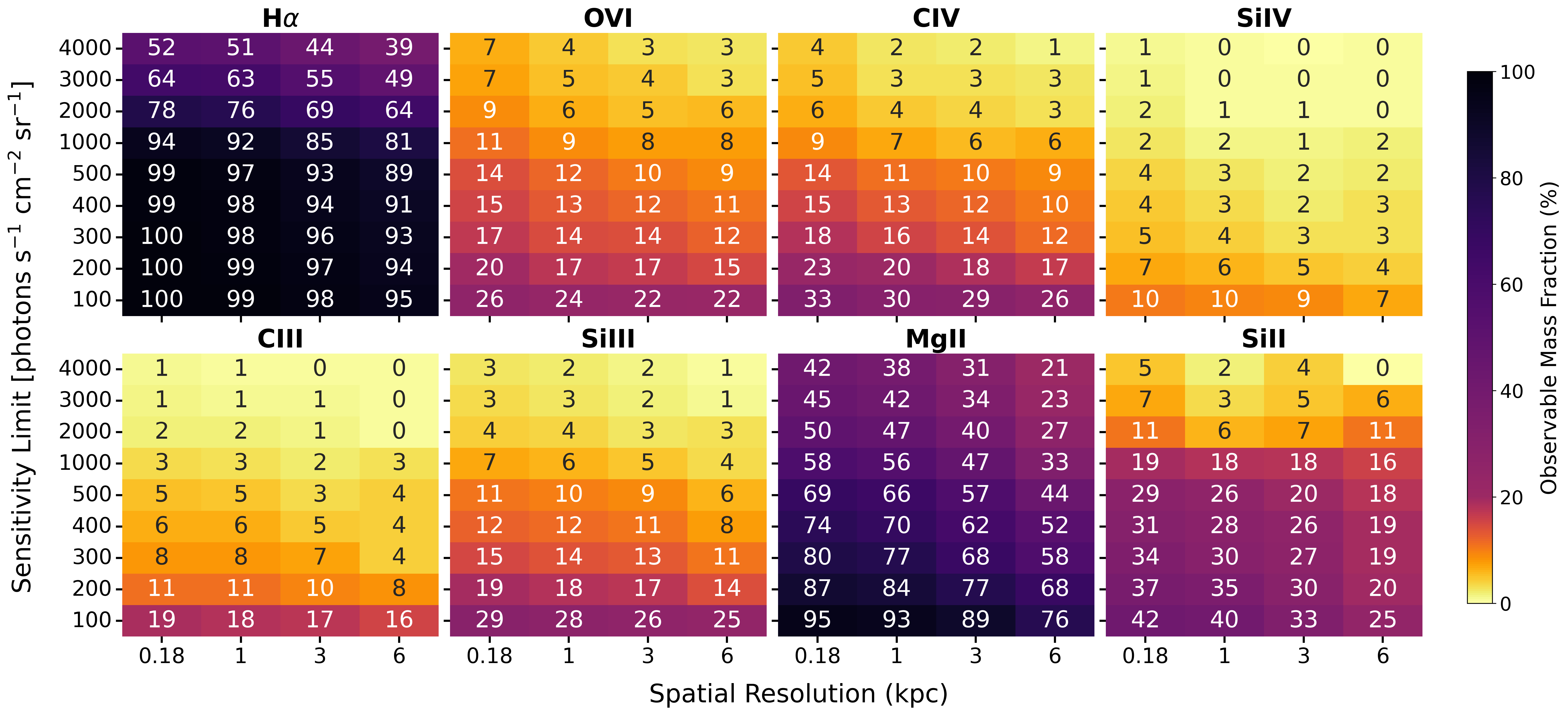}
\caption{Same as Figure~\ref{fig:heatmap-massfrac} but for a face-on view.}
\label{fig:heatmap-massfrac-face}
\end{center}
\end{figure*}

\section{Constructing CLOUDY Emissivity Tables }\label{sec:appendixC}

To generate the emissivity tables used in this work, we followed the pipeline described in \citet{Corlies2020} and implemented using the \textsc{CIAOLoop} package (\url{https://github.com/brittonsmith/cloudy_cooling_tools.git}). 
Below, we provide the key commands from the parameter file used to run \textsc{CIAOLoop} and produce the emissivity grids, along with a brief explanation of the output units and how they are converted to the values used in our analysis. Figure~\ref{fig:temp_number_density} shows 2D histograms of temperature versus number density for each ion, color-coded by emissivity; these plots highlight the physical conditions under which each ion emits most strongly.

\subsubsection{Example Parameter File}\label{appendixc-parameter-file}

The following configuration was used in addition to the standard run commands specifying the paths to \textsc{CLOUDY} and output directories.

\begin{lstlisting}
################## Line Map Parameters ####################

# Minimum and maximum temperatures (K)
coolingMapTmin = 1e3
coolingMapTmax = 1e8

# Number of temperature steps (log-spaced)
coolingMapTpoints = 51

coolingMapUseJeansLength = 1

# Emission lines to save

lineMapLine = H  1 6563
lineMapLine = Mg 2 2796
lineMapLine = Si 2 1260
lineMapLine = C  3 1910
lineMapLine = Si 3 1207
lineMapLine = C  4 1548
lineMapLine = Si 4 1394
lineMapLine = O  6 1032
lineMapLine = O  6 1038
############################################################
############ Commands executed for each run ###############

command failures 200 times map
command iterate to convergence max=2 error=0.20
command CMB redshift 0.00
command Table HM12 redshift 0.00
command set WeakHeatCool -20
command no H2 molecule
command no charge transfer
###########################################################
############# Parameters looped over ######################

# Loop over hydrogen number density (cm^-3)
loop [hden] (-6;2;0.5)
command metals 0 log
\end{lstlisting}

This configuration assumes a UV background from \citet{Haardt2012} (HM12) at $z=0$, solar metallicity (which is then scaled linearly by each cell’s metallicity during post-processing), temperature and density ranges as described in Section~\ref{sec:emissivity_cal}, and the set of emission lines analyzed in this paper (see Table~\ref{tab:emission_lines}). \vadd{We used \textsc{CLOUDY} version 23.01 for these analysis.}

\subsubsection{Emissivity Units and Conversion}\label{appendixc-unit}
\textsc{CLOUDY} outputs emissivities as $\text{emissivity}/n_{\rm H}^2$ in units of erg cm$^{3}$ s$^{-1}$.
To obtain the final surface brightness, we first multiply the tabulated emissivity values by $n_{\rm H}^2$ to recover the total emissivity. We then scale this by the metallicity of each simulation cell, and convert the result from energy to photon units using the method described in Appendix~\ref{sec:appendixB}. Finally, we use \textsc{yt} \citep{ytpaper} to project along the line of sight, integrating over the simulation cell depth $\Delta l$ (in cm) to compute the surface brightness in photons cm$^{-2}$ s$^{-1}$ sr$^{-1}$.

\subsubsection{Emissivity as a function of temperature and number density}

Figure~\ref{fig:temp_number_density} shows temperature–number density phase plots for the Maelstrom halo, color-coded by the emissivity of each ion. The emissivity values are calculated as described in Sections~\ref{sec:emissivity_cal}, \ref{appendixc-parameter-file}, and \ref{appendixc-unit}. Dark blue pixels indicate regions of parameter space where gas exists but does not emit significantly in the corresponding ion shown in each subplot.

These plots illustrate the thermodynamic conditions under which each ion emits most strongly. High-ionization lines like \ion{O}{6} exhibit peak emissivity at high temperatures ($T \sim 10^{5.5}$–$10^{6}$ K) and low densities, reflecting their origin in hot, diffuse CGM gas. In contrast, low-ionization species such as \ion{Mg}{2} and \ion{Si}{2} peak at cooler temperatures ($T \sim 10^{4}$ K) and higher densities, typically associated with denser clumps and filaments. Other ions (e.g., \ion{C}{3}, \ion{Si}{3}) trace a broader range of densities and temperatures, highlighting their role in bridging different phases of the CGM. These emissivity distributions motivate the use of multiple emission lines to map the full structure of the CGM.

\begin{figure*}[ht!]
\begin{center}
\includegraphics[width=0.95\textwidth]{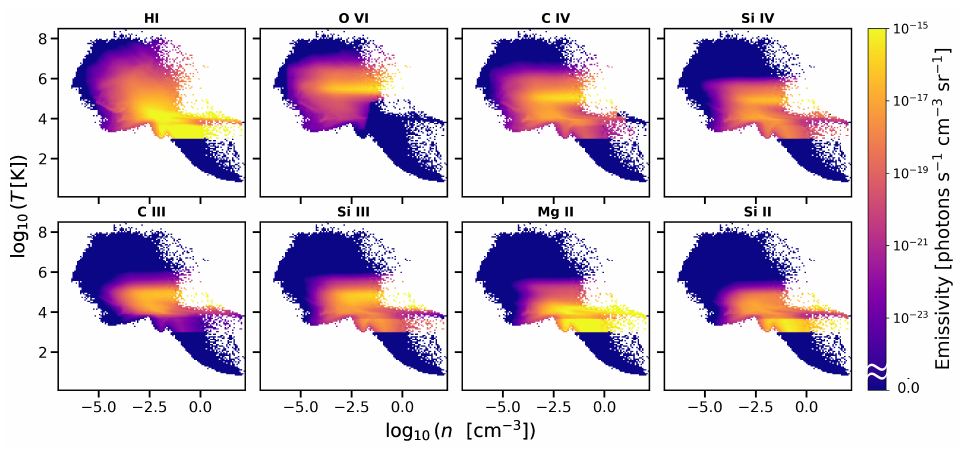}
\caption{2D histograms showing the emissivity distribution of each ion as a function of gas temperature and number density for the Maelstrom halo at $z = 0.5$. The dark blue background indicates all gas, while the colored pixels highlight regions where emission from that ion is present. The sharp cutoff at $\log_{10} T = 3$ arises because the minimum temperature included in the CLOUDY emissivity tables is $\log_{10} T = 3$, as this is the minimum temperature we expect to find in the CGM and we do not focus on ISM emission in this work.
}
\label{fig:temp_number_density}
\end{center}
\end{figure*}

\subsubsection{Velocity channel maps}\label{sec:appendixc04}
\vadd{
Figure \ref{fig:vchannelmap} shows channel maps of H$\alpha$ emission from the Maelstrom halo at $z=0.5$ , with each panel integrating emission over a spatial resolution of 0.183 kpc, corresponding to the native simulation resolution at this redshift. The maps are generated for an edge-on view of the galaxy, with the \ion{H}{1} disk removed. All panels show the same halo, projection, spatial scale, and color normalization; only the velocity bin changes. The large maps in the left column show projections of gas at a kinematic resolution of $\Delta v = 180 \: km \: s ^ {-1}$, while the six maps on the right show gas at a spectral resolution of $\Delta v = 30 \ km \ s ^ {-1}$ , spanning the same velocity range covered by a single bin in the left column.}

\vadd{At coarse kinematic resolution (left column), emission from structures spanning a wide range of line-of-sight velocities is blended into a single channel, preventing the isolation of distinct kinematic components. In contrast, the higher-resolution channel maps (right panels) reveal coherent spatial structures appearing in multiple adjacent velocity bins, demonstrating that gas with substantially different velocities can coexist along the same sightline. These maps illustrate that higher spectral resolution is required to begin separating spatially overlapping CGM structures kinematically and to enable meaningful interpretation of CGM dynamics, even before incorporating additional geometric or modeling context. }

\begin{figure*}[ht!]
\begin{center}
\includegraphics[width=0.7\textwidth]{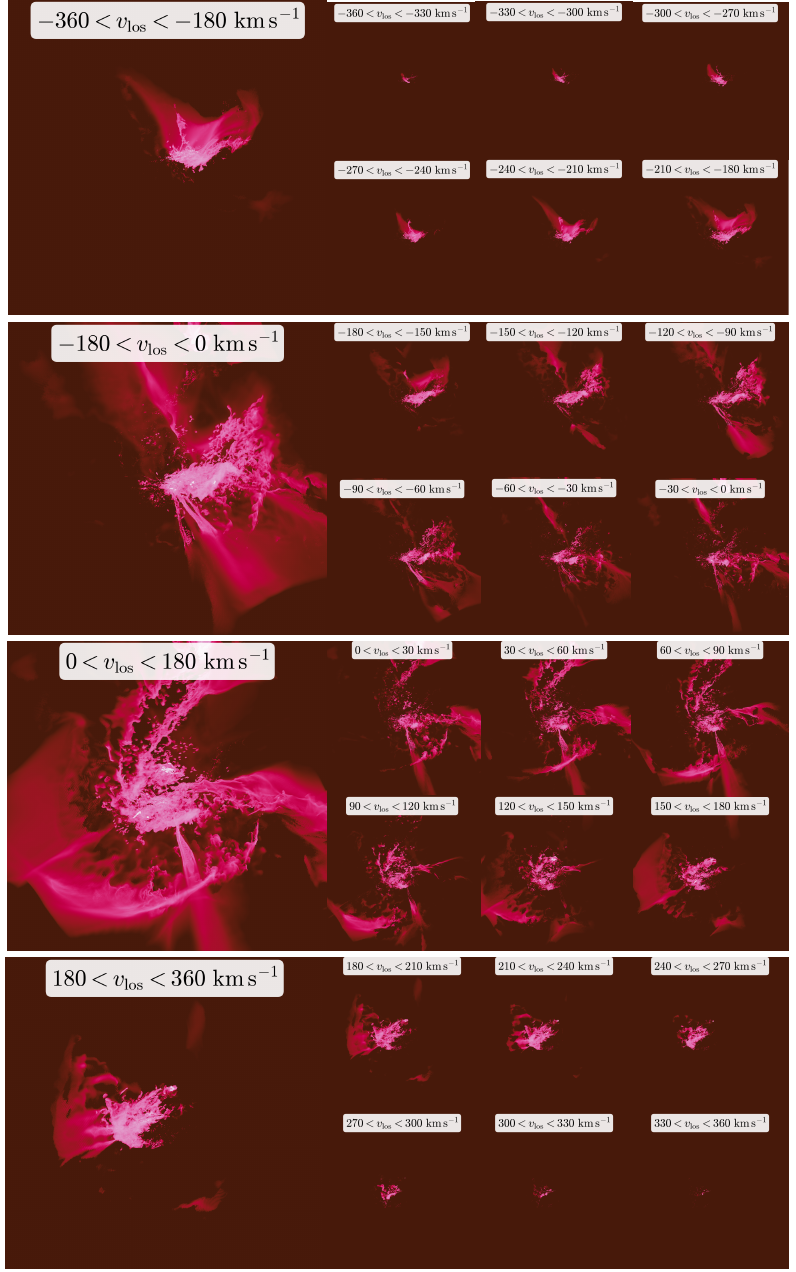}
\caption{Velocity channel maps illustrating the impact of spectral resolution on the spatial–kinematic structure of CGM emission. Channel maps of H$\alpha$ emission from the Maelstrom halo at  $z = 0.5$ are shown for two different kinematic resolutions. The maps are generated for an edge-on view of the galaxy, with the \ion{H}{1} disk removed. All panels use the same projection, spatial extent, and color scale, and are computed at the native simulation spatial resolution of 0.183 kpc. The left column shows emission integrated over large velocity bins ($\Delta v = 180 \: km \: s ^ {-1}$ ) while the right panels show emission in narrower velocity channels ($\Delta v = 30 \: km \: s ^ {-1}$) subdividing the same total velocity interval.
}
\label{fig:vchannelmap}
\end{center}
\end{figure*}

\bibliography{foggie-emission}{}

@INPROCEEDINGS{Harmoni2014,
       author = {{Thatte}, Niranjan A. and {Clarke}, Fraser and {Bryson}, Ian and {Schnetler}, Hermine and {Tecza}, Matthias and {Bacon}, Roland M. and {Remillieux}, Alban and {Mediavilla}, Evencio and {Herreros Linares}, Jos{\'e} Miguel and {Arribas}, Santiago and {Evans}, Christopher J. and {Lunney}, David W. and {Fusco}, Thierry and {O'Brien}, Kieran and {Tosh}, Ian A. and {Ives}, Derek J. and {Finger}, Gert and {Houghton}, Ryan and {Davies}, Roger L. and {Lynn}, James D. and {Allen}, Jamie R. and {Zieleniewski}, Simon D. and {Kendrew}, Sarah and {Ferraro-Wood}, Vanessa and {P{\'e}contal-Rousset}, Arlette and {Kosmalski}, Johan and {Richard}, Johan and {Jarno}, Aurelien and {Gallie}, Angus M. and {Montgomery}, David M. and {Henry}, David and {Zins}, G{\'e}rard and {Freeman}, David and {Garc{\'\i}a-Lorenzo}, Begona and {Rodr{\'\i}guez-Ramos}, Luis F. and {Revuelta}, Jorge S.~C. and {Hernandez Suarez}, Elvio and {Bueno-Bueno}, Alberto and {Gigante-Ripoll}, Jos{\'e} Vincente and {Garcia}, Adolfo and {Dohlen}, Kjetil and {Neichel}, Beno{\^\i}t.},
        title = "{HARMONI: the first light integral field spectrograph for the E-ELT}",
    booktitle = {Ground-based and Airborne Instrumentation for Astronomy V},
         year = 2014,
       editor = {{Ramsay}, Suzanne K. and {McLean}, Ian S. and {Takami}, Hideki},
       series = {Society of Photo-Optical Instrumentation Engineers (SPIE) Conference Series},
       volume = {9147},
        month = aug,
          eid = {914725},
        pages = {914725},
          doi = {10.1117/12.2055436},
       adsurl = {https://ui.adsabs.harvard.edu/abs/2014SPIE.9147E..25T}
}

@ARTICLE{Ticoras2026,
       author = {{Ticoras}, Mackenzie and {O'Shea}, Brian W. and {Kopenhafer}, Claire and {Lochhaas}, Cassandra and {Peeples}, Molly S. and {Tumlinson}, Jason and {Trapp}, Cameron and {Saeedzadeh}, Vida and {Augustin}, Ramona and {Lehner}, Nicolas and {Smith}, Britton D. and {Howk}, J. Christopher},
        title = "{Figuring Out Gas \& Galaxies in Enzo (FOGGIE). XV. Examining the Spatial and Kinematic Relationship between Circumgalactic Mg II and O VI}",
      journal = {arXiv e-prints},
         year = 2026,
        month = jan,
          eid = {arXiv:2601.02348},
        pages = {arXiv:2601.02348},
          doi = {10.48550/arXiv.2601.02348},
archivePrefix = {arXiv},
       eprint = {2601.02348},
 primaryClass = {astro-ph.GA},
       adsurl = {https://ui.adsabs.harvard.edu/abs/2026arXiv260102348T}
}

@ARTICLE{1956ApJ...124...20S,
       author = {{Spitzer}, Jr., Lyman},
        title = "{On a Possible Interstellar Galactic Corona.}",
      journal = {\apj},
         year = 1956,
        month = jul,
       volume = {124},
        pages = {20},
          doi = {10.1086/146200},
       adsurl = {https://ui.adsabs.harvard.edu/abs/1956ApJ...124...20S}
}

@ARTICLE{Taira2025,
       author = {{Taira}, Elias and {Kopenhafer}, Claire and {O'Shea}, Brian W. and {Manning}, Alexis and {Fuhrman}, Evelyn and {Peeples}, Molly S. and {Tumlinson}, Jason and {Smith}, Britton D.},
        title = "{Impacts of the Metagalactic Ultraviolet Background on Circumgalactic Medium Absorption Systems}",
      journal = {\apj},
         year = 2025,
        month = oct,
       volume = {991},
       number = {2},
          eid = {221},
        pages = {221},
          doi = {10.3847/1538-4357/adfc4e},
archivePrefix = {arXiv},
       eprint = {2503.11775},
 primaryClass = {astro-ph.GA},
       adsurl = {https://ui.adsabs.harvard.edu/abs/2025ApJ...991..221T}
}

@ARTICLE{Trapp25a,
       author = {{Trapp}, Cameron W. and {Peeples}, Molly S. and {Tumlinson}, Jason and {O'Shea}, Brian W. and {Lochhaas}, Cassandra and {Wright}, Anna C. and {Smith}, Britton D. and {Saeedzadeh}, Vida and {Acharyya}, Ayan and {Augustin}, Ramona and {Simons}, Raymond C.},
        title = "{Figuring Out Gas \& Galaxies In Enzo (FOGGIE). XII. The Formation and Evolution of Extended HI Galactic Disks and Warps with a Dynamic CGM}",
      journal = {arXiv e-prints},
         year = 2025,
        month = oct,
          eid = {arXiv:2511.00158},
        pages = {arXiv:2511.00158},
archivePrefix = {arXiv},
       eprint = {2511.00158},
 primaryClass = {astro-ph.GA},
       adsurl = {https://ui.adsabs.harvard.edu/abs/2025arXiv251100158T}
}

@ARTICLE{emerick19,
       author = {{Emerick}, Andrew and {Bryan}, Greg L. and {Mac Low}, Mordecai-Mark},
        title = "{Simulating an isolated dwarf galaxy with multichannel feedback and chemical yields from individual stars}",
      journal = {\mnras},
         year = 2019,
        month = jan,
       volume = {482},
       number = {1},
        pages = {1304-1329},
          doi = {10.1093/mnras/sty2689},
archivePrefix = {arXiv},
       eprint = {1807.07182},
 primaryClass = {astro-ph.GA},
       adsurl = {https://ui.adsabs.harvard.edu/abs/2019MNRAS.482.1304E}
}

@INPROCEEDINGS{chen2024,
       author = {{Chen}, Seery and {Lokhorst}, Deborah M. and {Pasha}, Imad and {Bowman}, William P. and {Liu}, Qing and {Shen}, Zili and {MacNichol}, Aiden and {Malakhov}, Evgeni I. and {Abraham}, Roberto G. and {van Dokkum}, Pieter},
        title = "{The Dragonfly Spectral Line Mapper: completion of the 120-lens array}",
    booktitle = {Ground-based and Airborne Telescopes X},
         year = 2024,
       editor = {{Marshall}, Heather K. and {Spyromilio}, Jason and {Usuda}, Tomonori},
       series = {Society of Photo-Optical Instrumentation Engineers (SPIE) Conference Series},
       volume = {13094},
        month = aug,
          eid = {130942J},
        pages = {130942J},
          doi = {10.1117/12.3019491},
archivePrefix = {arXiv},
       eprint = {2406.15101},
 primaryClass = {astro-ph.IM},
       adsurl = {https://ui.adsabs.harvard.edu/abs/2024SPIE13094E..2JC}
}

@ARTICLE{Abraham2014,
       author = {{Abraham}, Roberto G. and {van Dokkum}, Pieter G.},
        title = "{Ultra-Low Surface Brightness Imaging with the Dragonfly Telephoto Array}",
      journal = {\pasp},
         year = 2014,
        month = jan,
       volume = {126},
       number = {935},
        pages = {55},
          doi = {10.1086/674875},
archivePrefix = {arXiv},
       eprint = {1401.5473},
 primaryClass = {astro-ph.IM},
       adsurl = {https://ui.adsabs.harvard.edu/abs/2014PASP..126...55A}
}

@ARTICLE{Lokhorst2019,
       author = {{Lokhorst}, Deborah and {Abraham}, Roberto and {van Dokkum}, Pieter and {Wijers}, Nastasha and {Schaye}, Joop},
        title = "{On the Detectability of Visible-wavelength Line Emission from the Local Circumgalactic and Intergalactic Medium}",
      journal = {\apj},
         year = 2019,
        month = may,
       volume = {877},
       number = {1},
          eid = {4},
        pages = {4},
          doi = {10.3847/1538-4357/ab184e},
archivePrefix = {arXiv},
       eprint = {1904.07874},
 primaryClass = {astro-ph.GA},
       adsurl = {https://ui.adsabs.harvard.edu/abs/2019ApJ...877....4L}
}

@ARTICLE{chas2022,
       author = {{Melso}, Nicole and {Schiminovich}, David and {Smiley}, Brian and {Ong}, Hwei Ru and {Santiago}, B{\'a}rbara Cruvinel and {Sitaram}, Meghna and {Aleman}, Ignacio Cevallos and {Graber}, Sarah and {Murillo}, Marisa and {Rosenthal}, Marni and {Stelea}, Ioana},
        title = "{The Circumgalactic H{\ensuremath{\alpha}} Spectrograph (CH{\ensuremath{\alpha}}S). I. Design, Engineering, and Early Commissioning}",
      journal = {\apj},
         year = 2022,
        month = dec,
       volume = {941},
       number = {2},
          eid = {185},
        pages = {185},
          doi = {10.3847/1538-4357/ac9d9c},
archivePrefix = {arXiv},
       eprint = {2209.14999},
 primaryClass = {astro-ph.GA},
       adsurl = {https://ui.adsabs.harvard.edu/abs/2022ApJ...941..185M}
}

@ARTICLE{scipy2020,
       author = {{Virtanen}, Pauli and {Gommers}, Ralf and {Oliphant}, Travis E. and {Haberland}, Matt and {Reddy}, Tyler and {Cournapeau}, David and {Burovski}, Evgeni and {Peterson}, Pearu and {Weckesser}, Warren and {Bright}, Jonathan and {van der Walt}, St{\'e}fan J. and {Brett}, Matthew and {Wilson}, Joshua and {Millman}, K. Jarrod and {Mayorov}, Nikolay and {Nelson}, Andrew R.~J. and {Jones}, Eric and {Kern}, Robert and {Larson}, Eric and {Carey}, C.~J. and {Polat}, {\.I}lhan and {Feng}, Yu and {Moore}, Eric W. and {VanderPlas}, Jake and {Laxalde}, Denis and {Perktold}, Josef and {Cimrman}, Robert and {Henriksen}, Ian and {Quintero}, E.~A. and {Harris}, Charles R. and {Archibald}, Anne M. and {Ribeiro}, Ant{\^o}nio H. and {Pedregosa}, Fabian and {van Mulbregt}, Paul and {SciPy 1. 0 Contributors}},
        title = "{SciPy 1.0: fundamental algorithms for scientific computing in Python}",
      journal = {Nature Methods},
         year = 2020,
        month = feb,
       volume = {17},
        pages = {261-272},
          doi = {10.1038/s41592-019-0686-2},
archivePrefix = {arXiv},
       eprint = {1907.10121},
 primaryClass = {cs.MS},
       adsurl = {https://ui.adsabs.harvard.edu/abs/2020NatMe..17..261V}
}

@ARTICLE{ytpaper,
   author = {{Turk}, M.~J. and {Smith}, B.~D. and {Oishi}, J.~S. and {Skory}, S. and
     {Skillman}, S.~W. and {Abel}, T. and {Norman}, M.~L.},
    title = "{yt: A Multi-code Analysis Toolkit for Astrophysical Simulation Data}",
  journal = {The Astrophysical Journal Supplement Series},
archivePrefix = "arXiv",
   eprint = {1011.3514},
 primaryClass = "astro-ph.IM",
     year = 2011,
    month = jan,
   volume = 192,
      eid = {9},
    pages = {9},
      doi = {10.1088/0067-0049/192/1/9},
   adsurl = {https://ui.adsabs.harvard.edu/abs/2011ApJS..192....9T}
}

@ARTICLE{2019Natur.574..643R,
       author = {{Rupke}, David S.~N. and {Coil}, Alison and {Geach}, James E. and {Tremonti}, Christy and {Diamond-Stanic}, Aleksandar M. and {George}, Erin R. and {Hickox}, Ryan C. and {Kepley}, Amanda A. and {Leung}, Gene and {Moustakas}, John and {Rudnick}, Gregory and {Sell}, Paul H.},
        title = "{A 100-kiloparsec wind feeding the circumgalactic medium of a massive compact galaxy}",
      journal = {\nat},
         year = 2019,
        month = oct,
       volume = {574},
       number = {7780},
        pages = {643-646},
          doi = {10.1038/s41586-019-1686-1},
archivePrefix = {arXiv},
       eprint = {1910.13507},
 primaryClass = {astro-ph.GA},
       adsurl = {https://ui.adsabs.harvard.edu/abs/2019Natur.574..643R}
}

@ARTICLE{2022MNRAS.511.5782R,
       author = {{Reichardt Chu}, Bronwyn and {Fisher}, Deanne B. and {Nielsen}, Nikole M. and {Chisholm}, John and {Girard}, Marianne and {Kacprzak}, Glenn G. and {Bolatto}, Alberto and {Herrera-Camus}, Rodrigo and {Sandstrom}, Karin and {Li}, Miao and {Rickards Vaught}, Ryan and {McPherson}, Daniel K.},
        title = "{The DUVET Survey: Resolved maps of star formation-driven outflows in a compact, starbursting disc galaxy}",
      journal = {\mnras},
         year = 2022,
        month = apr,
       volume = {511},
       number = {4},
        pages = {5782-5796},
          doi = {10.1093/mnras/stac420},
archivePrefix = {arXiv},
       eprint = {2202.04672},
 primaryClass = {astro-ph.GA},
       adsurl = {https://ui.adsabs.harvard.edu/abs/2022MNRAS.511.5782R}
}

@ARTICLE{2011A&A...526A.118H,
       author = {{Heald}, G. and {J{\'o}zsa}, G. and {Serra}, P. and {Zschaechner}, L. and {Rand}, R. and {Fraternali}, F. and {Oosterloo}, T. and {Walterbos}, R. and {J{\"u}tte}, E. and {Gentile}, G.},
        title = "{The Westerbork Hydrogen Accretion in LOcal GAlaxieS (HALOGAS) survey. I. Survey description and pilot observations}",
      journal = {\aap},
         year = 2011,
        month = feb,
       volume = {526},
          eid = {A118},
        pages = {A118},
          doi = {10.1051/0004-6361/201015938},
archivePrefix = {arXiv},
       eprint = {1012.0816},
 primaryClass = {astro-ph.CO},
       adsurl = {https://ui.adsabs.harvard.edu/abs/2011A&A...526A.118H}
}

@ARTICLE{2008AJ....136.2563W,
       author = {{Walter}, Fabian and {Brinks}, Elias and {de Blok}, W.~J.~G. and {Bigiel}, Frank and {Kennicutt}, Jr., Robert C. and {Thornley}, Michele D. and {Leroy}, Adam},
        title = "{THINGS: The H I Nearby Galaxy Survey}",
      journal = {\aj},
         year = 2008,
        month = dec,
       volume = {136},
       number = {6},
        pages = {2563-2647},
          doi = {10.1088/0004-6256/136/6/2563},
archivePrefix = {arXiv},
       eprint = {0810.2125},
 primaryClass = {astro-ph},
       adsurl = {https://ui.adsabs.harvard.edu/abs/2008AJ....136.2563W}
}

@ARTICLE{2022ApJ...941..163R,
       author = {{Reichardt Chu}, Bronwyn and {Fisher}, Deanne B. and {Bolatto}, Alberto D. and {Chisholm}, John and {Fielding}, Drummond and {Berg}, Danielle and {Cameron}, Alex J. and {Glazebrook}, Karl and {Herrera-Camus}, Rodrigo and {Kacprzak}, Glenn G. and {Lenki{\'c}}, Laura and {Li}, Miao and {McPherson}, Daniel K. and {Nielsen}, Nikole M. and {Obreschkow}, Danail and {Rickards Vaught}, Ryan J. and {Sandstrom}, Karin},
        title = "{DUVET: Spatially Resolved Observations of Star Formation Regulation via Galactic Outflows in a Starbursting Disk Galaxy}",
      journal = {\apj},
         year = 2022,
       volume = {941},
       number = {2},
        pages = {163},
          doi = {10.3847/1538-4357/aca1bd},
archivePrefix = {arXiv},
       eprint = {2211.02063},
 primaryClass = {astro-ph.GA},
       adsurl = {https://ui.adsabs.harvard.edu/abs/2022ApJ...941..163R},
}

@ARTICLE{2021ApJ...919..112D,
       author = {{de la Cruz}, Lita M. and {Schneider}, Evan E. and {Ostriker}, Eve C.},
        title = "{Synthetic Absorption Lines from Simulations of Multiphase Gas in Galactic Winds}",
      journal = {\apj},
         year = 2021,
       volume = {919},
       number = {2},
        pages = {112},
          doi = {10.3847/1538-4357/ac04ac},
archivePrefix = {arXiv},
       eprint = {2105.12108},
 primaryClass = {astro-ph.GA},
       adsurl = {https://ui.adsabs.harvard.edu/abs/2021ApJ...919..112D},
}

@ARTICLE{2014ApJ...784..108B,
       author = {{Bordoloi}, Rongmon and {Lilly}, Simon J. and {Kacprzak}, Glenn G. and {Churchill}, Christopher W.},
        title = "{Modeling the Distribution of Mg II Absorbers around Galaxies Using Background Galaxies and Quasars}",
      journal = {\apj},
         year = 2014,
       volume = {784},
       number = {2},
        pages = {108},
          doi = {10.1088/0004-637X/784/2/108},
archivePrefix = {arXiv},
       eprint = {1307.6553},
 primaryClass = {astro-ph.CO},
       adsurl = {https://ui.adsabs.harvard.edu/abs/2014ApJ...784..108B},
}

@ARTICLE{2012MNRAS.426..801B,
       author = {{Bouch{\'e}}, Nicolas and {Hohensee}, William and {Vargas}, Rub{\'e}n and {Kacprzak}, Glenn G. and {Martin}, Crystal L. and {Cooke}, Jeff and {Churchill}, Christopher W.},
        title = "{Physical properties of galactic winds using background quasars}",
      journal = {\mnras},
         year = 2012,
       volume = {426},
       number = {2},
        pages = {801-815},
          doi = {10.1111/j.1365-2966.2012.21114.x},
archivePrefix = {arXiv},
       eprint = {1205.3620},
 primaryClass = {astro-ph.CO},
       adsurl = {https://ui.adsabs.harvard.edu/abs/2012MNRAS.426..801B},
}

@ARTICLE{2014ApJ...794..156R,
       author = {{Rubin}, Kate H.~R. and {Prochaska}, J. Xavier and {Koo}, David C. and {Phillips}, Andrew C. and {Martin}, Crystal L. and {Winstrom}, Laura O.},
        title = "{Evidence for Ubiquitous Collimated Galactic-scale Outflows along the Star-forming Sequence at z $\sim$ 0.5}",
      journal = {\apj},
         year = 2014,
       volume = {794},
       number = {2},
        pages = {156},
          doi = {10.1088/0004-637X/794/2/156},
archivePrefix = {arXiv},
       eprint = {1307.1476},
 primaryClass = {astro-ph.CO},
       adsurl = {https://ui.adsabs.harvard.edu/abs/2014ApJ...794..156R},
}

@ARTICLE{2000ApJ...537..720L,
       author = {{Lazarian}, A. and {Pogosyan}, D.},
        title = "{Velocity Modification of H I Power Spectrum}",
      journal = {\apj},
         year = 2000,
        month = jul,
       volume = {537},
       number = {2},
        pages = {720-748},
          doi = {10.1086/309040},
archivePrefix = {arXiv},
       eprint = {astro-ph/9901241},
 primaryClass = {astro-ph},
       adsurl = {https://ui.adsabs.harvard.edu/abs/2000ApJ...537..720L}
}

@INPROCEEDINGS{magpie2025,
       author = {{Vorobiev}, Dmitry},
        title = "{MaGPIE - A Comprehensive UV Spectroscopic Survey of the Baryon Cycle in the Local Universe}",
    booktitle = {246th Meeting of the American Astronomical Society},
         year = 2025,
       series = {American Astronomical Society Meeting Abstracts},
       volume = {246},
        month = jun,
          eid = {141.03},
        pages = {141.03},
       adsurl = {https://ui.adsabs.harvard.edu/abs/2025AAS...24614103V}
}

@ARTICLE{krishnarao2022,
       author = {{Krishnarao}, Dhanesh and {Fox}, Andrew J. and {D'Onghia}, Elena and {Wakker}, Bart P. and {Cashman}, Frances H. and {Howk}, J. Christopher and {Lucchini}, Scott and {French}, David M. and {Lehner}, Nicolas},
        title = "{Observations of a Magellanic Corona}",
      journal = {\nat},
         year = 2022,
        month = sep,
       volume = {609},
       number = {7929},
        pages = {915-918},
          doi = {10.1038/s41586-022-05090-5},
archivePrefix = {arXiv},
       eprint = {2209.15017},
 primaryClass = {astro-ph.GA},
       adsurl = {https://ui.adsabs.harvard.edu/abs/2022Natur.609..915K}
}

@ARTICLE{prochaska2017,
       author = {{Prochaska}, J. Xavier and {Werk}, Jessica K. and {Worseck}, G{\'a}bor and {Tripp}, Todd M. and {Tumlinson}, Jason and {Burchett}, Joseph N. and {Fox}, Andrew J. and {Fumagalli}, Michele and {Lehner}, Nicolas and {Peeples}, Molly S. and {Tejos}, Nicolas},
        title = "{The COS-Halos Survey: Metallicities in the Low-redshift Circumgalactic Medium}",
      journal = {\apj},
         year = 2017,
        month = mar,
       volume = {837},
       number = {2},
          eid = {169},
        pages = {169},
          doi = {10.3847/1538-4357/aa6007},
archivePrefix = {arXiv},
       eprint = {1702.02618},
 primaryClass = {astro-ph.GA},
       adsurl = {https://ui.adsabs.harvard.edu/abs/2017ApJ...837..169P}
}

@inproceedings{witt2025juniper,
  title={The Juniper CubeSat concept: mission overview},
  author={Witt, Emily M and Fleming, Brian T and McCandliss, Stephan and Ravi, Isu and Saeedzadeh, Vida and Tumlinson, Jason and Tuttle, Sarah and Lochhaas, Cassandra and Mingozzi, Matilde and Quijada, Manuel},
  booktitle={UV, X-Ray, and Gamma-Ray Space Instrumentation for Astronomy XXIV},
  volume={13625},
  pages={173--181},
  year={2025},
  organization={SPIE}
}

@INPROCEEDINGS{Aspera2021,
       author = {{Chung}, Haeun and {Vargas}, Carlos J. and {Hamden}, Erika and {McMahon}, Tom and {Gonzales}, Kerry and {Khan}, Aafaque R. and {Agarwal}, Simran and {Bailey}, Hop and {Behroozi}, Peter and {Brendel}, Trenton and {Choi}, Heejoo and {Connors}, Tom and {Corlies}, Lauren and {Corliss}, Jason and {Dettmar}, Ralf-J{\"u}rgen and {Dolana}, David and {Douglas}, Ewan S. and {Guzman}, John and {Hamara}, Dave and {Harris}, Walt and {Harshman}, Karl and {Hergenrother}, Carl and {Hoadley}, Keri and {Kidd}, John and {Kim}, Daewook and {Li}, Jessica S. and {Montoya}, Manny and {Sauve}, Corwynn and {Schiminovich}, David and {Selznick}, Sanford and {Siegmund}, Oswald and {Ward}, Michael and {Wolcott}, Ellie M. and {Zaritsky}, Dennis},
        title = "{Aspera: the UV SmallSat telescope to detect and map the warm-hot gas phase in nearby galaxy halos}",
    booktitle = {UV/Optical/IR Space Telescopes and Instruments: Innovative Technologies and Concepts X},
         year = 2021,
       editor = {{Barto}, Allison A. and {Breckinridge}, James B. and {Stahl}, H. Philip},
       series = {Society of Photo-Optical Instrumentation Engineers (SPIE) Conference Series},
       volume = {11819},
        month = aug,
          eid = {1181903},
        pages = {1181903},
          doi = {10.1117/12.2593001},
       adsurl = {https://ui.adsabs.harvard.edu/abs/2021SPIE11819E..03C}
}

@ARTICLE{kcwi2018,
       author = {{Morrissey}, Patrick and {Matuszewski}, Matuesz and {Martin}, D. Christopher and {Neill}, James D. and {Epps}, Harland and {Fucik}, Jason and {Weber}, Bob and {Darvish}, Behnam and {Adkins}, Sean and {Allen}, Steve and {Bartos}, Randy and {Belicki}, Justin and {Cabak}, Jerry and {Callahan}, Shawn and {Cowley}, Dave and {Crabill}, Marty and {Deich}, Willian and {Delecroix}, Alex and {Doppman}, Greg and {Hilyard}, David and {James}, Ean and {Kaye}, Steve and {Kokorowski}, Michael and {Kwok}, Shui and {Lanclos}, Kyle and {Milner}, Steve and {Moore}, Anna and {O'Sullivan}, Donal and {Parihar}, Prachi and {Park}, Sam and {Phillips}, Andrew and {Rizzi}, Luca and {Rockosi}, Constance and {Rodriguez}, Hector and {Salaun}, Yves and {Seaman}, Kirk and {Sheikh}, David and {Weiss}, Jason and {Zarzaca}, Ray},
        title = "{The Keck Cosmic Web Imager Integral Field Spectrograph}",
      journal = {\apj},
         year = 2018,
        month = sep,
       volume = {864},
       number = {1},
          eid = {93},
        pages = {93},
          doi = {10.3847/1538-4357/aad597},
archivePrefix = {arXiv},
       eprint = {1807.10356},
 primaryClass = {astro-ph.IM},
       adsurl = {https://ui.adsabs.harvard.edu/abs/2018ApJ...864...93M}
}

@INPROCEEDINGS{MUSE2008,
       author = {{McDermid}, R.~M. and {Bacon}, R. and {Bauer}, S. and {Boehm}, P. and {Boudon}, D. and {Brau-Nogu{\'e}}, S. and {Caillier}, P. and {Capoani}, L. and {Carollo}, C.~M. and {Champavert}, N. and {Contini}, T. and {Daguis{\'e}}, E. and {Delabre}, B. and {Devriendt}, J. and {Dreizler}, S. and {Dubois}, J. and {Dupieux}, M. and {Dupin}, J.~P. and {Emsellem}, E. and {Ferruit}, P. and {Franx}, M. and {Gallou}, G. and {Gerssen}, J. and {Guiderdoni}, B. and {Hahn}, T. and {Hofmann}, D. and {Jarno}, A. and {Kelz}, A. and {Koehler}, C. and {Kollatschny}, W. and {Kosmalski}, J. and {Laurent}, F. and {Lilly}, S.~J. and {Lizon}, J.~L. and {Loupias}, M. and {Manescau}, A. and {Monstein}, C. and {Nicklas}, H. and {Par{\`e}s}, L. and {Pasquini}, L. and {P{\'e}contal-Rousset}, A. and {P{\'e}contal}, E. and {Pello}, R. and {Petit}, C. and {Picat}, J. -P. and {Popow}, E. and {Quirrenbach}, A. and {Reiss}, R. and {Renault}, E. and {Roth}, M. and {Schaye}, J. and {Soucail}, G. and {Steinmetz}, M. and {Stroebele}, S. and {Stuik}, R. and {Weilbacher}, P. and {Wisotzki}, L. and {Wozniak}, H. and {de Zeeuw}, P.~T.},
        title = "{MUSE: A Second-Generation Integral-Field Spectrograph for the VLT}",
    booktitle = {2007 ESO Instrument Calibration Workshop},
         year = 2008,
       editor = {{Kaufer}, Andreas and {Kerber}, Florian},
        month = jan,
        pages = {325},
          doi = {10.1007/978-3-540-76963-7_44},
       adsurl = {https://ui.adsabs.harvard.edu/abs/2008eic..work..325M}
}

@ARTICLE{Nielsen2024,
       author = {{Nielsen}, Nikole M. and {Fisher}, Deanne B. and {Kacprzak}, Glenn G. and {Chisholm}, John and {Martin}, D. Christopher and {Reichardt Chu}, Bronwyn and {Sandstrom}, Karin M. and {Rickards Vaught}, Ryan J.},
        title = "{An emission map of the disk-circumgalactic medium transition in starburst IRAS 08339+6517}",
      journal = {Nature Astronomy},
         year = 2024,
        month = dec,
       volume = {8},
        pages = {1602-1609},
          doi = {10.1038/s41550-024-02365-x},
archivePrefix = {arXiv},
       eprint = {2311.00856},
 primaryClass = {astro-ph.GA},
       adsurl = {https://ui.adsabs.harvard.edu/abs/2024NatAs...8.1602N}
}

@ARTICLE{Chatzikos2023,
       author = {{Chatzikos}, M. and {Bianchi}, S. and {Camilloni}, F. and {Chakraborty}, P. and {Gunasekera}, C.~M. and {Guzm{\'a}n}, F. and {Milby}, J.~S. and {Sarkar}, A. and {Shaw}, G. and {van Hoof}, P.~A.~M. and {Ferland}, G.~J.},
        title = "{The 2023 Release of Cloudy}",
      journal = {\rmxaa},
         year = 2023,
        month = oct,
       volume = {59},
        pages = {327-343},
          doi = {10.22201/ia.01851101p.2023.59.02.12},
archivePrefix = {arXiv},
       eprint = {2308.06396},
 primaryClass = {astro-ph.GA},
       adsurl = {https://ui.adsabs.harvard.edu/abs/2023RMxAA..59..327C}
}

@ARTICLE{Bertone2010,
       author = {{Bertone}, Serena and {Schaye}, Joop and {Booth}, C.~M. and {Dalla Vecchia}, Claudio and {Theuns}, Tom and {Wiersma}, Robert P.~C.},
        title = "{Metal-line emission from the warm-hot intergalactic medium - II. Ultraviolet}",
      journal = {\mnras},
         year = 2010,
        month = oct,
       volume = {408},
       number = {2},
        pages = {1120-1138},
          doi = {10.1111/j.1365-2966.2010.17188.x},
archivePrefix = {arXiv},
       eprint = {1002.3393},
 primaryClass = {astro-ph.CO},
       adsurl = {https://ui.adsabs.harvard.edu/abs/2010MNRAS.408.1120B}
}

@ARTICLE{Cen&Ostriker2006,
       author = {{Cen}, Renyue and {Ostriker}, Jeremiah P.},
        title = "{Where Are the Baryons? II. Feedback Effects}",
      journal = {\apj},
         year = 2006,
        month = oct,
       volume = {650},
       number = {2},
        pages = {560-572},
          doi = {10.1086/506505},
archivePrefix = {arXiv},
       eprint = {astro-ph/0601008},
 primaryClass = {astro-ph},
       adsurl = {https://ui.adsabs.harvard.edu/abs/2006ApJ...650..560C}
}

@ARTICLE{Haardt2012,
       author = {{Haardt}, Francesco and {Madau}, Piero},
        title = "{Radiative Transfer in a Clumpy Universe. IV. New Synthesis Models of the Cosmic UV/X-Ray Background}",
      journal = {\apj},
         year = 2012,
        month = feb,
       volume = {746},
       number = {2},
          eid = {125},
        pages = {125},
          doi = {10.1088/0004-637X/746/2/125},
       adsurl = {https://ui.adsabs.harvard.edu/abs/2012ApJ...746..125H}
}

@ARTICLE{TridentRef,
    author = {{Hummels}, C.~B. and {Smith}, B.~D. and {Silvia}, D.~W.},
    title = "{Trident: A Universal Tool for Generating Synthetic Absorption Spectra from Astrophysical Simulations}",
    journal = {\apj},
    archivePrefix = "arXiv",
    eprint = {1612.03935},
    primaryClass = "astro-ph.IM",
    year = 2017,
    month = sep,
    volume = 847,
    eid = {59},
    pages = {59},
    doi = {10.3847/1538-4357/aa7e2d},
    adsurl = {http://adsabs.harvard.edu/abs/2017ApJ...847...59H}
}

@ARTICLE{2003ApJ...591..821O,
       author = {{Otte}, B. and {Murphy}, E.~M. and {Howk}, J.~C. and {Wang}, Q.~D. and {Oegerle}, W.~R. and {Sembach}, K.~R.},
        title = "{Probing O VI Emission in the Halos of Edge-on Spiral Galaxies}",
      journal = {\apj},
         year = 2003,
        month = jul,
       volume = {591},
       number = {2},
        pages = {821-826},
          doi = {10.1086/375535},
archivePrefix = {arXiv},
       eprint = {astro-ph/0304079},
 primaryClass = {astro-ph},
       adsurl = {https://ui.adsabs.harvard.edu/abs/2003ApJ...591..821O}
}

@ARTICLE{2022ApJ...941..185M,
       author = {{Melso}, Nicole and {Schiminovich}, David and {Smiley}, Brian and {Ong}, Hwei Ru and {Santiago}, B{\'a}rbara Cruvinel and {Sitaram}, Meghna and {Aleman}, Ignacio Cevallos and {Graber}, Sarah and {Murillo}, Marisa and {Rosenthal}, Marni and {Stelea}, Ioana},
        title = "{The Circumgalactic H{\ensuremath{\alpha}} Spectrograph (CH{\ensuremath{\alpha}}S). I. Design, Engineering, and Early Commissioning}",
      journal = {\apj},
         year = 2022,
        month = dec,
       volume = {941},
       number = {2},
          eid = {185},
        pages = {185},
          doi = {10.3847/1538-4357/ac9d9c},
archivePrefix = {arXiv},
       eprint = {2209.14999},
 primaryClass = {astro-ph.GA},
       adsurl = {https://ui.adsabs.harvard.edu/abs/2022ApJ...941..185M}
}

@ARTICLE{2019arXiv191206219T,
       author = {{The LUVOIR Team}},
        title = "{The LUVOIR Mission Concept Study Final Report}",
      journal = {arXiv e-prints},
         year = 2019,
        month = dec,
          eid = {arXiv:1912.06219},
        pages = {arXiv:1912.06219},
          doi = {10.48550/arXiv.1912.06219},
archivePrefix = {arXiv},
       eprint = {1912.06219},
 primaryClass = {astro-ph.IM},
       adsurl = {https://ui.adsabs.harvard.edu/abs/2019arXiv191206219T}
}

@ARTICLE{2018Natur.562..229W,
       author = {{Wisotzki}, L. and {Bacon}, R. and {Brinchmann}, J. and {Cantalupo}, S. and {Richter}, P. and {Schaye}, J. and {Schmidt}, K.~B. and {Urrutia}, T. and {Weilbacher}, P.~M. and {Akhlaghi}, M. and {Bouch{\'e}}, N. and {Contini}, T. and {Guiderdoni}, B. and {Herenz}, E.~C. and {Inami}, H. and {Kerutt}, J. and {Leclercq}, F. and {Marino}, R.~A. and {Maseda}, M. and {Monreal-Ibero}, A. and {Nanayakkara}, T. and {Richard}, J. and {Saust}, R. and {Steinmetz}, M. and {Wendt}, M.},
        title = "{Nearly all the sky is covered by Lyman-{\ensuremath{\alpha}} emission around high-redshift galaxies}",
      journal = {\nat},
         year = 2018,
        month = oct,
       volume = {562},
       number = {7726},
        pages = {229-232},
          doi = {10.1038/s41586-018-0564-6},
archivePrefix = {arXiv},
       eprint = {1810.00843},
 primaryClass = {astro-ph.GA},
       adsurl = {https://ui.adsabs.harvard.edu/abs/2018Natur.562..229W}
}

@ARTICLE{2025ApJ...986..190S,
       author = {{Shaban}, Ahmed and {Bordoloi}, Rongmon and {O'Meara}, John M. and {Sharon}, Keren and {Tejos}, Nicolas and {Lopez}, Sebastian and {Ledoux}, C{\'e}dric and {Barrientos}, L. Felipe and {Rigby}, Jane R.},
        title = "{Spatially Resolved Circumgalactic Medium around a Star-forming Galaxy Driving a Galactic Outflow at z {\ensuremath{\approx}} 0.8}",
      journal = {\apj},
         year = 2025,
        month = jun,
       volume = {986},
       number = {2},
          eid = {190},
        pages = {190},
          doi = {10.3847/1538-4357/add0b9},
archivePrefix = {arXiv},
       eprint = {2501.17940},
 primaryClass = {astro-ph.GA},
       adsurl = {https://ui.adsabs.harvard.edu/abs/2025ApJ...986..190S}
}

@ARTICLE{2020MNRAS.491.4442L,
       author = {{Lopez}, S. and {Tejos}, N. and {Barrientos}, L.~F. and {Ledoux}, C. and {Sharon}, K. and {Katsianis}, A. and {Florian}, M.~K. and {Rivera-Thorsen}, E. and {Bayliss}, M.~B. and {Dahle}, H. and {Fernandez-Figueroa}, A. and {Gladders}, M.~D. and {Gronke}, M. and {Hamel}, M. and {Pessa}, I. and {Rigby}, J.~R.},
        title = "{Slicing the cool circumgalactic medium along the major axis of a star-forming galaxy at z = 0.7}",
      journal = {\mnras},
         year = 2020,
        month = jan,
       volume = {491},
       number = {3},
        pages = {4442-4461},
          doi = {10.1093/mnras/stz3183},
archivePrefix = {arXiv},
       eprint = {1911.04809},
 primaryClass = {astro-ph.GA},
       adsurl = {https://ui.adsabs.harvard.edu/abs/2020MNRAS.491.4442L}
}

@ARTICLE{2021MNRAS.502.3733W,
       author = {{Wendt}, Martin and {Bouch{\'e}}, Nicolas F. and {Zabl}, Johannes and {Schroetter}, Ilane and {Muzahid}, Sowgat},
        title = "{MusE GAs FLOw and Wind V. The dust/metallicity-anisotropy of the circum-galactic medium}",
      journal = {\mnras},
         year = 2021,
        month = apr,
       volume = {502},
       number = {3},
        pages = {3733-3745},
          doi = {10.1093/mnras/stab049},
archivePrefix = {arXiv},
       eprint = {2009.08464},
 primaryClass = {astro-ph.GA},
       adsurl = {https://ui.adsabs.harvard.edu/abs/2021MNRAS.502.3733W}
}

@ARTICLE{2019ApJ...877....4L,
       author = {{Lokhorst}, Deborah and {Abraham}, Roberto and {van Dokkum}, Pieter and {Wijers}, Nastasha and {Schaye}, Joop},
        title = "{On the Detectability of Visible-wavelength Line Emission from the Local Circumgalactic and Intergalactic Medium}",
      journal = {\apj},
         year = 2019,
        month = may,
       volume = {877},
       number = {1},
          eid = {4},
        pages = {4},
          doi = {10.3847/1538-4357/ab184e},
archivePrefix = {arXiv},
       eprint = {1904.07874},
 primaryClass = {astro-ph.GA},
       adsurl = {https://ui.adsabs.harvard.edu/abs/2019ApJ...877....4L}
}

@ARTICLE{2019MNRAS.489.2417A,
       author = {{Augustin}, R. and {Quiret}, S. and {Milliard}, B. and {P{\'e}roux}, C. and {Vibert}, D. and {Blaizot}, J. and {Rasera}, Y. and {Teyssier}, R. and {Frank}, S. and {Deharveng}, J. -M. and {Picouet}, V. and {Martin}, D.~C. and {Hamden}, E.~T. and {Thatte}, N. and {Pereira Santaella}, M. and {Routledge}, L. and {Zieleniewski}, S.},
        title = "{Emission from the circumgalactic medium: from cosmological zoom-in simulations to multiwavelength observables}",
      journal = {\mnras},
         year = 2019,
        month = oct,
       volume = {489},
       number = {2},
        pages = {2417-2438},
          doi = {10.1093/mnras/stz2238},
archivePrefix = {arXiv},
       eprint = {1909.02575},
 primaryClass = {astro-ph.GA},
       adsurl = {https://ui.adsabs.harvard.edu/abs/2019MNRAS.489.2417A}
}

@ARTICLE{2025PASP..137h4103C,
       author = {{Chen}, Seery and {Lokhorst}, Deborah M. and {Pasha}, Imad and {Bowman}, William P. and {Janssens}, Steven R. and {Rhea}, Carter and {Liu}, Qing and {Shen}, Zili and {Abraham}, Roberto G. and {van Dokkum}, Pieter},
        title = "{First Light with the 120-lens Dragonfly Spectral Line Mapper}",
      journal = {\pasp},
         year = 2025,
        month = aug,
       volume = {137},
       number = {8},
          eid = {084103},
        pages = {084103},
          doi = {10.1088/1538-3873/adedb8},
       adsurl = {https://ui.adsabs.harvard.edu/abs/2025PASP..137h4103C}
}

@ARTICLE{2019NatAs...3..822M,
       author = {{Martin}, D. Christopher and {O'Sullivan}, Donal and {Matuszewski}, Mateusz and {Hamden}, Erika and {Dekel}, Avishai and {Lapiner}, Sharon and {Morrissey}, Patrick and {Neill}, James D. and {Cantalupo}, Sebastiano and {Prochaska}, Jason Xavier and {Steidel}, Charles and {Trainor}, Ryan and {Moore}, Anna and {Ceverino}, Daniel and {Primack}, Joel and {Rizzi}, Luca},
        title = "{Multi-filament gas inflows fuelling young star-forming galaxies}",
      journal = {Nature Astronomy},
         year = 2019,
        month = jul,
       volume = {3},
        pages = {822-831},
          doi = {10.1038/s41550-019-0791-2},
archivePrefix = {arXiv},
       eprint = {1904.11465},
 primaryClass = {astro-ph.GA},
       adsurl = {https://ui.adsabs.harvard.edu/abs/2019NatAs...3..822M}
}

@ARTICLE{2024ApJ...974..161M,
       author = {{Melso}, Nicole and {Schiminovich}, David and {Sitaram}, Meghna and {Cevallos-Aleman}, Ignacio and {Cruvinel Santiago}, B{\'a}rbara and {Smiley}, Brian and {Ong}, Hwei Ru},
        title = "{Very Extended Ionized Gas Discovered around NGC 1068 with the Circumgalactic H{\ensuremath{\alpha}} Spectrograph}",
      journal = {\apj},
         year = 2024,
        month = oct,
       volume = {974},
       number = {2},
          eid = {161},
        pages = {161},
          doi = {10.3847/1538-4357/ad6cd1},
archivePrefix = {arXiv},
       eprint = {2408.12597},
 primaryClass = {astro-ph.GA},
       adsurl = {https://ui.adsabs.harvard.edu/abs/2024ApJ...974..161M}
}

@ARTICLE{2025ApJS..278...58M,
       author = {{Miles}, Drew M. and {Picouet}, Vincent and {Lin}, Zeren and {Cevallos-Aleman}, Ignacio and {Schiminovich}, David and {Bray}, Nicolas and {Chevrier}, Charles-Antoine and {Davis}, Greyson and {Hamden}, Erika and {Hoadley}, Keri and {Martin}, D. Christopher and {Milliard}, Bruno and {Montel}, Johan and {Valls-Gabaud}, David and {Agarwal}, Simran and {Balard}, Philippe and {Blanchard}, Patrick and {Bradley}, Harrison and {Chung}, Haeun and {Crabill}, Marty and {Cruz Aguirre}, Fernando and {Deng}, Xihan and {Harmand}, Fabien and {Hourtolle}, Catherine and {Jones}, Olivia and {Kerkeser}, Nazende I. and {Khan}, Aafaque R. and {Kyne}, Gillian and {Li}, Jessica S. and {Melso}, Nicole and {Nikzad}, Shouleh and {Richard}, Julie and {Sitaram}, Meghna and {Termini}, Jared and {Valdivia}, Jean-Noel and {Vibert}, Didier and {Werneken}, Matthew},
        title = "{The 2023 Flight of the Faint Intergalactic-medium Redshifted Emission Balloon}",
      journal = {\apjs},
         year = 2025,
        month = jun,
       volume = {278},
       number = {2},
          eid = {58},
        pages = {58},
          doi = {10.3847/1538-4365/add15c},
       adsurl = {https://ui.adsabs.harvard.edu/abs/2025ApJS..278...58M}
}

@ARTICLE{2007ApJ...668..891G,
       author = {{Grimes}, J.~P. and {Heckman}, T. and {Strickland}, D. and {Dixon}, W.~V. and {Sembach}, K. and {Overzier}, R. and {Hoopes}, C. and {Aloisi}, A. and {Ptak}, A.},
        title = "{Feedback in the Local Lyman-break Galaxy Analog Haro 11 as Probed by Far-Ultraviolet and X-Ray Observations}",
      journal = {\apj},
         year = 2007,
        month = oct,
       volume = {668},
       number = {2},
        pages = {891-905},
          doi = {10.1086/521353},
archivePrefix = {arXiv},
       eprint = {0707.0693},
 primaryClass = {astro-ph},
       adsurl = {https://ui.adsabs.harvard.edu/abs/2007ApJ...668..891G}
}

@ARTICLE{2025ApJ...986...87H,
       author = {{Ha}, Triet and {Rupke}, David S.~N. and {Caraker}, Shane and {Harper}, Jack and {Coil}, Alison L. and {Li}, Miao and {Tremonti}, Christy A. and {Diamond-Stanic}, Aleksandar M. and {Geach}, James E. and {Hickox}, Ryan C. and {Johnson}, Sean D. and {Leung}, Gene C.~K. and {Moustakas}, John and {Perrotta}, Serena and {Rudnick}, Gregory H. and {Sell}, Paul H. and {Whalen}, Kelly E.},
        title = "{Deep Ultraviolet, Emission-line Imaging of the Makani Galactic Wind}",
      journal = {\apj},
         year = 2025,
        month = jun,
       volume = {986},
       number = {1},
          eid = {87},
        pages = {87},
          doi = {10.3847/1538-4357/add0b5},
archivePrefix = {arXiv},
       eprint = {2503.20042},
 primaryClass = {astro-ph.GA},
       adsurl = {https://ui.adsabs.harvard.edu/abs/2025ApJ...986...87H}
}

@ARTICLE{2024AJ....168...11K,
       author = {{Kim}, Jin-Ah and {Chung}, Haeun and {Vargas}, Carlos J. and {Hamden}, Erika},
        title = "{UV Cooling via O VI Emission in the Superwind of M82 Observed with the Far Ultraviolet Spectroscopic Explorer (FUSE)}",
      journal = {\aj},
         year = 2024,
        month = jul,
       volume = {168},
       number = {1},
          eid = {11},
        pages = {11},
          doi = {10.3847/1538-3881/ad4887},
archivePrefix = {arXiv},
       eprint = {2406.01742},
 primaryClass = {astro-ph.GA},
       adsurl = {https://ui.adsabs.harvard.edu/abs/2024AJ....168...11K}
}

@ARTICLE{2021ApJ...916....7C,
       author = {{Chung}, Haeun and {Vargas}, Carlos J. and {Hamden}, Erika},
        title = "{Revisiting FUSE O VI Emission in Galaxy Halos}",
      journal = {\apj},
         year = 2021,
        month = jul,
       volume = {916},
       number = {1},
          eid = {7},
        pages = {7},
          doi = {10.3847/1538-4357/ac04af},
archivePrefix = {arXiv},
       eprint = {2103.05008},
 primaryClass = {astro-ph.GA},
       adsurl = {https://ui.adsabs.harvard.edu/abs/2021ApJ...916....7C}
}

@INPROCEEDINGS{Chung2021,
       author = {{Chung}, Haeun and {Vargas}, Carlos J. and {Hamden}, Erika and {McMahon}, Tom and {Gonzales}, Kerry and {Khan}, Aafaque R. and {Agarwal}, Simran and {Bailey}, Hop and {Behroozi}, Peter and {Brendel}, Trenton and {Choi}, Heejoo and {Connors}, Tom and {Corlies}, Lauren and {Corliss}, Jason and {Dettmar}, Ralf-J{\"u}rgen and {Dolana}, David and {Douglas}, Ewan S. and {Guzman}, John and {Hamara}, Dave and {Harris}, Walt and {Harshman}, Karl and {Hergenrother}, Carl and {Hoadley}, Keri and {Kidd}, John and {Kim}, Daewook and {Li}, Jessica S. and {Montoya}, Manny and {Sauve}, Corwynn and {Schiminovich}, David and {Selznick}, Sanford and {Siegmund}, Oswald and {Ward}, Michael and {Wolcott}, Ellie M. and {Zaritsky}, Dennis},
        title = "{Aspera: the UV SmallSat telescope to detect and map the warm-hot gas phase in nearby galaxy halos}",
    booktitle = {UV/Optical/IR Space Telescopes and Instruments: Innovative Technologies and Concepts X},
         year = 2021,
       editor = {{Barto}, Allison A. and {Breckinridge}, James B. and {Stahl}, H. Philip},
       series = {Society of Photo-Optical Instrumentation Engineers (SPIE) Conference Series},
       volume = {11819},
        month = aug,
          eid = {1181903},
        pages = {1181903},
          doi = {10.1117/12.2593001},
       adsurl = {https://ui.adsabs.harvard.edu/abs/2021SPIE11819E..03C}
}

@ARTICLE{2014ApJ...786...54P,
       author = {{Peeples}, Molly S. and {Werk}, Jessica K. and {Tumlinson}, Jason and
         {Oppenheimer}, Benjamin D. and {Prochaska}, J. Xavier and {Katz}, Neal and
         {Weinberg}, David H.},
        title = "{A Budget and Accounting of Metals at z \textasciitilde 0: Results from the COS-Halos Survey}",
      journal = {\apj},
         year = "2014",
        month = "May",
       volume = {786},
          eid = {54},
        pages = {54},
          doi = {10.1088/0004-637X/786/1/54},
archivePrefix = {arXiv},
       eprint = {1310.2253},
 primaryClass = {astro-ph.CO},
       adsurl = {https://ui.adsabs.harvard.edu/\#abs/2014ApJ...786...54P}
}

@ARTICLE{2025arXiv251005667C,
       author = {{Cadiou}, Corentin and {Katz}, Harley and {Rey}, Martin P. and {Agertz}, Oscar and {Blaizot}, Jeremy and {Cameron}, Alex J. and {Choustikov}, Nicholas and {Devriendt}, Julien and {Hauk}, Uliana and {Jones}, Gareth C. and {Kimm}, Taysun and {Laseter}, Isaac and {Mart{\'\i}n {\'A}lvarez}, Sergio and {Matsumoto}, Kosei and {Nyhagen}, Camilla T. and {Pearce}, Autumn and {Rodr{\'\i}guez Montero}, Francisco and {Rosdahl}, Joki and {Rufo Pastor}, V{\'\i}ctor and {Sanati}, Mahsa and {Saxena}, Aayush and {Slyz}, Adrianne and {Stiskalek}, Richard and {Storck}, Anatole and {Yee}, Wonjae},
        title = "{MEGATRON: the impact of non-equilibrium effects and local radiation fields on the circumgalactic medium at cosmic noon}",
      journal = {arXiv e-prints},
         year = 2025,
        month = oct,
          eid = {arXiv:2510.05667},
        pages = {arXiv:2510.05667},
archivePrefix = {arXiv},
       eprint = {2510.05667},
 primaryClass = {astro-ph.GA},
       adsurl = {https://ui.adsabs.harvard.edu/abs/2025arXiv251005667C}
}

@ARTICLE{2025arXiv250616573L,
       author = {{Lehner}, Nicolas and {Howk}, J. Christopher and {Collins}, Lucy and {Sameer} and {Wakker}, Bart P. and {Augustin}, Ramona and {Barger}, Kathleen A. and {Berg}, Michelle A. and {Bordoloi}, Rongmon and {Brown}, Thomas M. and {Cashman}, Frances H. and {Faucher-Gigu{\`e}re}, Claude-Andr{\'e} and {Fox}, Andrew J. and {French}, David M. and {Gilbert}, Karoline M. and {Guhathakurta}, Puragra and {O'Meara}, John M. and {O'Shea}, Brian W. and {Peeples}, Molly S. and {Pisano}, D.~J. and {Prochaska}, J. Xavier and {Stern}, Jonathan and {Tumlinson}, Jason and {Werk}, Jessica K. and {Williams}, Benjamin F.},
        title = "{Project AMIGA: The Inner Circumgalactic Medium of Andromeda from Thick Disk to Halo}",
      journal = {arXiv e-prints},
         year = 2025,
        month = jun,
          eid = {arXiv:2506.16573},
        pages = {arXiv:2506.16573},
          doi = {10.48550/arXiv.2506.16573},
archivePrefix = {arXiv},
       eprint = {2506.16573},
 primaryClass = {astro-ph.GA},
       adsurl = {https://ui.adsabs.harvard.edu/abs/2025arXiv250616573L}
}

@ARTICLE{2012ApJ...760L...7K,
       author = {{Kacprzak}, Glenn G. and {Churchill}, Christopher W. and {Nielsen}, Nikole M.},
        title = "{Tracing Outflows and Accretion: A Bimodal Azimuthal Dependence of Mg II Absorption}",
      journal = {\apjl},
         year = 2012,
        month = nov,
       volume = {760},
       number = {1},
          eid = {L7},
        pages = {L7},
          doi = {10.1088/2041-8205/760/1/L7},
archivePrefix = {arXiv},
       eprint = {1205.0245},
 primaryClass = {astro-ph.CO},
       adsurl = {https://ui.adsabs.harvard.edu/abs/2012ApJ...760L...7K}
}

@ARTICLE{Hayes2016,
       author = {{Hayes}, Matthew and {Melinder}, Jens and {{\"O}stlin}, G{\"o}ran and {Scarlata}, Claudia and {Lehnert}, Matthew D. and {Mannerstr{\"o}m-Jansson}, Gustav},
        title = "{O VI Emission Imaging of a Galaxy with the Hubble Space Telescope: a Warm Gas Halo Surrounding the Intense Starburst SDSS J115630.63+500822.1}",
      journal = {\apj},
         year = 2016,
        month = sep,
       volume = {828},
       number = {1},
          eid = {49},
        pages = {49},
          doi = {10.3847/0004-637X/828/1/49},
archivePrefix = {arXiv},
       eprint = {1606.04536},
 primaryClass = {astro-ph.GA},
       adsurl = {https://ui.adsabs.harvard.edu/abs/2016ApJ...828...49H}
}

@ARTICLE{2023ARA&A..61..131F,
       author = {{Faucher-Gigu{\`e}re}, Claude-Andr{\'e} and {Oh}, S. Peng},
        title = "{Key Physical Processes in the Circumgalactic Medium}",
      journal = {\araa},
         year = 2023,
        month = aug,
       volume = {61},
        pages = {131-195},
          doi = {10.1146/annurev-astro-052920-125203},
archivePrefix = {arXiv},
       eprint = {2301.10253},
 primaryClass = {astro-ph.GA},
       adsurl = {https://ui.adsabs.harvard.edu/abs/2023ARA&A..61..131F}
}

@ARTICLE{Lehner2025,
       author = {{Lehner}, Nicolas and {Howk}, J. Christopher and {Collins}, Lucy and {Sameer} and {Wakker}, Bart P. and {Augustin}, Ramona and {Barger}, Kathleen A. and {Berg}, Michelle A. and {Bordoloi}, Rongmon and {Brown}, Thomas M. and {Cashman}, Frances H. and {Faucher-Gigu{\`e}re}, Claude-Andr{\'e} and {Fox}, Andrew J. and {French}, David M. and {Gilbert}, Karoline M. and {Guhathakurta}, Puragra and {O'Meara}, John M. and {O'Shea}, Brian W. and {Peeples}, Molly S. and {Pisano}, D.~J. and {Prochaska}, J. Xavier and {Stern}, Jonathan and {Tumlinson}, Jason and {Werk}, Jessica K. and {Williams}, Benjamin F.},
        title = "{Project AMIGA: The Inner Circumgalactic Medium of Andromeda from Thick Disk to Halo}",
      journal = {arXiv e-prints},
         year = 2025,
        month = jun,
          eid = {arXiv:2506.16573},
        pages = {arXiv:2506.16573},
          doi = {10.48550/arXiv.2506.16573},
archivePrefix = {arXiv},
       eprint = {2506.16573},
 primaryClass = {astro-ph.GA},
       adsurl = {https://ui.adsabs.harvard.edu/abs/2025arXiv250616573L}
}

@ARTICLE{Bordoloi2022Nature,
       author = {{Bordoloi}, Rongmon and {O'Meara}, John M. and {Sharon}, Keren and {Rigby}, Jane R. and {Cooke}, Jeff and {Shaban}, Ahmed and {Matuszewski}, Mateusz and {Rizzi}, Luca and {Doppmann}, Greg and {Martin}, D. Christopher and {Moore}, Anna M. and {Morrissey}, Patrick and {Neill}, James D.},
        title = "{Resolving the H I in damped Lyman {\ensuremath{\alpha}} systems that power star formation}",
      journal = {\nat},
         year = 2022,
        month = may,
       volume = {606},
       number = {7912},
        pages = {59-63},
          doi = {10.1038/s41586-022-04616-1},
archivePrefix = {arXiv},
       eprint = {2205.08554},
 primaryClass = {astro-ph.GA},
       adsurl = {https://ui.adsabs.harvard.edu/abs/2022Natur.606...59B}
}

@ARTICLE{Kusakabe2024,
       author = {{Kusakabe}, Haruka and {Mauerhofer}, Valentin and {Verhamme}, Anne and {Garel}, Thibault and {Blaizot}, Jeremy and {Wisotzki}, Lutz and {Richard}, Johan and {Boogaard}, Leindert A. and {Leclercq}, Floriane and {Guo}, Yucheng and {Claeyssens}, Adelaide and {Contini}, Thierry and {Herenz}, Edmund Christian and {Kerutt}, Josephine and {Maseda}, Michael V. and {Michel-Dansac}, Leo and {Nanayakkara}, Themiya and {Ouchi}, Masami and {Pessa}, Ismael and {Schaye}, Joop},
        title = "{The MUSE eXtremely Deep Field: Detections of circumgalactic SiII* emission at z>\raisebox{-0.5ex}\textasciitilde2}",
      journal = {arXiv e-prints},
         year = 2024,
        month = jun,
          eid = {arXiv:2406.04399},
        pages = {arXiv:2406.04399},
          doi = {10.48550/arXiv.2406.04399},
archivePrefix = {arXiv},
       eprint = {2406.04399},
 primaryClass = {astro-ph.GA},
       adsurl = {https://ui.adsabs.harvard.edu/abs/2024arXiv240604399K}
}

@ARTICLE{Augustin2021,
       author = {{Augustin}, Ramona and {P{\'e}roux}, C{\'e}line and {Hamanowicz}, Aleksandra and {Kulkarni}, Varsha and {Rahmani}, Hadi and {Zanella}, Anita},
        title = "{Clumpiness of observed and simulated cold circumgalactic gas}",
      journal = {\mnras},
         year = 2021,
        month = aug,
       volume = {505},
       number = {4},
        pages = {6195-6205},
          doi = {10.1093/mnras/stab1673},
archivePrefix = {arXiv},
       eprint = {2105.11480},
 primaryClass = {astro-ph.GA},
       adsurl = {https://ui.adsabs.harvard.edu/abs/2021MNRAS.505.6195A}
}

@ARTICLE{Augustin2025,
       author = {{Augustin}, Ramona and {Tumlinson}, Jason and {Peeples}, Molly S. and {O'Shea}, Brian W. and {Smith}, Britton D. and {Lochhaas}, Cassandra and {Wright}, Anna C. and {Acharyya}, Ayan and {Werk}, Jessica K. and {Lehner}, Nicolas and {Howk}, J. Christopher and {Corlies}, Lauren and {Simons}, Raymond C. and {O'Meara}, John M.},
        title = "{FOGGIE X: Characterizing the Small-Scale Structure of the CGM and its Imprint on Observables}",
      journal = {arXiv e-prints},
         year = 2025,
        month = jan,
          eid = {arXiv:2501.06551},
        pages = {arXiv:2501.06551},
          doi = {10.48550/arXiv.2501.06551},
archivePrefix = {arXiv},
       eprint = {2501.06551},
 primaryClass = {astro-ph.GA},
       adsurl = {https://ui.adsabs.harvard.edu/abs/2025arXiv250106551A}
}

@ARTICLE{astropy2022,
       author = {{Astropy Collaboration} and {Price-Whelan}, Adrian M. and {Lim}, Pey Lian and {Earl}, Nicholas and {Starkman}, Nathaniel and {Bradley}, Larry and {Shupe}, David L. and {Patil}, Aarya A. and {Corrales}, Lia and {Brasseur}, C.~E. and {N{\"o}the}, Maximilian and {Donath}, Axel and {Tollerud}, Erik and {Morris}, Brett M. and {Ginsburg}, Adam and {Vaher}, Eero and {Weaver}, Benjamin A. and {Tocknell}, James and {Jamieson}, William and {van Kerkwijk}, Marten H. and {Robitaille}, Thomas P. and {Merry}, Bruce and {Bachetti}, Matteo and {G{\"u}nther}, H. Moritz and {Aldcroft}, Thomas L. and {Alvarado-Montes}, Jaime A. and {Archibald}, Anne M. and {B{\'o}di}, Attila and {Bapat}, Shreyas and {Barentsen}, Geert and {Baz{\'a}n}, Juanjo and {Biswas}, Manish and {Boquien}, M{\'e}d{\'e}ric and {Burke}, D.~J. and {Cara}, Daria and {Cara}, Mihai and {Conroy}, Kyle E. and {Conseil}, Simon and {Craig}, Matthew W. and {Cross}, Robert M. and {Cruz}, Kelle L. and {D'Eugenio}, Francesco and {Dencheva}, Nadia and {Devillepoix}, Hadrien A.~R. and {Dietrich}, J{\"o}rg P. and {Eigenbrot}, Arthur Davis and {Erben}, Thomas and {Ferreira}, Leonardo and {Foreman-Mackey}, Daniel and {Fox}, Ryan and {Freij}, Nabil and {Garg}, Suyog and {Geda}, Robel and {Glattly}, Lauren and {Gondhalekar}, Yash and {Gordon}, Karl D. and {Grant}, David and {Greenfield}, Perry and {Groener}, Austen M. and {Guest}, Steve and {Gurovich}, Sebastian and {Handberg}, Rasmus and {Hart}, Akeem and {Hatfield-Dodds}, Zac and {Homeier}, Derek and {Hosseinzadeh}, Griffin and {Jenness}, Tim and {Jones}, Craig K. and {Joseph}, Prajwel and {Kalmbach}, J. Bryce and {Karamehmetoglu}, Emir and {Ka{\l}uszy{\'n}ski}, Miko{\l}aj and {Kelley}, Michael S.~P. and {Kern}, Nicholas and {Kerzendorf}, Wolfgang E. and {Koch}, Eric W. and {Kulumani}, Shankar and {Lee}, Antony and {Ly}, Chun and {Ma}, Zhiyuan and {MacBride}, Conor and {Maljaars}, Jakob M. and {Muna}, Demitri and {Murphy}, N.~A. and {Norman}, Henrik and {O'Steen}, Richard and {Oman}, Kyle A. and {Pacifici}, Camilla and {Pascual}, Sergio and {Pascual-Granado}, J. and {Patil}, Rohit R. and {Perren}, Gabriel I. and {Pickering}, Timothy E. and {Rastogi}, Tanuj and {Roulston}, Benjamin R. and {Ryan}, Daniel F. and {Rykoff}, Eli S. and {Sabater}, Jose and {Sakurikar}, Parikshit and {Salgado}, Jes{\'u}s and {Sanghi}, Aniket and {Saunders}, Nicholas and {Savchenko}, Volodymyr and {Schwardt}, Ludwig and {Seifert-Eckert}, Michael and {Shih}, Albert Y. and {Jain}, Anany Shrey and {Shukla}, Gyanendra and {Sick}, Jonathan and {Simpson}, Chris and {Singanamalla}, Sudheesh and {Singer}, Leo P. and {Singhal}, Jaladh and {Sinha}, Manodeep and {Sip{\H{o}}cz}, Brigitta M. and {Spitler}, Lee R. and {Stansby}, David and {Streicher}, Ole and {{\v{S}}umak}, Jani and {Swinbank}, John D. and {Taranu}, Dan S. and {Tewary}, Nikita and {Tremblay}, Grant R. and {Val-Borro}, Miguel de and {Van Kooten}, Samuel J. and {Vasovi{\'c}}, Zlatan and {Verma}, Shresth and {de Miranda Cardoso}, Jos{\'e} Vin{\'\i}cius and {Williams}, Peter K.~G. and {Wilson}, Tom J. and {Winkel}, Benjamin and {Wood-Vasey}, W.~M. and {Xue}, Rui and {Yoachim}, Peter and {Zhang}, Chen and {Zonca}, Andrea and {Astropy Project Contributors}},
        title = "{The Astropy Project: Sustaining and Growing a Community-oriented Open-source Project and the Latest Major Release (v5.0) of the Core Package}",
      journal = {\apj},
         year = 2022,
        month = aug,
       volume = {935},
       number = {2},
          eid = {167},
        pages = {167},
          doi = {10.3847/1538-4357/ac7c74},
archivePrefix = {arXiv},
       eprint = {2206.14220},
 primaryClass = {astro-ph.IM},
       adsurl = {https://ui.adsabs.harvard.edu/abs/2022ApJ...935..167A}
}

@ARTICLE{hunter2007,
       author = {{Hunter}, John D.},
        title = "{Matplotlib: A 2D Graphics Environment}",
      journal = {Computing in Science and Engineering},
         year = 2007,
        month = may,
       volume = {9},
       number = {3},
        pages = {90-95},
          doi = {10.1109/MCSE.2007.55},
       adsurl = {https://ui.adsabs.harvard.edu/abs/2007CSE.....9...90H}
}

@article{walt2011numpy,
  title={The NumPy array: a structure for efficient numerical computation},
  author={Walt, St{\'e}fan van der and Colbert, S Chris and Varoquaux, Gael},
  journal={Computing in Science \& Engineering},
  volume={13},
  number={2},
  pages={22--30},
  year={2011},
  publisher={IEEE}
}

@ARTICLE{Behroozi2013a,
       author = {{Behroozi}, Peter S. and {Wechsler}, Risa H. and {Wu}, Hao-Yi},
        title = "{The ROCKSTAR Phase-space Temporal Halo Finder and the Velocity Offsets of Cluster Cores}",
      journal = {\apj},
         year = 2013,
        month = jan,
       volume = {762},
       number = {2},
          eid = {109},
        pages = {109},
          doi = {10.1088/0004-637X/762/2/109},
archivePrefix = {arXiv},
       eprint = {1110.4372},
 primaryClass = {astro-ph.CO},
       adsurl = {https://ui.adsabs.harvard.edu/abs/2013ApJ...762..109B}
}

@ARTICLE{pontzen2018,
       author = {{Pontzen}, Andrew and {Tremmel}, Michael},
        title = "{TANGOS: The Agile Numerical Galaxy Organization System}",
      journal = {\apjs},
         year = 2018,
        month = aug,
       volume = {237},
       number = {2},
          eid = {23},
        pages = {23},
          doi = {10.3847/1538-4365/aac832},
archivePrefix = {arXiv},
       eprint = {1803.00010},
 primaryClass = {astro-ph.IM},
       adsurl = {https://ui.adsabs.harvard.edu/abs/2018ApJS..237...23P}
}

@ARTICLE{Bowen2016,
       author = {{Bowen}, David V. and {Chelouche}, Doron and {Jenkins}, Edward B. and {Tripp}, Todd M. and {Pettini}, Max and {York}, Donald G. and {Frye}, Brenda L.},
        title = "{The Structure of the Circumgalactic Medium of Galaxies: Cool Accretion Inflow Around NGC 1097}",
      journal = {\apj},
         year = 2016,
        month = jul,
       volume = {826},
       number = {1},
          eid = {50},
        pages = {50},
          doi = {10.3847/0004-637X/826/1/50},
archivePrefix = {arXiv},
       eprint = {1605.04907},
 primaryClass = {astro-ph.GA},
       adsurl = {https://ui.adsabs.harvard.edu/abs/2016ApJ...826...50B}
}

@ARTICLE{Acharyya2024,
       author = {{Acharyya}, Ayan and {Peeples}, Molly S. and {Tumlinson}, Jason and {Shea}, Brian W. O and {Lochhaas}, Cassandra and {Wright}, Anna C. and {Simons}, Raymond C. and {Augustin}, Ramona and {Smith}, Britton D. and {Hyeonmin Lee}, Eugene},
        title = "{Figuring Out Gas \& Galaxies In Enzo (FOGGIE) VIII: Complex and Stochastic Metallicity Gradients at z > 2}",
      journal = {arXiv e-prints},
         year = 2024,
        month = apr,
          eid = {arXiv:2404.06613},
        pages = {arXiv:2404.06613},
          doi = {10.48550/arXiv.2404.06613},
archivePrefix = {arXiv},
       eprint = {2404.06613},
 primaryClass = {astro-ph.GA},
       adsurl = {https://ui.adsabs.harvard.edu/abs/2024arXiv240406613A}
}

@ARTICLE{BrummelSmith2019,
       author = {{Brummel-Smith}, Corey and {Bryan}, Greg and {Butsky}, Iryna and {Corlies}, Lauren and {Emerick}, Andrew and {Forbes}, John and {Fujimoto}, Yusuke and {Goldbaum}, Nathan and {Grete}, Philipp and {Hummels}, Cameron and {Kim}, Ji-hoon and {Koh}, Daegene and {Li}, Miao and {Li}, Yuan and {Li}, Xinyu and {OShea}, Brian and {Peeples}, Molly and {Regan}, John and {Salem}, Munier and {Schmidt}, Wolfram and {Simpson}, Christine and {Smith}, Britton and {Tumlinson}, Jason and {Turk}, Matthew and {Wise}, John and {Abel}, Tom and {Bordner}, James and {Cen}, Renyue and {Collins}, David and {Crosby}, Brian and {Edelmann}, Philipp and {Hahn}, Oliver and {Harkness}, Robert and {Harper-Clark}, Elizabeth and {Kong}, Shuo and {Kritsuk}, Alexei and {Kuhlen}, Michael and {Larrue}, James and {Lee}, Eve and {Meece}, Greg and {Norman}, Michael and {Oishi}, Jeffrey and {Paschos}, Pascal and {Peruta}, Carolyn and {Razoumov}, Alex and {Reynolds}, Daniel and {Silvia}, Devin and {Skillman}, Samuel and {Skory}, Stephen and {So}, Geoffrey and {Tasker}, Elizabeth and {Wagner}, Rick and {Wang}, Peng and {Xu}, Hao and {Zhao}, Fen},
        title = "{ENZO: An Adaptive Mesh Refinement Code for Astrophysics (Version 2.6)}",
      journal = {The Journal of Open Source Software},
         year = 2019,
        month = oct,
       volume = {4},
       number = {42},
          eid = {1636},
        pages = {1636},
          doi = {10.21105/joss.01636},
       adsurl = {https://ui.adsabs.harvard.edu/abs/2019JOSS....4.1636B}
}

@ARTICLE{Bryan2014,
       author = {{Bryan}, Greg L. and {Norman}, Michael L. and {O'Shea}, Brian W. and
         {Abel}, Tom and {Wise}, John H. and {Turk}, Matthew J. and
         {Reynolds}, Daniel R. and {Collins}, David C. and {Wang}, Peng and
         {Skillman}, Samuel W. and {Smith}, Britton and {Harkness}, Robert P. and
         {Bordner}, James and {Kim}, Ji-hoon and {Kuhlen}, Michael and
         {Xu}, Hao and {Goldbaum}, Nathan and {Hummels}, Cameron and
         {Kritsuk}, Alexei G. and {Tasker}, Elizabeth and {Skory}, Stephen and
         {Simpson}, Christine M. and {Hahn}, Oliver and {Oishi}, Jeffrey S. and
         {So}, Geoffrey C. and {Zhao}, Fen and {Cen}, Renyue and {Li}, Yuan and
         {Enzo Collaboration}},
        title = "{ENZO: An Adaptive Mesh Refinement Code for Astrophysics}",
      journal = {\apjs},
         year = 2014,
        month = apr,
       volume = {211},
       number = {2},
          eid = {19},
        pages = {19},
          doi = {10.1088/0067-0049/211/2/19},
archivePrefix = {arXiv},
       eprint = {1307.2265},
 primaryClass = {astro-ph.IM},
       adsurl = {https://ui.adsabs.harvard.edu/abs/2014ApJS..211...19B}
}

@ARTICLE{Corlies2020,
       author = {{Corlies}, Lauren and {Peeples}, Molly S. and {Tumlinson}, Jason and
         {O'Shea}, Brian W. and {Lehner}, Nicolas and {Howk}, J. Christopher and
         {O'Meara}, John M. and {Smith}, Britton D.},
        title = "{Figuring Out Gas \& Galaxies in Enzo (FOGGIE). II. Emission from the z = 3 Circumgalactic Medium}",
      journal = {\apj},
         year = 2020,
        month = jun,
       volume = {896},
       number = {2},
          eid = {125},
        pages = {125},
          doi = {10.3847/1538-4357/ab9310},
archivePrefix = {arXiv},
       eprint = {1811.05060},
 primaryClass = {astro-ph.GA},
       adsurl = {https://ui.adsabs.harvard.edu/abs/2020ApJ...896..125C}
}

@ARTICLE{Hummels2019,
       author = {{Hummels}, Cameron B. and {Smith}, Britton D. and {Hopkins}, Philip F. and {O'Shea}, Brian W. and {Silvia}, Devin W. and {Werk}, Jessica K. and {Lehner}, Nicolas and {Wise}, John H. and {Collins}, David C. and {Butsky}, Iryna S.},
        title = "{The Impact of Enhanced Halo Resolution on the Simulated Circumgalactic Medium}",
      journal = {\apj},
         year = 2019,
        month = sep,
       volume = {882},
       number = {2},
          eid = {156},
        pages = {156},
          doi = {10.3847/1538-4357/ab378f},
archivePrefix = {arXiv},
       eprint = {1811.12410},
 primaryClass = {astro-ph.GA},
       adsurl = {https://ui.adsabs.harvard.edu/abs/2019ApJ...882..156H}
}

@ARTICLE{2018ApJ...866...36L,
       author = {{Lan}, Ting-Wen and {Mo}, Houjun},
        title = "{The Circumgalactic Medium of eBOSS Emission Line Galaxies: Signatures of Galactic Outflows in Gas Distribution and Kinematics}",
      journal = {\apj},
         year = 2018,
        month = oct,
       volume = {866},
       number = {1},
          eid = {36},
        pages = {36},
          doi = {10.3847/1538-4357/aadc08},
archivePrefix = {arXiv},
       eprint = {1806.05786},
 primaryClass = {astro-ph.GA},
       adsurl = {https://ui.adsabs.harvard.edu/abs/2018ApJ...866...36L}
}

@ARTICLE{Lehner2013,
       author = {{Lehner}, N. and {Howk}, J.~C. and {Tripp}, T.~M. and {Tumlinson}, J. and {Prochaska}, J.~X. and {O'Meara}, J.~M. and {Thom}, C. and {Werk}, J.~K. and {Fox}, A.~J. and {Ribaudo}, J.},
        title = "{The Bimodal Metallicity Distribution of the Cool Circumgalactic Medium at z <\raisebox{-0.5ex}\textasciitilde 1}",
      journal = {\apj},
         year = 2013,
        month = jun,
       volume = {770},
       number = {2},
          eid = {138},
        pages = {138},
          doi = {10.1088/0004-637X/770/2/138},
archivePrefix = {arXiv},
       eprint = {1302.5424},
 primaryClass = {astro-ph.CO},
       adsurl = {https://ui.adsabs.harvard.edu/abs/2013ApJ...770..138L}
}

@ARTICLE{Lehner2020,
       author = {{Lehner}, Nicolas and {Berek}, Samantha C. and {Howk}, J. Christopher and {Wakker}, Bart P. and {Tumlinson}, Jason and {Jenkins}, Edward B. and {Prochaska}, J. Xavier and {Augustin}, Ramona and {Ji}, Suoqing and {Faucher-Gigu{\`e}re}, Claude-Andr{\'e} and {Hafen}, Zachary and {Peeples}, Molly S. and {Barger}, Kat A. and {Berg}, Michelle A. and {Bordoloi}, Rongmon and {Brown}, Thomas M. and {Fox}, Andrew J. and {Gilbert}, Karoline M. and {Guhathakurta}, Puragra and {Kalirai}, Jason S. and {Lockman}, Felix J. and {O'Meara}, John M. and {Pisano}, D.~J. and {Ribaudo}, Joseph and {Werk}, Jessica K.},
        title = "{Project AMIGA: The Circumgalactic Medium of Andromeda}",
      journal = {\apj},
         year = 2020,
        month = sep,
       volume = {900},
       number = {1},
          eid = {9},
        pages = {9},
          doi = {10.3847/1538-4357/aba49c},
archivePrefix = {arXiv},
       eprint = {2002.07818},
 primaryClass = {astro-ph.GA},
       adsurl = {https://ui.adsabs.harvard.edu/abs/2020ApJ...900....9L}
}

@ARTICLE{Lochhaas2021,
       author = {{Lochhaas}, Cassandra and {Tumlinson}, Jason and {O'Shea}, Brian W. and {Peeples}, Molly S. and {Smith}, Britton D. and {Werk}, Jessica K. and {Augustin}, Ramona and {Simons}, Raymond C.},
        title = "{Figuring Out Gas \& Galaxies In Enzo (FOGGIE). V. The Virial Temperature Does Not Describe Gas in a Virialized Galaxy Halo}",
      journal = {\apj},
         year = 2021,
        month = dec,
       volume = {922},
       number = {2},
          eid = {121},
        pages = {121},
          doi = {10.3847/1538-4357/ac2496},
archivePrefix = {arXiv},
       eprint = {2102.08393},
 primaryClass = {astro-ph.GA},
       adsurl = {https://ui.adsabs.harvard.edu/abs/2021ApJ...922..121L}
}

@ARTICLE{Lochhaas2023,
       author = {{Lochhaas}, Cassandra and {Tumlinson}, Jason and {Peeples}, Molly S. and {O'Shea}, Brian W. and {Werk}, Jessica K. and {Simons}, Raymond C. and {Juno}, James and {Kopenhafer}, Claire and {Augustin}, Ramona and {Wright}, Anna C. and {Acharyya}, Ayan and {Smith}, Britton D.},
        title = "{Figuring Out Gas \& Galaxies in Enzo (FOGGIE). VI. The Circumgalactic Medium of L $^{{\ensuremath{*}}}$ Galaxies Is Supported in an Emergent, Nonhydrostatic Equilibrium}",
      journal = {\apj},
         year = 2023,
        month = may,
       volume = {948},
       number = {1},
          eid = {43},
        pages = {43},
          doi = {10.3847/1538-4357/acbb06},
archivePrefix = {arXiv},
       eprint = {2206.09925},
 primaryClass = {astro-ph.GA},
       adsurl = {https://ui.adsabs.harvard.edu/abs/2023ApJ...948...43L}
}

@ARTICLE{Lochhaas2025,
       author = {{Lochhaas}, Cassandra and {Peeples}, Molly S. and {O'Shea}, Brian W. and {Tumlinson}, Jason and {Corlies}, Lauren and {Saeedzadeh}, Vida and {Lehner}, Nicolas and {Wright}, Anna C. and {Werk}, Jessica K. and {Trapp}, Cameron W. and {Augustin}, Ramona and {Acharyya}, Ayan and {Smith}, Britton D.},
        title = "{Figuring Out Gas \& Galaxies In Enzo (FOGGIE) XI: Circumgalactic O VI Emission Traces Clumpy Inflowing Recycled Gas}",
      journal = {arXiv e-prints},
         year = 2025,
        month = oct,
          eid = {arXiv:2510.25844},
        pages = {arXiv:2510.25844},
archivePrefix = {arXiv},
       eprint = {2510.25844},
 primaryClass = {astro-ph.GA},
       adsurl = {https://ui.adsabs.harvard.edu/abs/2025arXiv251025844L}
}

@ARTICLE{Nielsen2013,
       author = {{Nielsen}, Nikole M. and {Churchill}, Christopher W. and {Kacprzak}, Glenn G.},
        title = "{MAGIICAT II. General Characteristics of the Mg II Absorbing Circumgalactic Medium}",
      journal = {\apj},
         year = 2013,
        month = oct,
       volume = {776},
       number = {2},
          eid = {115},
        pages = {115},
          doi = {10.1088/0004-637X/776/2/115},
archivePrefix = {arXiv},
       eprint = {1211.1380},
 primaryClass = {astro-ph.CO},
       adsurl = {https://ui.adsabs.harvard.edu/abs/2013ApJ...776..115N}
}

@ARTICLE{1969ApJ...156L..63B,
       author = {{Bahcall}, John N. and {Spitzer}, Jr., Lyman},
        title = "{Absorption Lines Produced by Galactic Halos}",
      journal = {\apjl},
         year = 1969,
        month = may,
       volume = {156},
        pages = {L63},
          doi = {10.1086/180350},
       adsurl = {https://ui.adsabs.harvard.edu/abs/1969ApJ...156L..63B}
}

@ARTICLE{2016ApJ...833..259B,
       author = {{Borthakur}, Sanchayeeta and {Heckman}, Timothy and {Tumlinson}, Jason and {Bordoloi}, Rongmon and {Kauffmann}, Guinevere and {Catinella}, Barbara and {Schiminovich}, David and {Dav{\'e}}, Romeel and {Moran}, Sean M. and {Saintonge}, Amelie},
        title = "{The Properties of the Circumgalactic Medium in Red and Blue Galaxies: Results from the COS-GASS+COS-Halos Surveys}",
      journal = {\apj},
         year = 2016,
        month = dec,
       volume = {833},
       number = {2},
          eid = {259},
        pages = {259},
          doi = {10.3847/1538-4357/833/2/259},
archivePrefix = {arXiv},
       eprint = {1609.06308},
 primaryClass = {astro-ph.GA},
       adsurl = {https://ui.adsabs.harvard.edu/abs/2016ApJ...833..259B}
}

@ARTICLE{2019MNRAS.484.2257Z,
       author = {{Zahedy}, Fakhri S. and {Chen}, Hsiao-Wen and {Johnson}, Sean D. and {Pierce}, Rebecca M. and {Rauch}, Michael and {Huang}, Yun-Hsin and {Weiner}, Benjamin J. and {Gauthier}, Jean-Ren{\'e}},
        title = "{Characterizing circumgalactic gas around massive ellipticals at z {\ensuremath{\sim}} 0.4 - II. Physical properties and elemental abundances}",
      journal = {\mnras},
         year = 2019,
        month = apr,
       volume = {484},
       number = {2},
        pages = {2257-2280},
          doi = {10.1093/mnras/sty3482},
archivePrefix = {arXiv},
       eprint = {1809.05115},
 primaryClass = {astro-ph.GA},
       adsurl = {https://ui.adsabs.harvard.edu/abs/2019MNRAS.484.2257Z}
}

@ARTICLE{Peeples2019,
       author = {{Peeples}, Molly S. and {Corlies}, Lauren and {Tumlinson}, Jason and
         {O'Shea}, Brian W. and {Lehner}, Nicolas and {O'Meara}, John M. and
         {Howk}, J. Christopher and {Earl}, Nicholas and {Smith}, Britton D. and
         {Wise}, John H. and {Hummels}, Cameron B.},
        title = "{Figuring Out Gas \& Galaxies in Enzo (FOGGIE). I. Resolving Simulated Circumgalactic Absorption at 2 {\ensuremath{\leq}} z {\ensuremath{\leq}} 2.5}",
      journal = {\apj},
         year = 2019,
        month = mar,
       volume = {873},
       number = {2},
          eid = {129},
        pages = {129},
          doi = {10.3847/1538-4357/ab0654},
archivePrefix = {arXiv},
       eprint = {1810.06566},
 primaryClass = {astro-ph.GA},
       adsurl = {https://ui.adsabs.harvard.edu/abs/2019ApJ...873..129P}
}

@ARTICLE{Planck2016,
       author = {{Planck Collaboration} and {Ade}, P.~A.~R. and {Aghanim}, N. and {Arnaud}, M. and {Ashdown}, M. and {Aumont}, J. and {Baccigalupi}, C. and {Banday}, A.~J. and {Barreiro}, R.~B. and {Bartlett}, J.~G. and {Bartolo}, N. and {Battaner}, E. and {Battye}, R. and {Benabed}, K. and {Beno{\^\i}t}, A. and {Benoit-L{\'e}vy}, A. and {Bernard}, J. -P. and {Bersanelli}, M. and {Bielewicz}, P. and {Bock}, J.~J. and {Bonaldi}, A. and {Bonavera}, L. and {Bond}, J.~R. and {Borrill}, J. and {Bouchet}, F.~R. and {Boulanger}, F. and {Bucher}, M. and {Burigana}, C. and {Butler}, R.~C. and {Calabrese}, E. and {Cardoso}, J. -F. and {Catalano}, A. and {Challinor}, A. and {Chamballu}, A. and {Chary}, R. -R. and {Chiang}, H.~C. and {Chluba}, J. and {Christensen}, P.~R. and {Church}, S. and {Clements}, D.~L. and {Colombi}, S. and {Colombo}, L.~P.~L. and {Combet}, C. and {Coulais}, A. and {Crill}, B.~P. and {Curto}, A. and {Cuttaia}, F. and {Danese}, L. and {Davies}, R.~D. and {Davis}, R.~J. and {de Bernardis}, P. and {de Rosa}, A. and {de Zotti}, G. and {Delabrouille}, J. and {D{\'e}sert}, F. -X. and {Di Valentino}, E. and {Dickinson}, C. and {Diego}, J.~M. and {Dolag}, K. and {Dole}, H. and {Donzelli}, S. and {Dor{\'e}}, O. and {Douspis}, M. and {Ducout}, A. and {Dunkley}, J. and {Dupac}, X. and {Efstathiou}, G. and {Elsner}, F. and {En{\ss}lin}, T.~A. and {Eriksen}, H.~K. and {Farhang}, M. and {Fergusson}, J. and {Finelli}, F. and {Forni}, O. and {Frailis}, M. and {Fraisse}, A.~A. and {Franceschi}, E. and {Frejsel}, A. and {Galeotta}, S. and {Galli}, S. and {Ganga}, K. and {Gauthier}, C. and {Gerbino}, M. and {Ghosh}, T. and {Giard}, M. and {Giraud-H{\'e}raud}, Y. and {Giusarma}, E. and {Gjerl{\o}w}, E. and {Gonz{\'a}lez-Nuevo}, J. and {G{\'o}rski}, K.~M. and {Gratton}, S. and {Gregorio}, A. and {Gruppuso}, A. and {Gudmundsson}, J.~E. and {Hamann}, J. and {Hansen}, F.~K. and {Hanson}, D. and {Harrison}, D.~L. and {Helou}, G. and {Henrot-Versill{\'e}}, S. and {Hern{\'a}ndez-Monteagudo}, C. and {Herranz}, D. and {Hildebrandt}, S.~R. and {Hivon}, E. and {Hobson}, M. and {Holmes}, W.~A. and {Hornstrup}, A. and {Hovest}, W. and {Huang}, Z. and {Huffenberger}, K.~M. and {Hurier}, G. and {Jaffe}, A.~H. and {Jaffe}, T.~R. and {Jones}, W.~C. and {Juvela}, M. and {Keih{\"a}nen}, E. and {Keskitalo}, R. and {Kisner}, T.~S. and {Kneissl}, R. and {Knoche}, J. and {Knox}, L. and {Kunz}, M. and {Kurki-Suonio}, H. and {Lagache}, G. and {L{\"a}hteenm{\"a}ki}, A. and {Lamarre}, J. -M. and {Lasenby}, A. and {Lattanzi}, M. and {Lawrence}, C.~R. and {Leahy}, J.~P. and {Leonardi}, R. and {Lesgourgues}, J. and {Levrier}, F. and {Lewis}, A. and {Liguori}, M. and {Lilje}, P.~B. and {Linden-V{\o}rnle}, M. and {L{\'o}pez-Caniego}, M. and {Lubin}, P.~M. and {Mac{\'\i}as-P{\'e}rez}, J.~F. and {Maggio}, G. and {Maino}, D. and {Mandolesi}, N. and {Mangilli}, A. and {Marchini}, A. and {Maris}, M. and {Martin}, P.~G. and {Martinelli}, M. and {Mart{\'\i}nez-Gonz{\'a}lez}, E. and {Masi}, S. and {Matarrese}, S. and {McGehee}, P. and {Meinhold}, P.~R. and {Melchiorri}, A. and {Melin}, J. -B. and {Mendes}, L. and {Mennella}, A. and {Migliaccio}, M. and {Millea}, M. and {Mitra}, S. and {Miville-Desch{\^e}nes}, M. -A. and {Moneti}, A. and {Montier}, L. and {Morgante}, G. and {Mortlock}, D. and {Moss}, A. and {Munshi}, D. and {Murphy}, J.~A. and {Naselsky}, P. and {Nati}, F. and {Natoli}, P. and {Netterfield}, C.~B. and {N{\o}rgaard-Nielsen}, H.~U. and {Noviello}, F. and {Novikov}, D. and {Novikov}, I. and {Oxborrow}, C.~A. and {Paci}, F. and {Pagano}, L. and {Pajot}, F. and {Paladini}, R. and {Paoletti}, D. and {Partridge}, B. and {Pasian}, F. and {Patanchon}, G. and {Pearson}, T.~J. and {Perdereau}, O. and {Perotto}, L. and {Perrotta}, F. and {Pettorino}, V. and {Piacentini}, F. and {Piat}, M. and {Pierpaoli}, E. and {Pietrobon}, D. and {Plaszczynski}, S. and {Pointecouteau}, E. and {Polenta}, G. and {Popa}, L. and {Pratt}, G.~W. and {Pr{\'e}zeau}, G.},
        title = "{Planck 2015 results. XIII. Cosmological parameters}",
      journal = {\aap},
         year = 2016,
        month = sep,
       volume = {594},
          eid = {A13},
        pages = {A13},
          doi = {10.1051/0004-6361/201525830},
archivePrefix = {arXiv},
       eprint = {1502.01589},
 primaryClass = {astro-ph.CO},
       adsurl = {https://ui.adsabs.harvard.edu/abs/2016A&A...594A..13P}
}

@ARTICLE{Simons2020,
       author = {{Simons}, Raymond C. and {Peeples}, Molly S. and {Tumlinson}, Jason and {O'Shea}, Brian W. and {Smith}, Britton D. and {Corlies}, Lauren and {Lochhaas}, Cassandra and {Zheng}, Yong and {Augustin}, Ramona and {Prasad}, Deovrat and {Snyder}, Gregory F. and {Tollerud}, Erik},
        title = "{Figuring Out Gas \& Galaxies in Enzo (FOGGIE). IV. The Stochasticity of Ram Pressure Stripping in Galactic Halos}",
      journal = {\apj},
         year = 2020,
        month = dec,
       volume = {905},
       number = {2},
          eid = {167},
        pages = {167},
          doi = {10.3847/1538-4357/abc5b8},
archivePrefix = {arXiv},
       eprint = {2004.14394},
 primaryClass = {astro-ph.GA},
       adsurl = {https://ui.adsabs.harvard.edu/abs/2020ApJ...905..167S}
}

@ARTICLE{Simons2024,
       author = {{Simons}, Raymond C. and {Peeples}, Molly S. and {Tumlinson}, Jason and {O'Shea}, Brian W. and {Lochhaas}, Cassandra and {Wright}, Anna C. and {Acharyya}, Ayan and {Augustin}, Ramona and {Hamilton-Campos}, Kathleen A. and {Smith}, Britton D. and {Lehner}, Nicolas and {Werk}, Jessica K. and {Zheng}, Yong},
        title = "{Figuring Out Gas \& Galaxies in Enzo (FOGGIE). IX: The Angular Momentum Evolution of Milky Way-like Galaxies and their Circumgalactic Gas}",
      journal = {arXiv e-prints},
         year = 2024,
        month = sep,
          eid = {arXiv:2409.17244},
        pages = {arXiv:2409.17244},
          doi = {10.48550/arXiv.2409.17244},
archivePrefix = {arXiv},
       eprint = {2409.17244},
 primaryClass = {astro-ph.GA},
       adsurl = {https://ui.adsabs.harvard.edu/abs/2024arXiv240917244S}
}

@ARTICLE{Smith2009,
       author = {{Smith}, Britton D. and {Turk}, Matthew J. and {Sigurdsson}, Steinn and {O'Shea}, Brian W. and {Norman}, Michael L.},
        title = "{Three Modes of Metal-Enriched Star Formation in the Early Universe}",
      journal = {\apj},
         year = 2009,
        month = jan,
       volume = {691},
       number = {1},
        pages = {441-451},
          doi = {10.1088/0004-637X/691/1/441},
archivePrefix = {arXiv},
       eprint = {0806.1653},
 primaryClass = {astro-ph},
       adsurl = {https://ui.adsabs.harvard.edu/abs/2009ApJ...691..441S}
}

@ARTICLE{Smith2017,
       author = {{Smith}, Britton D. and {Bryan}, Greg L. and {Glover}, Simon C.~O. and {Goldbaum}, Nathan J. and {Turk}, Matthew J. and {Regan}, John and {Wise}, John H. and {Schive}, Hsi-Yu and {Abel}, Tom and {Emerick}, Andrew and {O'Shea}, Brian W. and {Anninos}, Peter and {Hummels}, Cameron B. and {Khochfar}, Sadegh},
        title = "{GRACKLE: a chemistry and cooling library for astrophysics}",
      journal = {\mnras},
         year = 2017,
        month = apr,
       volume = {466},
       number = {2},
        pages = {2217-2234},
          doi = {10.1093/mnras/stw3291},
archivePrefix = {arXiv},
       eprint = {1610.09591},
 primaryClass = {astro-ph.CO},
       adsurl = {https://ui.adsabs.harvard.edu/abs/2017MNRAS.466.2217S}
}

@ARTICLE{Tumlinson2011,
       author = {{Tumlinson}, J. and {Thom}, C. and {Werk}, J.~K. and {Prochaska}, J.~X. and
         {Tripp}, T.~M. and {Weinberg}, D.~H. and {Peeples}, M.~S. and
         {O'Meara}, J.~M. and {Oppenheimer}, B.~D. and {Meiring}, J.~D. and
         {Katz}, N.~S. and {Dav{\'e}}, R. and {Ford}, A.~B. and {Sembach}, K.~R.},
        title = "{The Large, Oxygen-Rich Halos of Star-Forming Galaxies Are a Major Reservoir of Galactic Metals}",
      journal = {Science},
         year = 2011,
        month = nov,
       volume = {334},
       number = {6058},
        pages = {948},
          doi = {10.1126/science.1209840},
archivePrefix = {arXiv},
       eprint = {1111.3980},
 primaryClass = {astro-ph.CO},
       adsurl = {https://ui.adsabs.harvard.edu/abs/2011Sci...334..948T}
}

@ARTICLE{Tumlinson2013,
       author = {{Tumlinson}, Jason and {Thom}, Christopher and {Werk}, Jessica K. and
         {Prochaska}, J. Xavier and {Tripp}, Todd M. and {Katz}, Neal and
         {Dav{\'e}}, Romeel and {Oppenheimer}, Benjamin D. and
         {Meiring}, Joseph D. and {Ford}, Amanda Brady and {O'Meara}, John M. and
         {Peeples}, Molly S. and {Sembach}, Kenneth R. and {Weinberg}, David H.},
        title = "{The COS-Halos Survey: Rationale, Design, and a Census of Circumgalactic Neutral Hydrogen}",
      journal = {\apj},
         year = 2013,
        month = nov,
       volume = {777},
       number = {1},
          eid = {59},
        pages = {59},
          doi = {10.1088/0004-637X/777/1/59},
archivePrefix = {arXiv},
       eprint = {1309.6317},
 primaryClass = {astro-ph.CO},
       adsurl = {https://ui.adsabs.harvard.edu/abs/2013ApJ...777...59T}
}

@ARTICLE{Tumlinson2017,
       author = {{Tumlinson}, Jason and {Peeples}, Molly S. and {Werk}, Jessica K.},
        title = "{The Circumgalactic Medium}",
      journal = {\araa},
         year = 2017,
        month = aug,
       volume = {55},
       number = {1},
        pages = {389-432},
          doi = {10.1146/annurev-astro-091916-055240},
archivePrefix = {arXiv},
       eprint = {1709.09180},
 primaryClass = {astro-ph.GA},
       adsurl = {https://ui.adsabs.harvard.edu/abs/2017ARA&A..55..389T}
}

@ARTICLE{vandeVoort2019,
       author = {{van de Voort}, Freeke and {Springel}, Volker and {Mandelker}, Nir and {van den Bosch}, Frank C. and {Pakmor}, R{\"u}diger},
        title = "{Cosmological simulations of the circumgalactic medium with 1 kpc resolution: enhanced H I column densities}",
      journal = {\mnras},
         year = 2019,
        month = jan,
       volume = {482},
       number = {1},
        pages = {L85-L89},
          doi = {10.1093/mnrasl/sly190},
archivePrefix = {arXiv},
       eprint = {1808.04369},
 primaryClass = {astro-ph.GA},
       adsurl = {https://ui.adsabs.harvard.edu/abs/2019MNRAS.482L..85V}
}

@ARTICLE{Werk2014,
       author = {{Werk}, Jessica K. and {Prochaska}, J. Xavier and {Tumlinson}, Jason and
         {Peeples}, Molly S. and {Tripp}, Todd M. and {Fox}, Andrew J. and
         {Lehner}, Nicolas and {Thom}, Christopher and {O'Meara}, John M. and
         {Ford}, Amanda Brady and {Bordoloi}, Rongmon and {Katz}, Neal and
         {Tejos}, Nicolas and {Oppenheimer}, Benjamin D. and {Dav{\'e}}, Romeel and
         {Weinberg}, David H.},
        title = "{The COS-Halos Survey: Physical Conditions and Baryonic Mass in the Low-redshift Circumgalactic Medium}",
      journal = {\apj},
         year = 2014,
        month = sep,
       volume = {792},
       number = {1},
          eid = {8},
        pages = {8},
          doi = {10.1088/0004-637X/792/1/8},
archivePrefix = {arXiv},
       eprint = {1403.0947},
 primaryClass = {astro-ph.CO},
       adsurl = {https://ui.adsabs.harvard.edu/abs/2014ApJ...792....8W}
}

@ARTICLE{Wright2024,
       author = {{Wright}, Anna C. and {Tumlinson}, Jason and {Peeples}, Molly S. and {O'Shea}, Brian W. and {Lochhaas}, Cassandra and {Corlies}, Lauren and {Smith}, Britton D. and {Binh}, Nguyen and {Augustin}, Ramona and {Simons}, Raymond C.},
        title = "{Figuring Out Gas and Galaxies in Enzo (FOGGIE). VII. The (Dis)assembly of Stellar Halos}",
      journal = {\apj},
         year = 2024,
        month = jul,
       volume = {970},
       number = {1},
          eid = {70},
        pages = {70},
          doi = {10.3847/1538-4357/ad49a3},
archivePrefix = {arXiv},
       eprint = {2309.10039},
 primaryClass = {astro-ph.GA},
       adsurl = {https://ui.adsabs.harvard.edu/abs/2024ApJ...970...70W}
}

@ARTICLE{Zheng2020,
       author = {{Zheng}, Yong and {Peeples}, Molly S. and {O'Shea}, Brian W. and
         {Simons}, Raymond C. and {Lochhaas}, Cassandra and {Corlies}, Lauren and
         {Tumlinson}, Jason and {Smith}, Britton D. and {Augustin}, Ramona},
        title = "{Figuring Out Gas \& Galaxies in Enzo (FOGGIE). III. The Mocky Way: Investigating Biases in Observing the Milky Way's Circumgalactic Medium}",
      journal = {\apj},
         year = 2020,
        month = jun,
       volume = {896},
       number = {2},
          eid = {143},
        pages = {143},
          doi = {10.3847/1538-4357/ab960a},
archivePrefix = {arXiv},
       eprint = {2001.07736},
 primaryClass = {astro-ph.GA},
       adsurl = {https://ui.adsabs.harvard.edu/abs/2020ApJ...896..143Z}
}
\bibliographystyle{aasjournalv7}

\end{document}